\documentclass{esub2acm}
\usepackage{latexsym}
\usepackage{color}
\usepackage{xspace} 
\usepackage{epsfig}
\usepackage{array}
\usepackage[mathscr]{eucalr}
\usepackage{url} 
\usepackage{amsmath,amscd,amssymb}

\usepackage[english]{varioref}

{\obeyspaces %
\gdef\sepspaces{\def {\ }}}
\def\EX{%
\parindent=0pt\parskip=0pt\tt\obeylines\obeyspaces\sepspaces}

{\ignorespaces\EX}%
{}

\newcommand{\hermes}{\textsl{HERMES}\index{HERMES@\textsl{HERMES}}\xspace}
\newcommand{\impact}{\textsl{IMPACT}\index{IMPACT@\textsl{IMPACT}}\xspace}

\newcommand{\tsimmis}{\textsl{TSIMMIS}\index{TSIMMIS@\textsl{TSIMMIS}}\xspace}
\newcommand{\sims}{\textsl{SIMS}\index{SIMS@\textsl{SIMS}}\xspace}

\renewcommand{\iff}{\textit{if and only if}\xspace}
\newcommand{\iffdef}{\xspace\textit{iff}\xspace}

\newcommand{\defeq}{\ensuremath{=_{\mathit{def}}}\xspace}

\newcommand{\true}{\ensuremath{\protect\mathbf{true}}\xspace}

\newcommand{\ag}[1]{\ensuremath{\protect\mathscr{#1}\xspace}}

\newcommand{\type}[1]{{\ensuremath{\mathtt{#1}}}\xspace}

\newcommand{\var}[1]{\ensuremath{\protect\mathtt{#1}}\xspace} 
\renewcommand{\root}[1]{\ensuremath{\mathit{root(\mathtt{#1})}}\xspace} 
\newcommand{\cc}[3]{\ensuremath{{\ag{#1}}\mathit{\,:\,}\mathit{#2}(#3)}\xspace}                                                            

\newcommand{\IN}[2]{\ensuremath{\textrm{\normalfont\bfseries{in}}\protect\mathbf{(}\protect\mathtt{#1}\protect\mathbf{,\:}\protect\mathtt{#2}\protect\mathbf{)}}}

\newcommand{\str}[1]{\text{{\EX "}#1{\EX "}}}

\newcommand{\op}{\textsf{Op}\index{\textsf{Op}}\index{Op@\textsf{Op}}\xspace}

\newcommand{\agprog}{\ensuremath{\protect\mathcal{P}}\index{P@\ensuremath{\protect\mathcal{P}}}\index{\ensuremath{\protect\mathcal{P}}}\xspace}

\newcommand{\trans}[1]{\ensuremath{\protect\mathfrak{Trans}({#1})}\index{\ensuremath{\protect\mathfrak{Trans}}}\index{Trans@\ensuremath{\protect\mathfrak{Trans}}}\xspace}

\newcommand{\transz}{\ensuremath{\protect\mathfrak{Trans}}\index{\ensuremath{\protect\mathfrak{Trans}}}\index{Trans@\ensuremath{\protect\mathfrak{Trans}}}\xspace}

\newcommand{\nop}[1]{{}}

\newcommand{\hide}[1]{}

\newcommand{\NP}{\ensuremath{\protect\mathrm{NP}}\index{\ensuremath{\protect\mathrm{NP}}}\index{NP@\ensuremath{\protect\mathrm{NP}}}\xspace}

\newcommand{\coNP}{\ensuremath{\protect\mathrm{co\textrm{-}NP}}\index{\ensuremath{\protect\mathrm{co\textrm{-}NP}}}\index{co-NP@\ensuremath{\protect\mathrm{co\textrm{-}NP}}}\xspace}

\newcommand{\ie}{\textsf{ie}\xspace}
\newcommand{\iv}{\textsf{inv}\xspace}
\newcommand{\ic}{\textsf{ic}\xspace}
\newcommand{\INS}{\textsl{INS}\xspace}
\newcommand{\cost}{\textbf{cost}\xspace}
\newcommand{\gain}{\textbf{gain}\xspace}
\newcommand{\coacost}{\textbf{coalesced\_cost}\xspace}
\newcommand{\weight}{\textsl{weight}\xspace}
\newcommand{\eval}{\textsl{eval}\xspace}
\newcommand{\ccost}{\textsl{\ensuremath{\mathcal{C}\_cost}}\xspace}
\newcommand{\fincost}{\textsl{fin\_cost}\xspace}
\newcommand{\ccccost}{\textsl{IdCom\_cost}\xspace}
\newcommand{\mergecost}{\textsl{Merge\_cost}\xspace}

\newcommand{\alg}[1]{
\renewcommand{\baselinestretch}{1}
\begin{center}
\begin{tabular}[htb]{c}
\fcolorbox{black}{mygray}{
\begin{minipage}{1.0\textwidth}
\smallskip
#1

\end{minipage}
}
\end{tabular}
\end{center}
}

\newcommand{\NIL}{\textbf{NIL}\xspace}

\newcommand{\cali}{\ensuremath{\mathcal{I}}\xspace}
\newcommand{\kw}[1]{\underline{\textbf{#1}}}
\def\pq{{\phantom.\hspace{0.1in}}}
\def\pqq{{\phantom.\hspace{0.1in}\hspace{0.1in}}}
\def\pqqq{{\phantom.\hspace{0.1in}\hspace{0.1in}\hspace{0.1in}}}
\def\pqqqq{{\phantom.\hspace{0.1in}\hspace{0.1in}\hspace{0.1in}\hspace{0.1in}}}
\def\pqqqqq{{\phantom.\hspace{0.1in}\hspace{0.1in}\hspace{0.1in}\hspace{0.1in}\hspace{0.1in}}}

\newcommand\chkimp{\textbf{Chk\_Imp}\xspace}
\newcommand\chktaut{\textbf{Chk\_Taut}\xspace}
\newcommand\chkent{\textbf{Chk\_Ent}\xspace}
\definecolor{mygray}{gray}{0.85}

\newtheorem{theorem}{Theorem}[section]

\newtheorem{corollary}[theorem]{Corollary}
\newtheorem{definition}[theorem]{Definition}
\newtheorem{proposition}[theorem]{Proposition}
\newtheorem{lemma}[theorem]{Lemma}
\newtheorem{Remark}{Remark}[section]
\newtheorem{Convention}{Convention}[section]

\newtheorem{Example}[theorem]{Example}

\newenvironment{example}[0]{\begin{Example}\rm}{\end{Example}}

\newenvironment{convention}[0]{\begin{Convention}\rm}
        {\end{Convention}}

\begin{document}

\title{Improving Performance of Heterogeneous Agents}
  
\author{Fatma \"{O}zcan, V.S.~Subrahmanian\\
  University of Maryland, Dept. of CS\\
  College Park, MD 20752, USA\\ $\{$fatma,vs$\}$@cs.umd.edu \and
  J\"{u}rgen Dix\\ The University of Manchester, Dept.~of CS\\ 
Oxford Road, Manchester, M13 9PL, UK\\ dix@cs.man.ac.uk}

\begin{abstract}
  With the increase in agent-based applications, there are now agent
  systems that support \emph{concurrent} client accesses. The ability
  to process large volumes of simultaneous requests is critical in
  many such applications.  In such a setting, the traditional approach
  of serving these requests one at a time via queues (e.g.
  \textsf{FIFO} queues, priority queues) is insufficient.  Alternative
  models are essential to improve the performance of such
  \emph{heavily loaded} agents. In this paper, we propose a set of
  \emph{cost-based algorithms} to \emph{optimize} and \emph{merge}
  multiple requests submitted to an agent. In order to merge a set of
  requests, one first needs to identify commonalities among such
  requests.  First, we provide an \emph{application independent
    framework} within which an agent developer may specify
  relationships (called \emph{invariants}) between requests.  Second,
  we provide two algorithms (and various accompanying heuristics)
  which allow an agent to automatically rewrite requests so as to
  avoid redundant work---these algorithms take invariants associated
  with the agent into account.  Our algorithms  are independent
  of any specific agent framework.  For an implementation, we 
  implemented both these algorithms on top of the \impact agent development
  platform, and on top of a (non-\impact) geographic database agent.
  Based on these implementations, we conducted experiments and show
  that our algorithms are considerably more efficient than methods
  that use the $A^*$ algorithm.
\end{abstract}


\begin{bottomstuff}
  The first and third authors gratefully acknowledge support from the Army
  Research Laboratory 
  under contract number DAAL01-97-K0135, and by DARPA/AFRL under
  grant number  F306029910552.
\end{bottomstuff}

\markboth{F. \"{O}zcan, V.S.~Subrahmanian and J. Dix}
     {Improving Performance of Heavily-Loaded Agents}

\maketitle

\section{Motivation and Introduction}
\label{sec:intro}
A \emph{heavily loaded} agent is one that experiences a large volume
of service requests and/or has a large number of conditions to track
on behalf of various users.  The traditional model for servicing
requests is via one kind of queue or the other (e.g. \textsf{FIFO},
\textsf{LIFO}, priority queue, etc.). For instance, a company may
deploy a \emph{PowerPoint} agent \ag{ppt} that automatically creates
PowerPoint presentations for different users based on criteria they
have registered earlier. The finance director may get the latest
budget data presented to him, a shop worker may get information on the
latest work schedules for him, and the CEO
may get information on stock upheavals.

If the \ag{ppt} agent has thousands of such presentations to create
for different users, it may well choose to exploit ``redundancies''
among the various requests to enhance its own performance. Hence,
rather than sequentially creating a presentation for the CEO, then one
for the finance director, then one for the marketing manager, then one
for the shop manager, etc., it may notice that the finance director
and CEO both want some relevant financial data---this data can be
accessed and a PowerPoint page created for it once, instead of twice.
Likewise, a heterogeneous database agent \ag{hdb} tracking inventory
information for thousands of users may well wish to exploit the
commonality between queries such as 
\begin{quote}
\emph{Find all suppliers who can
  provide 1000 automobile engines by June 25, 2003} and \emph{Find all
  suppliers who can provide 1500 VX2 automobile engines by June  21,
  2003.}
\end{quote}  In this case, the latter query can be executed by using the
answer returned by the first query, rather than by executing the
second query from scratch.  This may be particularly valuable when the
\ag{hdb} agent has to access multiple remote supplier databases---by
leveraging the \emph{common aspects} of such requests, the \ag{hdb}
agent can greatly reduce load on the network and the time taken to
jointly process these two requests.

The same problem occurs in yet another context.  \cite{subetal99}
have described a framework called \impact within which software
agents may be built on top of arbitrary data structures and software
packages.  In their framework, an agent manipulates a set of data
structures (including a message box) via a set of well defined
functions.  The state of the agent at a given point in time consists
of a set of objects in the agent's data structures.  The agent also
has a set of integrity constraints.  When the agent state changes
(this may happen if a message is received from another agent, a shared
workspace is written by another agent or entity, a clock tick occurs,
etc.), the agent must take some actions that cause the state to again
be consistent with the integrity constraints. Hence, each agent has an
associated set of actions (with the usual preconditions and effects),
and an \emph{agent program} which specifies under what conditions an
agent is permitted to take an action, under what conditions it is
obliged to take an action, under what conditions it is forbidden from
taking an action, and under what conditions an action is in fact
taken.  \cite{esr99} have shown how (under some restrictions) such an
agent program may be \emph{compiled} into a set of conditions to be
evaluated at run-time over the agent's state.  When the agent state
changes, then for each action $\alpha$, one such condition needs to be
evaluated over the state in order to determine which instances of that
action (if any) need to be performed.  Hence, numerous such conditions
need to be simultaneously evaluated so that the agent can decide what
actions to take so as to restore consistency of the state with the
integrity constraints.

Therefore, in this paper, we consider the following technical problem.
Suppose an agent is built on top of heterogeneous data structures
(e.g.  using methods such as those described in various agent
frameworks such as
\cite{eite-etal-99a,subetal99,dixsubpic99,dixkrasub00,dixnansub00}).
\begin{quote}
  \emph{Suppose the agent is confronted with a set $S$ of requests.
    How should the agent process these requests so as to reduce the
    overall load on itself?}
\end{quote}
In the case of the \ag{ppt} agent for example, this capability will
allow the agent to recognize the fact that many presentations
requested by different clients require common financial data to be
computed and/or analyzed, and hence, performing this \emph{once}
instead of \emph{many times} will most certainly enhance performance.
Likewise, in the case of the \ag{hdb} agent, merging the two queries
about automobile engines presented earlier allows the agent to reduce
load on itself, thus allowing it to respond to other queries faster
than by queuing.

The paper is organized as follows: First, we provide the basic
definitions and some preliminary results that will be employed
throughout the paper in Section~\ref{sec:prelim}. Then, we present our
architecture in Section~\ref{sec:arch}. In Sections
\ref{sec:development} and~\ref{sec:deployment}, we discuss the
development phase and the deployment phase components, respectively.
The experiments are discussed in Section~\ref{sec:exps}.  Finally,
Section~\ref{sec:related} presents related work and Section
\ref{sec:conc} concludes the paper.
 
 \section{Preliminaries}
\label{sec:prelim}
All agents manipulate some set ${\cal T}$ of data types and manipulate
these types via some set of functions (application program interface
functions).  The input/output types of functions are known.  If \ag{d}
is the name of a data structure (or even a software package), and $f$
is an $n$-ary function defined in that package, then
\[\cc{d}{f}{a_1,\ldots ,a_n}\] is a \emph{code call}.  This code call says
\begin{quote}\emph{Execute function $f$ as defined in data structure/package \ag{d}
    on the stated list of arguments.}
\end{quote}
We assume this code call returns as output, a \emph{set} of
objects---if an atomic object is returned, it can be coerced into a
set.  For instance, if we consider a commonly used data structure
called a \emph{quad-tree} \cite{sa89} for geographic reasoning,
\cc{quadtree}{range}{\langle20,30\rangle ,\var{T},40)} may be a code call that
says \emph{find all objects in the quadtree the root of which is
  pointed to by $\var{T}$ which are within 40 units of location
  $\langle20,30\rangle$}---this query returns a set of points.

An \emph{atomic code call condition} is an expression of the form
\[\IN{\textit{t}}{\cc{d}{f}{\textit{a}_1,\ldots ,\textit{a}_n}}\] which
succeeds if $\textit{t}$ is in the set of answers returned by the code
call in question.  For example,
$\IN{\textit{t}}{\cc{excel}{chart}{\textit{excelFile}, \textit{rec},
    \textit{date}}}$ is an atomic code call condition that succeeds if
$\textit{t}$ is a chart plotting \textit{rec} with respect to
\textit{date} in the \textit{excelFile}.

We assume that for each type $\tau$ manipulated by the agent, there is a
set $\root{\tau}$ of ``root'' variable symbols ranging over $\tau$.  In
addition, suppose $\tau$ is a complex record type having fields
$\type{f_1},\ldots,\type{f_n}$.  Then, for every variable
$\var{X}$ of type $\tau$, we
require that $\var{X.f_i}$ be a variable of type $\tau_i$ where $\tau_i$
is the type of field $\type{f_i}$.  In the same vein, if $\type{f_i}$
itself has a sub-field $\type{g}$ of type $\gamma$, then $\var{X.f_i.g}$
is a variable of type $\gamma$, and so on.  The variables, $\var{X.f_i}$,
$\var{X.f_i.g}$, etc.~are called \emph{path variables}. For any path
variable $\var{Y}$ of the form $\var{X.path}$, where $\var{X}$ is a
root variable, we refer to $\var{X}$ as the root of $\var{Y}$, denoted
by \root{Y}; for technical convenience, \root{X}, where $\var{X}$ is a
root variable, refers to itself. If $S$ is a set of variables, then
\root{S} = $\{ \root{X} \mid \var{X} \in S \}$.

\begin{convention}From now on, we use lower case letters
  ($a, b, c, c_1, \ldots $) to denote constants and upper case letters
  ($\var{X}, \var{Y}, \var{Z}, \var{X_1}, \ldots$) to denote variables.
  When it is clear from context, we will also use lower case letters
  like $s, t$ as metavariables ranging over constants, variables or
  terms.
\end{convention}

A \emph{code call condition} (ccc) may now be defined as follows:
 \begin{enumerate}
 \item Every atomic code call condition is a code call condition.
 \item If ${s}$ and ${t}$ are either variables or objects,
   then ${s}={t}$ is an (equality) code call condition.
 \item If ${s}$ and ${t}$ are either integers/real valued
   objects, or are variables over the integers/reals, then ${s} <
   {t},\ {s} > {t},\ {s}\leq {t},$ and ${s}\geq
   {t}$ are (inequality) code call conditions.
 \item If $\chi_1$ and $\chi_2$ are code call conditions, then $\chi_1\,\&\,
   \chi_2$ is a code call condition.
\end{enumerate}
Code call conditions provide a simple, but powerful language syntax to
access heterogeneous data structures and legacy software code.
\begin{example}[Sample ccc] \label{ex1}
  The code call condition
 \[\begin{array}{l}
\IN{\var{FinanceRec}}{\cc{rel}{select}{\textit{financeRel},\textit{date},\str{=},\str{11/15/99}}}\
  \&\, \\
 \var{FinanceRec.sales} \geq 10K \ \&\\
 \IN{\var{C}}{\cc{excel}{chart}{\textit{excelFile},\var{FinanceRec}, \textit{day}}}\ \& \\
\IN{\var{Slide}}{\cc{ppt}{include}{\var{C},\str{presentation.ppt}}}
\end{array}\]
is a complex condition that accesses and merges data across a
relational database, an Excel file, and a PowerPoint file.  It first
selects all financial records associated with \str{11/15/99}: this is
done with the variable \var{FinanceRec} in the first line.  It then
filters out those records having sales more than $10K$ (second line).
Using the remaining records, an Excel chart is created with day of
sale on the $x$-axis and the resulting chart is included in the
PowerPoint file \str{presentation.ppt} (fourth line).
\end{example}
In the above example, it is very important that the first code call
be evaluable. If, for example the constant $financeRel$ were a
variable, then
\[\cc{rel}{select}{\var{FinanceRel},date,\str{=},\str{11/15/99}}\]
would not be evaluable, unless there were another condition
instantiating this variable.  In order to come up with a notion of
\emph{evaluability}, we need the following notion.
\begin{definition}[Dependent ccc's]
For an atomic code call condition of the form
  \IN{X_i}{cc_i} we define $\root{cc_i} =\{\root{Y} \mid \var{Y} ~\text{occurs in}
  ~cc_i\}$ and $\root{X_i} = \{ \root{Y}) \mid \var{Y} ~\text{occurs in}
  ~\var{X_i} \}$. For an (in-)equality code call condition
  $ccc_{\text{in/eq}}$
we define $\text{var}(ccc_{\text{in/eq}})= \{ \root{Y}) \mid \var{Y} ~\text{occurs in}
  ~ccc_{\text{in/eq}} \}$.
  
  A code call condition $\chi_j$ is said to be \emph{dependent on}
  $\chi_i$ \iffdef the following holds:
\begin{enumerate}
\item \textbf{Case 1:}  $\chi_i$ is of the form  \IN{X_i}{cc_i}.
\begin{enumerate}
\item If $\chi_j$ is an atomic code call condition of the form \IN{X_j}{
    cc_j} then\\ $\root{X_i} \subseteq \root{cc_j}$.
\item If $\chi_j$ is an equality or inequality code call condition of
    the form $\var{s_1} ~op ~\var{s_2}$, then either $\var{s_1}$ is a
    variable and $\root{s_1} \in \root{X_i}$ or $\var{s_2}$ is a
    variable and $\root{\var{s_2}} \in \root{X_i}$ or both.
\end{enumerate}
\item  \textbf{Case 2:} $\chi_i$ is an (in-)equality code call 
  condition.
\begin{enumerate}
\item If $\chi_j$ is an atomic code call condition of the form \IN{X_j}{
    cc_j} then\\ $\text{var}(\chi_i) \subseteq \root{cc_j}$.
\item If $\chi_j$ is an equality or inequality code call condition of
    the form $\var{s_1} ~op ~\var{s_2}$, then either $\var{s_1}$ is a
    variable and $\root{s_1} \in \text{var}(\chi_i)$ or $\var{s_2}$ is a
    variable and $\root{\var{s_2}} \in \text{var}(\chi_i)$ or both.
\end{enumerate}
\end{enumerate}
 \end{definition}

 \begin{example}[Dependency among ccc's]
   The  ccc $\chi_1:\var{FinanceRec.sales} \geq
   10K$ is dependent on the atomic code call condition
   \[\chi_2:\IN{\var{FinanceRec}}{\cc{rel}{select}{\textit{financeRel},
       \textit{date}, \str{=}, \str{11/15/99}}},\] because
   $\root{\var{FinanceRec.sales}} \in \root{\var{FinanceRec}}$.
   Similarly, the atomic code call condition
   $\chi_3:\IN{\var{C}}{\cc{excel}{chart}{\textit{excelFile}, \var{FinanceRec},
       \textit{day}}}$ is dependent on the atomic code call condition $\chi_2$,
   as the root variable $\var{FinanceRec}$ which appears as an
   argument in the code call of $\chi_3$ is instantiated in $\chi_2$.
\end{example}

\begin{definition}[Code Call Evaluation Graph (cceg) of a ccc]
  A \emph{code call evaluation graph} for a code call condition $\chi =
  \chi_1 \& ... \& \chi_n$, $n \geq 1$ where each $\chi_i$ is either an
  atomic, equality or inequality code call condition, is a directed
  graph $cceg(\chi)=(V,E)$ where:

 \begin{enumerate}
\item $V \defeq \{ \chi_i \mid 1 \leq i \leq n \}$,
\item $E \defeq \{ \langle \chi_i, \chi_j\rangle  \mid \chi_j ~\text{is ~dependent ~on $\chi_i$ ~and} ~1 \leq
  i \neq j \leq n \}$.
\end{enumerate}
 \end{definition}

\begin{figure}[htb]
 \begin{center}
\input{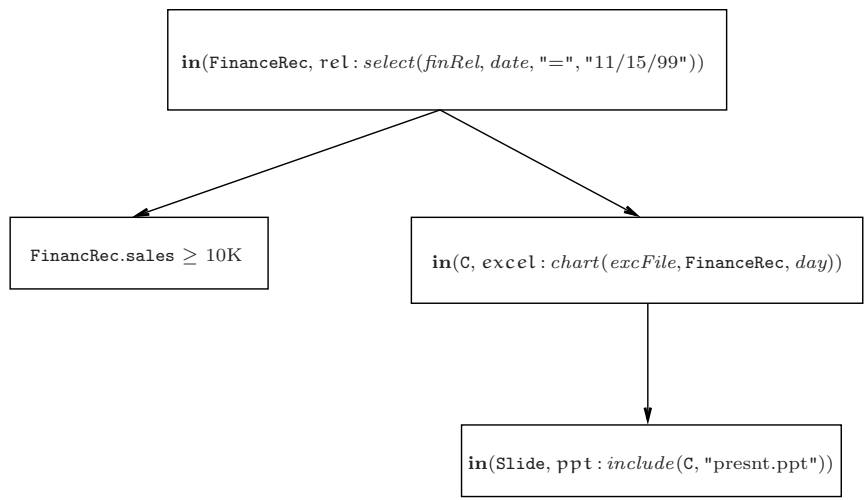}
\caption{The code call evaluation graph of Example~\ref{ex1}}
\label{cceg}
 \end{center}
 \end{figure}

 \begin{example}
   Figure~\ref{cceg} shows an example code call evaluation graph for
   the code call condition of Example~\ref{ex1}.
\end{example}
If $finRel$ were a variable \var{FinRel}, then the ccc would depend
on the equality ccc $\var{FinRel}=finRel$.

 Using the dependency relation on the constituents of a code call
 condition, we are now able to give a precise description of an
 \emph{evaluable} ccc.
 \begin{definition}[Evaluability of a ccc,
   \text{var$_{\text{base}}$}(ccc)]
\label{evaluable}
   A code call evaluation graph is \emph{evaluable} \iffdef 
\begin{enumerate}
\item It is acyclic.
\item For all nodes $\chi_i$, with in-degree $0$ the following holds:
\begin{enumerate}
\item If $\chi_i$ is an atomic code call condition of the form
  $\IN{\var{X_i}}{\cc{d}{f}{d_1,\ldots,d_n}}$, then each $d_i, 1 \leq i \leq
  n$, is ground.
\item If $\chi_i$ is an equality or inequality code call condition of
  the form $\var{s_1} ~op ~\var{s_2}$, then either $\var{s_1}$ or
  $\var{s_2}$ or both are constants.
 \end{enumerate}
\end{enumerate}

A code call condition ccc is \emph{evaluable} \iffdef it has an
evaluable code call evaluation graph.

For an evaluable ccc, we denote by $\text{var$_{\text{base}}$}(ccc)$
the set of all variables ocurring in nodes having in-degree $0$.  The
set $\text{var}(ccc)$ of all variables ocurring in ccc may be a
superset of $\text{var$_{\text{base}}$}(ccc)$.
 \end{definition}

 \begin{example}
   The code call evaluation graph in Figure~\ref{cceg} is evaluable
   because the atomic code call condition of the only node with
   in-degree 0 has ground arguments in its code call and it contains
   no cycles.
 \end{example}
 
 In~\cite{esr99} the notion of a \emph{safe} code call was defined to
 provide the necessary means to check if a given code call is
 evaluable. It defines a linear ordering of atomic, equality and
 inequality code calls within a given code call condition in such a
 way that when executed from left to right the code call condition is
 executable. Before tying our new notion of \emph{graph-evaluability}
 to the notion of \emph{safety}, we recapitulate the definition of
 safety from~\cite{esr99}.

 \begin{definition}[Safe Code Call (Condition)]
\label{safety}\mbox{}\\
A code call \cc{d}{f}{arg_1,\ldots, arg_n} is \emph{safe} \iffdef each
$arg_i$ is ground.  A code call condition $\chi_1\,\&\ldots \&\, \chi_n$, $n
\geq 1$, is \emph{safe} \iffdef there exists a permutation $\pi$ of
$\chi_1,\ldots,\chi_n$ such that for every $i=1,\ldots,n$ the following holds:

 \begin{enumerate}
 \item If $\chi_{\pi(i)}$ is an equality/inequality $\var{s_1}
   ~op~\var{s_2}$, then
\begin{itemize}
\item at least one of \var{s_1},\var{s_2} is a constant or a variable
  \var{X} such that $\root{\var{X}}$ belongs to $RV_{\pi(i)} = \{
  root(\var{Y}) \mid \exists j<i \text{~s.t. $Y$ ~occurs ~in} ~\chi_{\pi(j)}
  \}$;
\item if \var{s_i} is neither a constant nor a variable \var{X} such
  that $\root{X} \in RV_{\pi(i)}$, then \var{s_i} is a root variable.
\end{itemize} 
\item If $\chi_{\pi(i)}$ is an atomic code call condition of the form
  $\IN{\var{X_{\pi(i)}}}{cc_{\pi(i)}}$, then the root of each variable
  \var{Y} occurring in $\mathtt{cc}_{\pi(i)}$ belongs to $RV_{\pi(i)}$,
  and either \var{X_{\pi(i)}} is a root variable, or \root{X_{\pi(i)}}
  is from $RV_{\pi(i)}$.
\end{enumerate}
We call the permutation $\pi$ with the above properties a \emph{witness
to the safety}. 
\end{definition}

Intuitively, a code call is safe, if we can reorder the atomic code
call conditions occurring in it in a way such that we can evaluate
these atoms left to right, assuming that root variables are
incrementally bound to objects.

\begin{example}
  Consider the code call condition
   \[\begin{array}{l}
\IN{\var{FinanceRec}}{\cc{rel}{select}{\textit{financeRel}, data,
       \str{=}, \str{11/15/99}}}\ \&  \\
   \IN{\var{C}}{\cc{excel}{chart}{\textit{excelFile}, \var{FinanceRec},
       \textit{day}}}. 
\end{array}\]
This code call condition is safe as it meets both of the safety
requirements.  However, the following code call condition is not safe:
 \[\begin{array}{l}\IN{\var{FinanceRec}}{\cc{rel}{select}{\textit{financeRel}, data,
       \str{=}, \str{11/15/99}}}\ \& \\
   \IN{\var{C}}{\cc{excel}{chart}{\var{ExcelFile}, \var{FinanceRec},
       \textit{day}}}. 
\end{array}\]
This is because, there is no permutation of these two atomic code call
conditions which allows safety requirement 1 to be met for the
variable $\var{ExcelFile}$.
\end{example}

As a cceg is acyclic for any evaluable graph, ccegs determine a
partial ordering $\preceq$ on the $\chi_i$'s: \[\chi_i \preceq \chi_j \ \iff \ 
\langle\chi_i,\chi_j \rangle \in E.\] Hence, we may abuse notation and talk about
\emph{topological sorts}~\cite{knuth} of a graph to mean the
topological sort of the associated partial ordering.  Recall that
given a partially ordered set $(S,\leq)$, a topological sorting of that
set yields a linearly ordered set $(S,\preceq)$ such that $(\forall x,y\in S) x\leq
y\to x\preceq y$.  In the same vein, a topological sort of a directed
acyclic graph (dag) is a linear ordering of nodes in the graph, such
that if there exists an edge $\langle v_1, v_2\rangle$ in the graph, then
$v_1$ precedes $v_2$ in the topological sort.

 \begin{theorem}
\label{minwitness}
$\pi$ is a witness to the safety of $\chi$ \iff $\pi$ is a valid topological
sort of the cceg of $\chi$.
\end{theorem}

The algorithm \textbf{Create-cceg} (Figure~\ref{alg:cceg}) takes a code
call condition $\chi$ and creates an evaluable code call evaluation graph 
if $\chi$ is evaluable---otherwise it returns \NIL.

 \begin{figure}[htb]
\alg{
\par\noindent \textbf{Create-cceg}($\chi$) \\
\begin{tabbing}
/* nnnnnnn\=mmmmmmmmmmmmmmmmmmmmm\= \kill
/* \textbf{Input}:\>$\chi :\, \chi_1 \,\&\, \chi_2\, \&\, ...\, \&\, \chi_n$ \>*/ \\
/* \textbf{Output}:\> \NIL,  if $\chi$ is not evaluable \>*/ \\
/* \> a cceg $CCEG=(V,E)$, if $\chi$ is  evaluable \>*/ \\
\end{tabbing}
\begin{tabbing}
$LLL\,$\=:=\,M\,$\{M|$\,\,\=\kill
$L$\>:=\,$\{\chi_1, \chi_2,.., \chi_n\}$; \\
$L'$\>:=\,$\emptyset$; \\
$Var$\>:=\,$\emptyset$; \\
$E$\>:=\,$\emptyset$;\\
$V$\>:=\,$\{ \chi_i \mid$\>$ 1 \leq i \leq n \}$;\\
$Ok$\>:=\,$\{\chi_i\mid$\> $\chi_i$ is either of the form
\IN{\var{X}}{\cc{d}{f}{args}} where args is ground or\\
\>\>of the form $\var{s_1} ~op ~\var{s_2}$, where  either
  \var{s_1} or \var{s_2} or both are constants $\}$; \\
for all pairs $\langle \chi_i, \chi_j\rangle$, $\chi_i, \chi_j \in Ok$ such that $\chi_j$ is dependent on $\chi_i$\\ 
create an edge $\langle \chi_i,\chi_j\rangle $ and add it to $E$;\\
$Var$\>:=\,$Var\, \cup \,\{ \root{X_i} \mid \IN{\var{X_i}}{\cc{d}{f}{args}} \in Ok \}$; \\
$L$\>:=\,$L - Ok$;\\
$L'$\>:=\,$L'\, \cup\, Ok;$ \\
\kw{while} ($L$ is not empty) \kw{do} \\
\> $\Psi:=\,\{\chi_i\mid\ \,\chi_i \in L$ and all variables in $\chi_i$ are
in $Var$ and\\
\>\>\, \ $\exists \chi_j \in L` ~\text{such that $\chi_i$ depends on $\chi_j$} \}$; \\
\> \kw{if} card ($\Psi$) = 0 \kw{then} \kw{Return} \NIL; \\
\> \kw{else} \\
\> $Var :=\, Var \,\cup \,\{\root{X_i}\mid \IN{\var{X_i}}{\cc{d}{f}{args}}\in\Psi \}$; \\
\> for all pairs $\langle\chi_i, \chi_j\rangle$, $\chi_j \in \Psi$, such that 
   $\chi_j$ is dependent on $\chi_i \in L'$\\ 
\> create an edge $\langle\chi_i,\chi_j\rangle$ and add it to $E$;\\
\> $L \,:=\, L - \Psi$; \\
\> $L' :=\, L' \,\cup\, \Psi$;
\end{tabbing}
\kw{Return} $(V,E)$; \\
\kw{End-Algorithm} 
}
\caption{\textbf{Create-cceg} Algorithm}
\label{alg:cceg}
\end{figure}

The following example demonstrates the working of this algorithm for
the code call condition of Example~\ref{ex1}.

 \begin{example} 
   Let
\[\begin{array}{l}
\chi_1:
   \IN{\var{FinanceRec}}{\cc{rel}{select}{financialRel, \textit{date},
       \str{=}, \str{11/15/99}}},\\
\chi_2: \var{FinanceRec.sales} \geq
   10K,\\
\chi_3: \IN{\var{C}}{\cc{excel}{chart}{\textit{excelFile},
       \var{FinanceRec}, \textit{day}}}, \text{ and}\\
\chi_4:
   \IN{\var{Slide}}{\cc{ppt}{include}{\var{C},
       \str{presentation.ppt}}}. 
\end{array}\]
First, $L = \{\chi_1, \chi_2, \chi_3, \chi_4\}$, $L'=Var=E=\emptyset$.  We first
create a node for each of the four code call conditions.  $Ok =
\{\chi_1, \chi_2 \}$, as all arguments in the code call of $\chi_1$ are
ground, and 10K is a constant in $\chi_2$. Next, we create the edge
$(\chi_1,\chi_2)$. Because $\chi_2$ depends on $\chi_1$.  Then, $L= \{\chi_3,
\chi_4\}$, $L' = \{\chi_1, \chi_2 \}$ and $Var = \{\var{FinanceRec}\}$. In
the first iteration of the while loop $\Psi = \{\chi_3\}$ as $\chi_3$
depends on $\chi_1$, and all variables in $\chi_3$ (\var{FinanceRec}) are
in $Var$. $Var$ becomes $\{\var{FinanceRec}, \var{C}\}$ and we create
the edge $(\chi_1, \chi_3)$.  Now, $L = \{\chi_4\}$, $L'= \{ \chi_1, \chi_2,
\chi_3\}$.  In the second iteration of the while loop $\Psi = \{\chi_4\}$,
since $\chi_4$ depends on $\chi_3$ and all variables in $\chi_4$ (namely
$\{\var{C}\}$) are in $Var$.  This time, $Var$ becomes
$\{\var{FinanceRec}, \var{C}, \var{Slide}\}$, and we add the edge $\langle
\chi_3, \chi_4\rangle $ to the graph.  Now $L$ becomes the empty set and the
algorithm returns the code call evaluation graph given in
Figure~\ref{cceg}.
\end{example}

\begin{convention}Throughout the rest of this paper, we assume that
  all code call conditions considered are evaluable and that the graph
  associated with each code call condition has been generated.
\end{convention}

The \textbf{Create-cceg} algorithm runs in $O(n^3)$ time, where $n$ is
the number of constituents $\chi_i$ of $\chi$. The number of iterations of
the while loop is bounded by $n$, and the body of the while loop can
be executed in quadratic time.

We have conducted experiments to evaluate the execution time of the
\textbf{Create-cceg} algorithm. Those experiments are described in
detail in Section \ref{exp:cceg}.

 \begin{definition}[State of an agent]
The state of an agent is a set of ground code call conditions.
 \end{definition}
 
 When an agent developer builds an agent, she specifies several
 parameters. One of these parameters must include some
 \emph{domain-specific} information, explicitly laying out what
 inclusion and equality relations are known to hold of code calls.
 Such information is specified via \emph{invariants}.

\begin{definition}[Invariant Expression]\mbox{}\\[-.5cm]
\begin{itemize}
\item Every evaluable code call condition is an invariant
  expression. We call such expressions \emph{atomic}.
\item If $\ie_1$ and $\ie_2$ are invariant expressions, then $(\ie_1 \cup
  \ie_2)$ and $(\ie_1 \cap \ie_2)$ are invariant expressions. (We
  will often omit the parentheses.) 
\end{itemize}
  \end{definition}

 \begin{example}
Two examples of invariant expressions are:
 \[\begin{array}{l}
   \IN{\var{StudentRec}}{\cc{rel}{select}{\textit{courseRel}, \textit{exam},
      \str{=}, \textit{midterm1}}} \,\&\,\\
  \IN{\var{C}}{\cc{excel}{chart}{\textit{excelFile}, \var{StudentRec},
       \textit{grade}}}\\
\\
\IN{\var{X}}{\cc{spatial}{horizontal}{\var{T},\var{B},\var{U}}}
  \, \cup
 (\IN{\var{Y}}{\cc{spatial}{horizontal}{\var{T'},\var{B'},\var{U'}}}\, 
 \cup \\
  \IN{\var{Z}}{\cc{spatial}{horizontal}{\var{T'},\var{B'},\var{U}}}).
 \end{array}\]
 \end{example}

What is the meaning, i.e.~the \emph{denotation} of such expressions? The
first invariant represents the set of all objects
$\textit{c}$ such that  
\[\begin{array}{l}
\IN{\var{StudentRec}}{\cc{rel}{select}{\textit{courseRel}, \textit{exam},
      \str{=}, \textit{midterm1}}} \,\& \\
  \IN{\textit{c}}{\cc{excel}{chart}{\textit{excelFile}, \var{StudentRec},
       \textit{grade}}}
\end{array}\] holds: we are looking for instantiations of \var{C}.
 Note that under this viewpoint, the intermediate 
  variable \var{StudentRec} which is needed in order to instantiate
\var{C} to an object $\textit{c}$ does not matter. There might just as well be
situations where we are interested in pairs $\langle c,studentrec\rangle$ instead 
of just $c$.
Therefore a notion of denotation must be flexible enough to allow
this.

Let us now consider the invariant
 \[\begin{array}{l}
   \IN{\var{StudentRec}}{\cc{rel}{select}{\textit{courseRel},  \textit{exam},
      \str{=}, \var{TypeofExam}}} \,\&\,\\
  \IN{\var{C}}{\cc{excel}{chart}{\textit{excelFile}, \var{StudentRec},
       \textit{grade}}}\\
 \end{array}\]
 where the object \textit{midterm1} has been replaced by the variable
 \var{TypeofExam} which is now a base variable.  Then we might be
 interested in all $c$'s that result if an instantiation of
 \var{TypeofExam} is given, i.e.~for different instantiations of
 \var{TypeofExam} we get different $c$'s. Thus we have to distinguish
 carefully between various sorts of variables: \emph{base} variables
 (defined in Definition~\ref{basevar}), \emph{auxiliary} variables and
 the \emph{main} variables defining the set of objects of interest.
 \begin{definition}[Denotation of an Invariant Expression]\label{denot}
   Let $\ie$ be an invariant expression with
   $\text{var}(\ie)=\text{var$_{\text{base}}$}(\ie)\cup \{\var{V}_1, \ldots ,
   \var{V}_n\}$.  The denotation of  $\ie$ with
   respect to a state $S$, an assignment $\mathbf{\theta}$ of the
   variables in $\text{var$_{\text{base}}$}(\ie)$ and a sequence $\langle
   \var{V}_{i_1}, \ldots , \var{V}_{i_k}\rangle$ (where $ \var{V}_{i_1}, \ldots ,\var{V}_{i_k}\} \subseteq \{\var{V}_1, \ldots
   , \var{V}_n\} $) is defined as follows: 
 \begin{itemize}
 \item Let 
\[\begin{array}{lll}
[\ie]_{S,\mathbf{\theta}} := & \{ \,\langle o_{\pi(1)}, \ldots , o_{\pi(n_k)}  \rangle  \mid &
(\ie\,\mathbf{\theta})\tau  ~\text{is ground and is true in state} ~S,\\
& &  \text{$\pi$ is a permutation on $\{1, \ldots , n\}$,  $n_k\leq n$},\\
& &  \text{$\tau$ is a  grounding substitution,}\\
& & \text{$\tau$ is of the form } [\var{V}_1/o_1, \ldots ,\var{V}_n/o_n]\ \}\\
\end{array}\]
\item $ [\ie_1 \cap \ie_2 ]_{S,\mathbf{\theta}}  := [\ie_1]_{S,\mathbf{\theta}} \cap [\ie_2]_{S,\mathbf{\theta}} $,
\item $ [\ie_1 \cup \ie_2 ]_{S,\mathbf{\theta}} := [\ie_1]_{S,\mathbf{\theta}} \cup [\ie_2]_{S,\mathbf{\theta}} $.
 \end{itemize}
 \end{definition}
 The variables in $\{ \var{V}_{\pi(1)}, \ldots , \var{V}_{\pi(n_k)} \} $ are
 called \emph{main variables} while all remaining variables $\{
 \var{V}_{\pi(n_k+1)}, \ldots ,\var{V}_{\pi(n)} \}$ are called
 \emph{auxiliary}.  The substitution $\tau$ is defined on the set of
 \emph{main variables} (in our example above it is the set
 $\{\var{C}\}$). The set of auxiliary variables consists of
 $\{\var{StudentRec}\}$ and the only base variable is
 \var{TypeofExam}. Taking the first viewpoint in our example above,
 $\tau $ would be defined on $\{\var{C},\var{StudentRec} \}$.
 
 As usual, we abuse notation and say that $\ie_1 \subseteq \ie_2$ if
 $[\ie_1]_{S,\mathbf{\theta}} \subseteq [\ie_2]_{S,\mathbf{\theta}}$ for all $S$ and
 all assignments $\mathbf{\theta}$. Similarly, we say that $\ie_1 = \ie_2$
 if $[\ie_1]_{S,\mathbf{\theta}} = [\ie_2]_{S,\mathbf{\theta}}$ for all $S$
 and all assignments $\mathbf{\theta}$. Now we are ready to define an
 invariant.


\begin{definition}[Invariant Condition (\ic)]
  An \emph{invariant condition atom} is a statement of the form $t_1\:
  \op\: ~t_2$ where $\op \in \{ \leq, \geq, <, >, =\}$ and each of $t_1$,
  $t_2$ is either a variable or a constant. An invariant condition
  (IC) is defined inductively as follows:
\begin{enumerate}
\item Every invariant condition atom is an \ic.
\item If $C_1$ and $C_2$ are \ic's, then $C_1 \land C_2$ and $C_1 \lor C_2$
  are \ic's.
\end{enumerate}
\end{definition}

\begin{definition}[Invariant \iv,
   \text{var$_{\text{base}}$}(\iv), $\textsf{INV}_{\text{simple}}$,
$\textsf{INV}_{\text{ordinary}}$, \textsf{INV}] \label{basevar}
   An \emph{invariant}, denoted by \iv, is a statement of the
   form
\begin{equation}\label{eq:invariant}
\ic  \Longrightarrow ~~\ie_1 ~~\Re  ~~\ie_2 
\end{equation}
where 
\begin{enumerate}
\item \ic is an invariant condition, 
 all  variables occuring in  \ic are among
  $\text{var$_{\text{base}}$}(\ie_1) \cup  \text{var$_{\text{base}}$}(\ie_2)$.
\item $\Re \in \{ =, \subseteq \}$, and
\item $\ie_1,\ie_2$ are invariant expressions.
\end{enumerate}
If $\ie_1$ and $\ie_2$ both contain solely \emph{atomic} code call
conditions, then we say that $\iv$ is a simple invariant.

If $\ic$ is a conjunction of invariant condition atoms,
then we say that $\iv$ is an ordinary invariant.

We denote by \text{var$_{\text{base}}$}(\iv) the set of all variables
of \iv that need to be instantiated in order to evaluate \iv in the
current state:
$\text{var$_{\text{base}}$}(\iv):=\text{var$_{\text{base}}$}(\ie_1) \cup
\text{var$_{\text{base}}$}(\ie_2)$.

The set of all invariants is denoted by \textsf{INV}. The set of all
simple invariants is denoted by $\textsf{INV}_{\text{simple}}$ and the
set of all ordinary invariants is denoted by
$\textsf{INV}_{\text{ordinary}}$.
\end{definition}

An invariant expresses \emph{semantic knowledge} about a domain.
Invariants used by each of our two example agents---\ag{ppt} and
\ag{hdb} are given below.

 \begin{example}
   The following are valid invariant conditions:
   $\textit{val}_1 \leq \textit{val}_2$, $\var{Rel_1} = \var{Rel_2}$.
   Note that such expressions can be evaluated over a given state $S$. 
   Only the two
   relations $ \leq$ and $\geq$ require that the constants occurring on the
   right or left hand sides must be of the appropriate type: these
   relations must be defined over each state $S$.

   The invariant
\[\begin{array}{c}
  \var{File}=\var{File'} \land \var{Rec}=\var{Rec'} \land
  \var{Col}=\var{Col'}\\
 \Longrightarrow\\
 \IN{\var{C}}{\cc{excel}{chart\_one}
    {\var{File},\var{Rec},\var{Col}}} =
  \IN{\var{C'}}{\cc{excel}{chart\_two}
    {\var{File'},\var{Rec'},\var{Col'}}}
\end{array}
\]
says  that these two code call conditions are equivalent if their
arguments unify. Note that the code calls involved are different.
  The invariant,
\begin{center}
$ \var{Rel}=\var{Rel'} \land  \var{Attr}=\var{Attr'}\,\land
\var{Op}=\var{Op'}= \str{\ensuremath{\leq}}\,\land \var{Val} < \var{Val'}$\\ 
$\Longrightarrow $\\
$\IN{\var{X}}{\cc{rel}{select}{\var{Rel}, \var{Attr}, \var{Op}, 
\var{Val}}} \subseteq \IN{\var{Y}}{\cc{rel}{select}{\var{Rel'}, 
\var{Attr'}, \var{Op'}, \var{Val'}}}$ 
\end{center}
says that the code call condition
$\IN{\var{X}}{\cc{rel}{select}{\var{Rel}, \var{Attr},
    \var{Op},\var{Val}}}$ can be evaluated by using the results of the
code call condition
\[\IN{\var{Y}}{\cc{rel}{select}{\var{Rel'},\var{Attr'},
    \var{Op'}, \var{Val'}}}\] if the above conditions are satisfied.
Note that this expresses \emph{semantic information} that is not
available on the syntactic level: the operator   \str{\ensuremath{\leq}}
is related to the relation symbol \str{\ensuremath{<}}. 
\end{example}

\begin{convention}\label{conv1}Throughout the rest of this paper, we 
  assume that we have the code calls
  \cc{ag}{addition}{\var{X},\var{Y}} and
  \cc{ag}{subtraction}{\var{X},\var{Y}} available for all agents
  \ag{ag}.  These code calls return the sum, (resp.~the difference) of
  \var{X} and \var{Y}, where \var{X} and \var{Y} range over the reals
  or the integers.  We also assume we have code calls \cc{ag}{ge_0}{X}
  (resp.~ \cc{ag}{geq\_0}{X}) available which returns $1$ if \var{X}
  is strictly greater (resp.~greater or equal) than $0$ and $0$
  otherwise.
\end{convention}

By stating invariants, we focus interest on states where the
invariants hold. This is like in classical predicate logic, where we
write down axioms and thereby constrain the set of models---we are
only interested in the class of models satisfying the axioms. We
therefore have to define formally what it means for a state $S$ to
satisfy an invariant \iv.
\begin{definition}[Satisfaction, $S\models\iv$, $\cali \models \iv$, \text{Taut}]\label{satisfaction}\mbox{}\\
  A state $S$ \emph{satisfies} the invariant $\iv$ having the form
  shown in Formula~(\ref{eq:invariant}) above with respect to an
  assignment $\mathbf{\theta}$ \iffdef for every ground instance
  $(\iv\,\mathbf{\theta})\,\tau $ of $\iv\,\mathbf{\theta}$, it is the case that
  either $(\text{\ic}\,\mathbf{\theta})\,\tau $ evaluates to false, or
  $\,(\ie_1\,\mathbf{\theta})\,\tau \,\Re\, (\ie_2\,\mathbf{\theta})\,\tau $ is
  true in $S$.
  
  We say that a set of invariants $\cali$ \emph{entails} an invariant
  \iv \iffdef all states $S$ and assignments $\mathbf{\theta}$ satisfying
  $\cali$ also satisfy $\iv$. We write $\cali \models \iv$.  We call an
  invariant \iv a tautology, if \iv is true in all states $S$ for all
  assignments $\mathbf{\theta}$: \[\text{Taut}\defeq \{\iv\, | \, \models \iv\}.\]
  
  From now on we do not mention explicitly the assignment
  $\mathbf{\theta}$ and we write simply $S\models\iv$.
\end{definition}
It is worth noting that there are indeed trivial invariants that are
satisfied in all states: such invariants are like tautologies in
classical logic (therefore their name in the last definition). For
example the following invariant is true in all states whatsoever (note
the difference from the similar invariant above):
\begin{center}
  \var{File}=\var{File'} $\land$ \var{Rec}=\var{Rec'} $\land$
  \var{Col}=\var{Col'} $\Longrightarrow$\\
$ \IN{\var{C}}{\cc{excel}{chart}
    {\var{File},\var{Rec},\var{Col}}} =
  \IN{\var{C'}}{\cc{excel}{chart}
    {\var{File'},\var{Rec'},\var{Col'}}}$
\end{center}
The reason that this last invariant is a tautology, is that for the
same set of instances of \var{Y} for a code call \cc{d}{f}{\var{Y}},
we always get the same set representing the atomic code call condition
\IN{X}{\cc{d}{f}{Y}}.
\begin{theorem}\label{elimcondlists}\mbox{}\\
  There is a translation $\mathfrak{Trans}$ which associates with each
  conjunction \ic of invariant
  condition atoms,  and invariant expression \ie another
  invariant expression 
  $\mathfrak{Trans}(\ic,\ie_1)$ such that the following holds for all
  states $S$, assignments $\mathbf{\theta}$  and invariants 
$\ic \Longrightarrow ~~\ie_1 ~~\Re ~~\ie_2 $
\[\begin{array}{c}
(S,\mathbf{\theta})\models (\ic \Longrightarrow ~~\ie_1 ~~\Re  ~~\ie_2)\\
\iff\\
 (S,\mathbf{\theta})\models \text{true} \Longrightarrow
 ~~\mathfrak{Trans}(\ic,\ie_1)  ~~\Re
 ~~\mathfrak{Trans}(\ic,\ie_2).
\end{array} \]
\end{theorem}

\begin{corollary}[Eliminating Invariant Conditions, 
  $\mathfrak{Trans}$]\label{cor:trans}
Let $\iv : \ic  \Longrightarrow ~~~\ie_1 ~~\Re  ~~\ie_2$
be an arbitrary invariant.
Then, the following holds for all states $S$ and assignments 
$\mathbf{\theta}$

\[\begin{array}{c}
(S,\mathbf{\theta})\models (\ic \Longrightarrow ~~\ie_1 ~~\Re  ~~\ie_2) \\
\iff\\
(\forall C_i, 1 \leq i \leq m) ~~(S,\mathbf{\theta})\models \text{true} 
\Longrightarrow \mathfrak{Trans}(C_i,\ie_1)  ~\Re
 ~\mathfrak{Trans}(C_i, \ie_2).
\end{array} \]
where the $C_i$, $1 \leq i \leq m$, are the disjuncts in the DNF of \ic.
\end{corollary}

 \section{Architecture}
\label{sec:arch}
Let us suppose now that we have a set $\mathcal{I}$ of invariants, and
a set $\mathcal{S}$ of data structures that are manipulated by the
agent.  How exactly should a set $\mathcal{C}$ of code call conditions
be merged together?  And what needs to be done to support this?  Our
architecture contains two parts:
\begin{enumerate}
\item[(i)] a \emph{development time} phase stating what the agent
  developer must specify when building her agent, and what algorithms
  are used to operate on that specification, and
\item[(ii)] a \emph{deployment time} phase which specifies how the
  above development-time specifications are used when the agent is in
  fact running autonomously.
\end{enumerate}
We describe each of these pieces below.

 \subsection{Development Time Phase}
When the agent developer builds her agent, the following
things need to be done.
 \begin{enumerate}
\item First, the agent developer specifies a set $\mathcal{I}$ of
  invariants.
\item Suppose $\mathcal{C}$ is a set of CCCs to be evaluated by the
  agent.  Each code call condition $\chi\in \mathcal{C}$ is represented
  via an evaluable cceg.  Let $\INS(\mathcal{C})$ represent the set of
  all nodes in ccegs of $\chi$s in $\mathcal{C}$.  That is,
  \[\INS(\mathcal{C}) = \{ v_i \mid \exists \chi \in \mathcal{C} ~s.t. ~v_i ~is
  ~in ~\chi's ~cceg\}.\] This can be done by a topological sort of the
  cceg for each $\chi \in \mathcal{C}$.
\item Additional invariants can be derived from the initial set
  $\cali$ of invariants. This requires the ability to check whether a
  set $\mathcal{I}$ of invariants implies an inclusion relationship
  between two invariant expressions.  We will provide a \emph{generic}
  test called $\chkimp$ for implication checking between invariants.
  Although the set of invariants entailed by $\cali$ is defined by
  Definition \ref{satisfaction}, the set of invariants actually
  derived by the $\chkimp$ test will depend on the set of axioms used
  in the test. Hence, some $\chkimp$ tests will be sound, but not
  complete.  On the other hand, some tests will be ``more complete''
  than others, because the set of invariants derived by them will be a
  superset of the set of invariants derived by others. ``More
  complete'' tests may use a larger set of axioms, hence will be more
  expensive to compute. The agent developer can select a test that is
  appropriate for her agent. Given an arbitrary (but fixed) $\chkimp$
  test, we will provide an algorithm called
  \textbf{Compute-Derived-Invariants} that calculates the set of
  derivable invariants from the initial set $\cali$ of invariants and
  needs to be executed just once.
\end{enumerate}

 \subsection{Deployment Time Phase}
 Once the agent has been ``developed'' and deployed and is running, it
 will need to continuously determine how to merge a set $\mathcal{C}$
 of code call conditions.  This will be done as follows:

 \begin{enumerate}
 \item The system must identify three types of relationships between
   nodes in $\INS(\mathcal{C})$. 

\begin{description}
\item[Identical ccc's:] First, we'd like to identify nodes
  $\chi_1,\chi_2\in \INS(\mathcal{C})$ which are ``equivalent'' to one
  another, i.e.  $\chi_1=\chi_2$ is a logical consequence of the set of
  invariants $\mathcal{I}$.  This requires a definition of
  \emph{equivalence} of two code call conditions w.r.t.  a set of
  invariants.  This strategy is useful because we can replace the two
  nodes $\chi_1,\chi_2$ by a single node.  This avoids redundant
  computation of both $\chi_1$ and $\chi_2$.
\item[Implied ccc's:] Second, we'd like to identify nodes $\chi_1,\chi_2\in
  \INS(\mathcal{C})$ which are not equivalent in the above sense, but
  such that either $\chi_1\subseteq \chi_2$ or $\chi_2\subseteq \chi_1$ hold, but not both.
  Suppose $\chi_1\subseteq \chi_2$. Then we can compute $\chi_2$ first, and then
  compute $\chi_1$ from the answer returned by computing $\chi_2$.  This
  way of computing $\chi_1,\chi_2$ may be faster than computing them
  separately.
\item[Overlapping ccc's:] Third, we'd like to identify nodes
  $\chi_1,\chi_2\in \INS(\mathcal{C})$ for which the preceding two
  conditions do not hold, but $\chi_1\,\,\&\,\,\chi_2$ is consistent with
  $\INS(\mathcal{C})$. In this case, we might be able to compute the
  answer to $\chi_1\,\lor\, \chi_2$.  From the answer to this, we may
  compute the answer to $\chi_1$ and the answer to $\chi_2$.  This way of
  computing $\chi_1,\chi_2$ may be faster than computing them separately.
\end{description}
We will provide an algorithm, namely \textbf{Improved-CSI}, which will
use the set of derived invariants returned by the
\textbf{Compute-Derived-Invariants} algorithm above, to detect
commonalities (equivalent, implied and overlapping code call
conditions) among members of $\mathcal{C}. $

\begin{example}
  The two code call conditions
  $\IN{\var{X}}{\cc{spatial}{vertical}{\var{T},\var{L},\var{R}}}$ and
  $\IN{\var{Y}}{\cc{spatial}{vertical}{\var{T'},\var{L'},\var{R'}}}$
  are equivalent to one another if their arguments are unifiable.  The
  results of evaluating the code call condition
  \[\IN{\var{Z}}{\cc{spatial}{range}{\var{T},40,50,25}}\] is a subset of
  the results of evaluating the code call condition
  \[\IN{\var{W}}{\cc{spatial}{range}{\var{T'},40,50,50}}\] if \var{T} =
  \var{T'}. Note that
  \cc{spatial}{range}{\var{T},\var{X},\var{Y},\var{Z}}
  returns all points in \var{T} that are  \var{Z} units away from the
  point $\langle \var{X},\var{Y}\rangle $.
In this case, we can compute the results of the former
  code call condition by executing a selection on the results of the
  latter rather than executing the former from scratch. Finally,
  consider the following two code call conditions:
  \[\begin{array}{l}
\IN{\var{X}}{\cc{spatial}{horizontal}{\textit{map},100,200}},\\
\IN{\var{Y}}{\cc{spatial}{horizontal}{\textit{map},150,250}}. 
\end{array}\] 
Here \cc{spatial}{horizontal}{\textit{map},a,b} returns all points
(\var{X}, \var{Y}) in $\textit{map}$ such that $a \leq \var{Y} \leq b$.
Obviously, the results of neither of these two code call conditions
are subset of the results of the other.  However, the results of these
two code call conditions \emph{overlap} with one another. In this
case, we can execute the code call condition
$\IN{\var{Z}}{\cc{spatial}{horizontal}{\textit{map},100,250}}$.  Then,
we can compute the results of the two code call conditions by
executing selections on the results of this code call condition.
\end{example}

\item We will then provide two procedures to merge sets of code call
  conditions, \textbf{BFMerge} and \textbf{DFMerge}, that take as
  input, \emph{(i)} the set $\mathcal{C}$ and \emph{(ii)} the output
  of the \textbf{Improved-CSI} algorithm above, and \emph{(iii)} a
  cost model for agent code call condition evaluations.  Both these
  algorithms are parameterized by heuristics and we propose three
  alternative heuristics.  Then we evaluate our six implementations (3
  heuristics times 2 algorithms) and also compare it with an $A^{*}$
  based approach.
\end{enumerate}
\section{Development Phase}
\label{sec:development}
Prior to deployment of the agent, once the agent developer has defined
a set of invariants, we compute a set of derived invariants from it.
These derived invariants are stored.  Once deployed, when the agent is
confronted with a set of requests from other agents, it can examine
these stored derived invariants for a ``pattern match'' which then
enables it to classify invariants into one of the three categories
listed (equivalent, implied or overlapping invariants).

Consider the case when $\cali$ contains the two invariants:
\begin{eqnarray}
\var{V}_1 \leq \var{V}_2 & \Longrightarrow & \IN{X}{\cc{d_1}{f_1}{V_1}} = \IN{Y}{\cc{d_2}{f_2}{V_2}}.
\label{iv1:eq}\\
\var{V}_3 \leq \var{V}_4 & \Longrightarrow & \IN{Z}{\cc{d_2}{f_2}{V_3}} \subseteq \IN{W}{\cc{d_3}{f_3}{V_4}}.
\label{iv2:eq}
\end{eqnarray}
Clearly from these two invariants, we can infer
the invariant
\begin{eqnarray}
\var{V}_1 \leq \var{V}_2\,\,\land \var{V}_2\leq \var{V}_4 & \Longrightarrow & \IN{X}{\cc{d_1}{f_1}{V_1}}\subseteq \IN{W}{\cc{d_3}{f_3}{V_4}}
\label{iv4:eq}
\end{eqnarray}

Algorithm \textbf{Combine\_1} (Figure~\ref{alg:combine1}) combines two
invariants. The algorithm uses a \textbf{simplify} routine which
simplifies a conjunction of invariant conditions and checks if the
resulting invariant condition is inconsistent or not. If so, it
returns \NIL.  The \textbf{Combine\_1} algorithm makes use of two
important algorithms: \chkimp ~~and \chkent, which we will discuss in
detail later.  The $\chkimp$ algorithm checks if one invariant
expression implies another, while the $\chkent$ algorithm checks if
some member of a set of invariants entails an other invariant. Let us
first define the table which is implemented by the \textbf{Combine\_1}
algorithm.

\begin{definition}[Combine1]\label{def:comb1}
  Let $\iv_1 : \ic_1 ~\Longrightarrow ~\ie_1 ~\Re_1 ~\ie_1'$ and $\iv_2 :
  \ic_2 ~\Longrightarrow ~\ie_2 ~\Re_2 ~\ie_2'$. Then, the following table
  provides the resulting derived relation of the form
\[ \textbf{simplify}(\ic_1 \,\land \ic_2)~~\Longrightarrow ~~\ie_1 ~~\Re_1 ~~\ie_2'\]
when $\iv_1$ and $\iv_2$ are
combined. The ``*'' denotes a ``don't care'' condition in this table.
\begin{table}[htb]
\begin{center}
{\small
\begin{tabular}{|c|c|c|c||c|} \hline \hline
$\Re_1$ & $\Re_2$ & $\chkimp(\ie_1',\ie_2)$ &
$\chkimp(\ie_2,\ie_1')$ & derived\_rel \\ \hline \hline
* & * & False & * & \NIL \\ \hline
= & = & True & True & = \\ \hline
= & $\subseteq$ & True & * & $\subseteq$  \\ \hline
$\subseteq$ & = & True & * & $\subseteq$  \\ \hline
$\subseteq$ & $\subseteq$ & True & * & $\subseteq$  \\ \hline
\end{tabular}
\caption{Summary of Combining Two Invariants}
\label{summary}
}
\end{center}
\end{table}
The \textbf{simplify} routine checks whether $\ic_1 \,\land \ic_2$ is 
inconsistent. If so, it returns \emph{false}, if not it returns an
equivalent (perhaps simplified) formula for $\ic_1 \,\land \ic_2$
(the precise realization of  \textbf{simplify} is not important here
and leaves enough freedom for the actual implementation):

\[ \textbf{simplify}(\ic)
 = \left\{ \begin{array}{ll}
         \text{\emph{true}}, & \text{if  for all  \ensuremath{(S,\mathbf{\theta})\models \ic},}\\
     \text{\emph{false}}, & \text{if for all
       \ensuremath{(S,\mathbf{\theta})\models \lnot \ic},}\\
         \phi\footnotemark
         , & 
         \text{otherwise.}
\end{array} \right.
\]\footnotetext{where $\phi$ is any formula equivalent to $\ic$, 
i.e.~for all states $(S,\mathbf{\theta})$: \ensuremath{(S,\mathbf{\theta})\models \phi
             \leftrightarrow \ic }.}
\end{definition}
Figure~\ref{alg:combine1} implements a slightly generalized version of
the last definition. Namely, we assume that there is given a set \cali
of invariants and we are considering states satisfying these
invariants. This is an additional parameter. For simplicity, assume
that $\Re_1=\Re_2=\str{$\subseteq$}$. The idea is that although the subset
relation $\ie_1'\subseteq\ie_2$ might not hold in general (i.e.~in all
states) it could be implied by the invariants in $\cali$ (i.e.~holds
in all states satisfying \cali). That is, if there is $\ic^* ~\Longrightarrow~
\ie_1^* ~\subseteq ~\ie_2^* \ \in \cali$ s.t.  ($\ic^{**} ~\Longrightarrow~ \ie_1' ~\subseteq
~\ie_1^* \ \in \cali$, $\ic^{***} ~\Longrightarrow~ \ie_2^* ~\subseteq ~\ie_2 \ \in \cali$,
and $(\ic_1\land \ic_2)\, \to\, (\ic^* \land \ic^{**} \land \ic^{***})$). Under
these conditions, we can derive the invariant $\textbf{simplify}(\ic_1
\,\land \ic_2)~~\Longrightarrow ~~\ie_1 ~~\subseteq ~~\ie_2'$.
\begin{figure}[htb]
\alg{
\par\noindent \textbf{Combine\_1($\iv_1,\iv_2, \cali)$} \\
/* $\iv_1 : \ic_1 ~~\Longrightarrow ~~\ie_1 ~~\Re_1 ~~\ie_1'$ */ \\
/* $\iv_2 : \ic_2 ~~\Longrightarrow ~~\ie_2 ~~\Re_2 ~~\ie_2'$ */ \\

\kw{if} ~~(\chkimp($\ie_1',\ie_2$) = \emph{false}) ~~\kw{and} \\
 \pq~(there is no $\ic^* ~\Longrightarrow~ \ie_1^* ~\subseteq ~\ie_2^* \ \in \cali$
s.t.\\
 \pq~($\ic^{**} ~\Longrightarrow~ \ie_1' ~\subseteq  ~\ie_1^* \ \in \cali$,
 ~ ~$\ic^{***} ~\Longrightarrow~ \ie_2^* ~\subseteq  ~\ie_2 \ \in \cali$,\\
 \pq~ $(\ic_1\land \ic_2)\, \to\, (\ic^* \land \ic^{**} \land \ic^{***})$),~~\kw{then} \\
\pq ~~\kw{Return} ~\NIL; \\
\kw{if} ~~($\Re_1 = \Re_2 = \str{=}$) ~~\kw{then} \\
\pq \kw{if} ~~(\chkimp($\ie_2,\ie_1') = true$) ~~\kw{or} \\
 \pq~(there is  $\ic^* ~\Longrightarrow~ \ie_1^* ~\subseteq ~\ie_2^* \ \in \cali$
s.t.\\
 \pq~($\ic^{**} ~\Longrightarrow~ \ie_2 ~\subseteq  ~\ie_1^* \ \in \cali$,
 ~ ~$\ic^{***} ~\Longrightarrow~ \ie_2^* ~\subseteq  ~\ie_1' \ \in \cali$,\\
 \pq~ $(\ic_1\land \ic_2)\, \to\, (\ic^* \land \ic^{**} \land \ic^{***})$)),~~\kw{then} \\
\pqq $relation :=\, (\ie_1 = \ie_2'); $ \\
\pq \kw{else} ~$relation :=\, (\ie_1 \subseteq \ie_2');$\\
\kw{else} ~$relation :=\, (\ie_1 \subseteq \ie_2');$\\
$derived_ic :=\, \textbf{simplify}(\ic_1 \,\land \ic_2)$; \\
\kw{if} ~~($derived_ic$ = \emph{false}) ~~\kw{then} ~~\kw{Return} ~\NIL; \\
~$derived\_inv :=\, (derived_ic \Longrightarrow relation); $ \\
\kw{if} (\text{there is $\iv \in \cali$ with} ~~(\chkent(\iv, $derived\_inv) = true$) ~~\kw{then} \\
\pq ~~\kw{Return}~ \NIL;  \\
\kw{else} ~~\kw{Return} ~$derived\_inv; $\\
\kw{End-Algorithm}
}
\caption{\textbf{Combine\_1} Algorithm}
\label{alg:combine1}
\end{figure}

We introduce three notions,  $\chkimp$, $\chktaut$ and $\chkent$ of
increasing complexity. The first notion, $\chkimp$, is a relation
between invariant expressions. 
\begin{definition}[Implication: \chkimp, $\ie_1 \to \ie_2$]\label{impl}\mbox{}\\
  An invariant expression $\ie_1$ is said to imply another invariant
  expression $\ie_2$, denoted by $\ie_1 \to \ie_2$, \iffdef it is the case
  that $[\ie_1]_{S,\mathbf{\theta}} \subseteq
  [\ie_2]_{S,\mathbf{\theta}}$ for all $S$ and all   assignments $\mathbf{\theta}$.
  
  $\chkimp$ is said to be an \emph{implication check} algorithm if it
takes two invariant expressions $\ie_1,\ie_2$ and returns a boolean
output.  We say that $\chkimp$ is \emph{sound} \iffdef whenever
$\chkimp$($\ie_1,\ie_2$)=true, then $\ie_1$ implies $\ie_2$.  We say
$\chkimp$ is \emph{complete} \iffdef $\chkimp(\ie_1,\ie_2)$ = true \iff
$\ie_1$ implies $\ie_2$.
 
If $\chkimp_1,\chkimp_2$ are both sound, and for all $\ie_1,\ie_2$,
$\chkimp_1(\ie_1,\ie_2)=$ true implies that $\chkimp_2(\ie_1,\ie_2)=$
true, then we say that $\chkimp_2$ is \emph{more complete} than
$\chkimp_1$.
\end{definition}

\begin{definition}[\chktaut, \chkent as Relations between Invariants]
\label{taut,ent}
  $\chktaut$ is said to be a \emph{tautology check} algorithm if it
takes a single  invariant $\iv$ and returns a boolean
output.  $\chktaut$ is \emph{sound} \iffdef whenever
$\chktaut$($\iv$)=true, then $\iv\in \text{Taut}$ (see
Definition~\ref{satisfaction}). 
$\chktaut$ is \emph{complete} \iffdef $\chktaut(\iv)$ = true \iff
$\iv\in \text{Taut}$.
 
$\chkent$ is said to be an \emph{entailment check} algorithm if it
takes two invariants $\iv_1,\iv_2$ and returns a boolean output.  We
say that $\chkent$ is \emph{sound} \iffdef whenever
$\chkent$($\iv_1,\iv_2$)=true, then $\iv_1$ entails $\iv_2$ ($\iv_1 \models
\iv_2$).  We say $\chkent$ is \emph{complete} \iffdef
$\chkent(\iv_1,\iv_2)$ = \text{true} \iff $\iv_1$ entails $\iv_2$.

Similarly to Definition~\ref{impl}, we use the notion of \emph{being
  more complete} for tautology as well as for entailment check
algorithms.
\end{definition}

\begin{lemma}[Relation between \chkimp, \chktaut and \chkent]\label{reduction}\mbox{}\\[-.3cm]
\begin{enumerate}
\item[(1)] $\chkimp$ can be reduced to \chktaut: \[\chkimp(\ie_1,\ie_2)
  \ \iff \ \chktaut(\text{true} \Longrightarrow \ie_1 \subseteq
  \ie_2).\]
\item[(2)]  $\chktaut$ can be reduced to \chkimp:
\[\begin{array}{c}
\chktaut((C_1 \lor C_2 \lor \ldots \lor C_m) \Longrightarrow \ie_1\subseteq \ie_2)\\
\iff\\
\forall C_i, 1 \leq i \leq m,  \chkimp(\mathfrak{Trans}(C_i,\ie_1),\mathfrak{Trans}(C_i,\ie_2)).
\end{array}\]
\item[(3)] $\chktaut$ is an instance of $\chkent$.
\end{enumerate}
\end{lemma}
Thus in general, implication checking between invariant expressions is
a special case of tautology checking of invariants. Conversely,
checking tautologies is an instance of implication checking. Note that
checking simple invariants is reduced to checking implications of
non-simple invariant expressions.

It is also obvious that checking for tautologies is a special case of
the entailment problem.

The following results tell us that the implementation of the $\chkimp$
routine used in the \textbf{Combine\_1} algorithm is undecidable in
general. Even if we restrict to finite domains, it is still
intractable.
\begin{proposition}[Undecidability  of $\chkimp$, $\chktaut$, $\chkent$]\label{undec}
  \mbox{}\\ Suppose we consider arbitrary datatypes.  Then the problem
  of checking whether an arbitrary invariant expression $\ie_1$
  implies another invariant expression $\ie_2$ is undecidable. The
  same holds for checking tautologies of invariants or entailment
  between invariants.
\end{proposition}

\begin{proposition}[\coNP Completeness of Checking Implication]\label{conp}\mbox{}\\
  Suppose all datatypes have a finite domain (i.e.~each datatype has
  only finitely many values of that datatype).  Then the problem of
  checking whether an arbitrary invariant expression $\ie_1$ implies
  another invariant expression $\ie_2$ is \coNP complete. The same
  holds for the problem of checking whether an invariant is a tautology.
\end{proposition}

As the problem of checking implication (and hence equivalence)
between invariant expressions is \coNP complete, in this paper, we decided
to study the tradeoffs involved in using sound, but perhaps incomplete
implementations of implication checking.

There are clearly many ways of implementing the algorithm $\chkimp$
that are sound, but not complete.  In this paper, we propose a generic
algorithm to implement $\chkimp$, where the complexity can be
controlled by two input parameters---an \emph{axiomatic inference
system} and a \emph{threshold}.

\begin{itemize}
\item The \emph{axiomatic inference system} used by $\chkimp$ includes
  some axioms and inference rules. By selecting the axioms and
  inference rules, the agent developer is controlling the branching
  factor of the search space.
\item The second parameter called the \emph{threshold} is either an
  integer or $\infty$, and determines the maximum depth of the search
  tree.  If it is $\infty$, then the generic algorithm does not have an
  upper bound on the number of rule applications, and terminates
  either when it proves the implication or there is no further rule
  that is applicable(i.e.~\emph{failure}). When it is an integer value,
  the algorithm reports \emph{failure} if it cannot prove the
  implication by using the threshold number of rule applications.
\end{itemize}
We have conducted experiments with different instances of these two
parameters. Those experiments are discussed in detail in
Section~\ref{exps:compile}.

It is important to note that the set of all derived invariants
obtained from $\cali$ may be very large because they contain
``redundant'' constraints. For instance, using our example $\cali$
above, every invariant of the form
\[\begin{array}{c}
\var{V_1}\leq \var{V_2}\,\,\land \var{V_2}\leq \var{V_4} \\ \Longrightarrow\\
 \IN{X}{\cc{d_1}{f_1}{V_1}}
\subseteq \IN{W}{\cc{d_3}{f_3}{V_4}} \,\cup \IN{T}{\cc{d_4}{f_4}{V_5}} \ldots
\end{array}\]
would be entailed from $\cali$---however, these invariants are
\emph{redundant} as they are entailed by the single invariant (\ref{iv4:eq}).  

As we have seen above (Propositions \ref{undec},~\ref{conp}), such an
entailment test between invariants is either undecidable or
intractable. It would be much better if we had a purely syntactical
test (which must be necessarily incomplete) of checking such
implications.

The following lemma shows that entailment between two invariants can,
under certain assumptions, be reduced to a syntactical test.
\begin{lemma}\label{subsumption}
  Let $\iv_1 : \ic_1 \Longrightarrow ~\ie_1 \subseteq \ie_1'$ and $\iv_2 : \ic_2 \Longrightarrow
  ~\ie_2 \subseteq \ie_2'$ be two simple invariants, i.e.~$\ie_1$ has the form
\IN{X}{\cc{d_1}{f_1}{\ldots}}, $\ie_1'$ has the form
\IN{X}{\cc{d_1'}{f_1'}{\ldots}},    $\ie_2$ has the form
\IN{Y}{\cc{d_2}{f_2}{\ldots}} and  $\ie_2'$ has the form
\IN{Y}{\cc{d_2'}{f_2'}{\ldots}}. 

If $\iv_1 \models \iv_2$ and  $\iv_2$ is not a tautology
($\not \models \iv_2$), then the following holds:
\begin{enumerate}
\item \ag{d_1}=\ag{d_2} and $\mathit{f_1}=\mathit{f_2}$,
\item \ag{d_1'}=\ag{d_2'} and $\mathit{f_1'}=\mathit{f_2'}$,
\item In all states that do not satisfy $\iv_2$, it holds
  ``$\ic_2 \to \ic_1$''. I.e.~each counterexample for $\iv_2$ is also 
  a counterexample for  $\iv_1$.
\end{enumerate}
\end{lemma}

\begin{corollary}[Sufficient Condition for \chkent]\label{notred}\mbox{}\\
  There is a sufficient condition for $\chkent(\iv_1,\iv_2)$ based on
  \chkimp and \chktaut: First check whether $\iv_2 \in \text{Taut}$. If
  yes, $\chkent(\iv_1,\iv_2)$ holds. If not, check whether $\ic_2 \to
  \ic_1$ holds in all states (i.e.~$\chkimp(\ic_2,\ic_1)$).  If yes,
  $\chkent(\iv_1,\iv_2)$ holds.
\end{corollary}

In this  paper, we use  the  following sound but incomplete  $\chkent$
algorithm. Let $\iv_1 : \ic_1 \Longrightarrow \ie_1 ~~\Re_1 ~~\ie_1'$
and $\iv_2 :  \ic_2  \Longrightarrow \ie_2 ~~\Re_2 ~~\ie_2'$.   Then,
$\chkent(\iv_1,\iv_2) = true$ \iffdef
\begin{enumerate}
\item For all states $S$: $S \models \ic_2 \to \ic_1$,
\item ($\Re_1 = \str{\ensuremath{\subseteq}}$ and $\Re_2 =
  \str{\ensuremath{\subseteq}}$) or ($\Re_1 = \str{=}$ and $\Re_2 =
  \str{\ensuremath{\subseteq}}$),
\item $\ie_2 \to\ie_1$,
\item $\ie_1' \to\ie_2'$.
 \end{enumerate}

 \subsection{Computing All Derived Invariants}
 In this section, we define how given a set $\cali$, the set of all
invariants that are entailed by \cali may be computed using the
 selected $\chkimp$ and $\chkent$ algorithms.

The \textbf{Compute-Derived-Invariants} algorithm presented in
Figure~\ref{alg:cdi} takes as input a set of invariants $\cali$, and
returns a set of invariants $\cali^*$, such that every invariant
in $\cali^*$ is entailed by $\cali$.  Although the
\textbf{Compute-Derived-Invariants} algorithm has exponential running
time, it is executed only once at registration-time, and hence the
worst case complexity of the algorithm is acceptable.

\begin{figure}[htb]
\alg{
\par\noindent \textbf{Compute-Derived-Invariants}($\cali$) \\
$X :=\, \cali$;\\
$change :=\, true$;\\
$Done :=\, \emptyset$;\\
\kw{while} $change$ \kw{do}\\
\pq $change :=\, false$\\
\pq \kw{forall} $\iv_i\in X$ \kw{do}\\
\pqq \kw{forall} $\iv_j\in X-\{\iv_i\}$ s.t. $(\iv_i,\iv_j)\notin Done$
\kw{do}\\
\pqqq $derived\_inv_1 :=\, \textbf{combine\_1}(\iv_i,\iv_j,X)$; \\
\pqqq \kw{if} $derived\_\iv_1$ != \NIL ~~\kw{then}\\
\pqqqq $X :=\, X\,\cup\, \{ derived\_inv_1 \}$; $change :=\, true$; \\
\pqqq $derived\_inv_2 :=\, \textbf{combine\_1}(\iv_j,\iv_i,X)$; \\
\pqqq \kw{if} $derived\_inv_2$ != \NIL ~~\kw{then}\\
\pqqqq $X :=\, X\,\cup\, \{ derived\_inv_2 \}$;
$change :=\, true$; \\
\pqqq $derived\_inv_3 :=\, \textbf{combine\_2}(\iv_j,\iv_i)$; \\
\pqqq \kw{if} $derived\_inv_3$ != \NIL ~~\kw{then}\\
\pqqqq $X :=\, X\,\cup\, \{ derived\_inv_3 \}$;
$change :=\, true$; \\
\pqqq $derived\_inv_4 :=\, \textbf{combine\_3}(\iv_j,\iv_i)$; \\
\pqqq \kw{if} $derived\_inv_4$ != \NIL ~~\kw{then}\\
\pqqqq $X :=\, X\,\cup\, \{ derived\_inv_4 \}$;
$change :=\, true$; \\
\pqqq $Done :=\, Done\,\cup\, \{ (\iv_i,\iv_j),(\iv_j,\iv_i)\}$;\\
\kw{Return} $X$.\\
\kw{End-Algorithm}
}
\caption{\textbf{Compute-Derived-Invariants} Algorithm}
\label{alg:cdi}
\end{figure}

\begin{lemma}\label{lemma1} For all $\cali$: 
$\{\iv_1,\iv_2 \}\,\cup\, \cali \models \textbf{Combine\_1}(\iv_1, \iv_2,\cali)$.
\end{lemma}

\textbf{Combine\_1} does not derive all invariants that are logically
entailed by $\cali$. For example from ``$\text{true}\Rightarrow \ie_1 \subseteq
\ie_2$'' and ``$\text{true}\Rightarrow \ie_2 \subseteq \ie_1$'' we can infer
``$\text{true}\Rightarrow \ie_1 = \ie_2$''. We call this procedure, slightly
generalized, \textbf{Combine\_2}.  It is illustrated in
Figure~\ref{alg:combine2}.  The \textbf{unify} routine takes two
invariant expressions and returns the most general unifier if the two
are unifiable, and returns \NIL $\: $if they are not unifiable.

\begin{figure}[htb]
\alg{
\par\noindent \textbf{Combine\_2($\iv_1,\iv_2)$} \\
/* $\iv_1 : \ic_1 ~~\Longrightarrow ~~\ie_1 ~~\Re_1 ~~\ie_1'$ */ \\
/* $\iv_2 : \ic_2 ~~\Longrightarrow ~~\ie_2 ~~\Re_2  ~~\ie_2'$ */ \\

\kw{if} ~~($\Re_1 = \Re_2 = \str{\ensuremath{\subseteq}}$) ~~\kw{then} \\
\pq $\theta := unify(\ie_2, \ie_1')$;\\
\pq $\gamma := unify(\ie_2', \ie_1)$;\\
\pq \kw{if} ($\theta$ != \NIL) and ($\gamma$ != \NIL) \kw{then} \\
\pqq $derived_ic :=\,\textbf{simplify}((\ic_1 \land \ic_2)\theta\gamma)$; \\
\pqq \kw{if} ~~($derived_ic$ = \emph{false}) ~~\kw{then} ~~\kw{Return} ~\NIL; \\
\pqq $derived\_inv :=\, (derived_ic \Longrightarrow (\ie_1)\theta\gamma = (\ie_1')\theta\gamma); $ \\
\pqq \kw{Return} ~$derived\_inv; $\\
\pq \kw{else} ~~\kw{Return} ~\NIL.\\
\kw{else} ~~\kw{Return} ~\NIL.\\
\kw{End-Algorithm}
}
\caption{\textbf{Combine\_2} Algorithm}
\label{alg:combine2}
\end{figure}

Another set we need is the set of all invariant tautologies
\[\text{Taut}\defeq \{\text{true}\Rightarrow \ie_1 \subseteq \ie_2: \, \chkimp(\ie_1, \ie_2) \}. \]
Obviously, all tautologies are satisfied in all states and the
invariant computed in the \textbf{Combine\_2} Algorithm (if it exists)
is entailed by the invariants it is computed from.

\begin{lemma}\label{lemma1-2}
$\{\iv_1,\iv_2 \} \models \textbf{Combine\_2}(\iv_1, \iv_2)$.
\end{lemma}

However, the above sets are still not sufficient. Consider the situation
$\iv_1 : x<0 ~~\Longrightarrow ~~\ie_1 ~~\Re_1 ~~\ie_1'$, and
$\iv_2 : x\geq 0 ~~\Longrightarrow  ~~\ie_1 ~~\Re_1 ~~\ie_1'$. 
Then, we can conclude $\true  ~~\Longrightarrow ~~\ie_1 ~~\Re_1 ~~\ie_1'$.
However, neither \textbf{Combine\_1} nor \textbf{Combine\_2} is able to 
compute this invariant. As a result, we define the final routine,
\textbf{Combine\_3}, given in Figure \ref{alg:combine3}, to capture these cases.

\begin{figure}[htb]
\alg{
\par\noindent \textbf{Combine\_3($\iv_1,\iv_2)$} \\
/* $\iv_1 : \ic_1 ~~\Longrightarrow ~~\ie_1 ~~\Re_1 ~~\ie_1'$ */ \\
/* $\iv_2 : \ic_2 ~~\Longrightarrow ~~\ie_2 ~~\Re_2  ~~\ie_2'$ */ \\

\kw{if} ~~($\Re_1 = \Re_2$) ~~\kw{then} \\
\pq $\theta := \textrm{unify}(\ie_1, \ie_2)$;\\
\pq $\gamma := \textrm{unify}(\ie_1', \ie_2')$;\\
\pq \kw{if} ($\theta$ != \NIL) and ($\gamma$ != \NIL) \kw{then} \\
\pqq $derived_ic := \textbf{simplify}((\ic_1 \lor \ic_2)\theta\gamma)$; \\
\pqq $derived\_inv := derived_ic \Longrightarrow ((\ie_1)\theta\gamma ~\Re_1 ~(\ie_1')\theta\gamma)$;\\
\pqq \kw{Return} ~$derived\_inv$;\\
\kw{Return}  ~\NIL\\
\kw{End-Algorithm} 
}
\caption{\textbf{Combine\_3} Algorithm}
\label{alg:combine3}
\end{figure}

We emphasize in the \textbf{Combine\_3} Algorithm our use of the
\textbf{simplify} routine introduced just after
Definition~\ref{def:comb1}. Our example is captured because $x<0 \ \lor
\ x\geq 0$ is simplified to true. By recursively applying
\textbf{Combine\_3}, one can also handle more complicated intervals
like $x<0 \ \lor \ (x\geq 0 \, \land \, x<1) \ \lor \ x\geq 1$.

\begin{lemma}\label{lemma1-3} 
$\{\iv_1,\iv_2 \} \models \textbf{Combine\_3}(\iv_1, \iv_2)$.
\end{lemma}

\begin{definition}[Operator $C_{\cali}$]
  We associate with any set $\cali$ of invariants, a mapping
  $C_{\cali}: \textsf{INV} \to \textsf{INV}$ 
which maps sets of invariants to sets of invariants, as follows:
\[ \begin{array}{lll@{}llll}
C_{\cali}(X) &\defeq &\{&\textbf{Combine\_1}(\iv_1,\iv_2,X\cup \cali) &\mid&
\iv_1, \iv_2 \in X \cup \cali \}& \cup\\
          &      &\{& \textbf{Combine\_2}(\iv_1,\iv_2) &\mid&
\iv_1, \iv_2 \in X \cup \cali \}& \cup\\ 
&      &\{& \textbf{Combine\_3}(\iv_1,\iv_2) &\mid& \iv_1, \iv_2 \in X \cup \cali \}& \cup\\
  &      & \multicolumn{5}{l}{\cali \cup X}.
\end{array}\]
\end{definition}

\begin{definition}[Powers of $C_{\cali}$]
The powers of $C_{\cali}(X)$ are defined as follows:
\[ \begin{array}{lll}
C_{\cali} \uparrow^0 &:= \text{Taut}\\
C_{\cali} \uparrow^{(i+1)} &:= C_{\cali}( C_{\cali} \uparrow^i)\\
C_{\cali} \uparrow^\omega &:= \bigcup_{i \geq 0} (C_{\cali} \uparrow^i)
 \end{array}\]
 \end{definition}

\begin{proposition}[Monotonicity of $C_{\cali}$]
\label{theo3}
If $X_1 \subseteq X_2$, then $C_{\cali}(X_1) \subseteq C_{\cali}(X_2)$.
\end{proposition}

 \begin{lemma}
\label{theo5}
$C_{\cali}(C_{\cali} \uparrow^\omega) \subseteq C_{\cali} \uparrow^\omega$.
 \end{lemma}

 \begin{lemma}
\label{theo6}
$C_{\cali} \uparrow^\omega \subseteq \{ \iv \mid \cali \models \iv \}$.
 \end{lemma}
 
 What we are really interested in is a \emph{converse} of the last
 lemma, namely that all invariants that follow from $\cali$ can be
 derived. Strictly speaking, this is not the case: we already noticed
 that there are many redundant invariants that follow from $\cali$ but
 are subsumed by others. Such ``redundant'' invariants contribute
 little.  We show below that whenever an invariant is entailed from
 $\cali$ as a whole, it is already entailed by another variant in
 $C_{\cali} \uparrow^\omega$.  This is the statement of our main
 Corollary~\ref{corol1}.


\begin{theorem}[All Entailed \str{\ensuremath{\subseteq}}-Invariants 
are Subsumed in $C_{\cali} \uparrow^\omega$]
\label{theo7}\mbox{}\\
Suppose $\cali \models \iv$. We assume further that all the invariants
are simple and that $\Re = \str{\ensuremath{\subseteq}}$ in all
invariants.  Then, there is $\iv^{\prime} \in C_{\cali}
\uparrow^\omega$ such that $\iv^{\prime}$ entails $\iv$.
\end{theorem}

\begin{corollary}[All Entailed Invariants are Subsumed in $C_{\cali} \uparrow^\omega$]
\label{corol1}
We are now considering arbitrary simple invariants, i.e.~$\Re = \{
\subseteq, = \}$.  If $\cali \models \iv$, then there exists
$\iv^{\prime} \in C_{\cali} \uparrow^\omega$ such that $\iv^{\prime}$
entails $\iv$.
\end{corollary}

The following corollary tells us that if the
implementation of $\chkimp$ and $\chkent$ algorithms used are
complete, then the \textbf{Compute-Derived-Invariants} algorithm
correctly computes all derived invariants.

 \begin{corollary}[Development-Time Check]\label{cor2}
  Suppose $\chkimp$ is a complete implication check, and $\chkent$ is
  a complete subsumption check algorithm.  Then, the set of invariants
  returned by the \textbf{Compute-Derived-Invariants} has the following
  properties:

 \begin{enumerate}
\item Every invariant returned by it is implied by $\cali$ and
\item If an invariant is implied by $\cali$, then there is an
  invariant $\iv^{\prime}$ returned by the \textbf{Compute-Derived-Invariants}
  algorithm that entails $\iv$.
\end{enumerate}
 \end{corollary}

Our results above apply to simple invariants only. The reason is that
in Table~\ref{summary} only a subset of all possible derivable
invariants are listed. For example even if the \chkimp tests do not hold,
then there are still the following nontrivial invariants entailed:
\begin{enumerate}
\item $\text{derived-}\iv_1 : \text{\ic}_1 \,\land \text{\ic}_2 \Longrightarrow
  (\ie_1 \cap \ie_2) ~\Re ~(\ie_1' \cap \ie_2')$
\item $\text{derived-}\iv_2 : \text{\ic}_1 \,\land \text{\ic}_2 \Longrightarrow (\ie_1 \cup \ie_2) ~\Re ~(\ie_1'
  \cup \ie_2')$
\end{enumerate} 

In fact, our framework can be easily extended as follows.
Let $iv_1 : \text{\ic}_1
\Longrightarrow \ie_1 ~\Re_1 ~\ie_1'$ and $iv_2 : \text{\ic}_2
\Longrightarrow \ie_2 ~\Re_2 ~\ie_2'$. In addition to the derived
invariant returned by the \textbf{Combine\_1} algorithm, the new extended
\textbf{XCombine\_1} also returns the derived invariants determined by
Tables~\ref{xsum1} and \ref{xsum2}.

\begin{table}[htb]
\begin{center}
\begin{tabular}{|c|c|c|c|} \hline
$\Re_1$ & $\Re_2$ & \textbf{simplify}($\ic_1, \ic_2$) & derived\_inv \\ \hline
\hline
* & * & \NIL & \NIL \\ \hline 
$\subseteq$ & $\subseteq$ & $\ic'$ & $\ic' \Longrightarrow (\ie_1 \cap \ie_2)
\subseteq (\ie_1' \cap \ie_2')$ \\ 
& & & $\ic' \Longrightarrow (\ie_1 \cap \ie_2) \subseteq (\ie_1 \cap \ie_2')$ \\    
& & & $\ic' \Longrightarrow (\ie_1' \cap \ie_2) \subseteq (\ie_1' \cap \ie_2')$
 \\   
& & & $\ic' \Longrightarrow (\ie_1 \cap \ie_2) \subseteq (\ie_1' \cap \ie_2)$ \\ 
& & & $\ic' \Longrightarrow (\ie_1 \cap \ie_2') \subseteq (\ie_1' \cap \ie_2')$
 \\ \hline 
$\subseteq$ & = & $\ic'$ & $\ic' \Longrightarrow (\ie_1 \cap \ie_2)
\subseteq (\ie_1' \cap \ie_2')$ \\ 
& & & $\ic' \Longrightarrow (\ie_1 \cap \ie_2) = (\ie_1 \cap \ie_2')$ \\
& & & $\ic' \Longrightarrow (\ie_1' \cap \ie_2) = (\ie_1' \cap \ie_2')$ \\
& & & $\ic' \Longrightarrow (\ie_1 \cap \ie_2) \subseteq (\ie_1' \cap \ie_2)$ \\ 
& & & $\ic' \Longrightarrow (\ie_1 \cap \ie_2') \subseteq (\ie_1' \cap \ie_2')$
 \\ \hline
= & $\subseteq$ & $\ic'$ & $\ic' \Longrightarrow (\ie_1 \cap \ie_2)
\subseteq (\ie_1' \cap \ie_2')$ \\ 
& & & $\ic' \Longrightarrow (\ie_1 \cap \ie_2) \subseteq (\ie_1 \cap \ie_2')$ \\
& & & $\ic' \Longrightarrow (\ie_1' \cap \ie_2) \subseteq (\ie_1' \cap \ie_2')$
 \\
& & & $\ic' \Longrightarrow (\ie_1 \cap \ie_2) = (\ie_1' \cap \ie_2)$
 \\
& & & $\ic' \Longrightarrow (\ie_1 \cap \ie_2') = (\ie_1' \cap \ie_2')$ \\ \hline 
= & = & $\ic'$ & $\ic' \Longrightarrow (\ie_1 \cap \ie_2) = (\ie_1' \cap \ie_2
')$ \\ 
& & & $\ic' \Longrightarrow (\ie_1 \cap \ie_2) = (\ie_1 \cap \ie_2')$ \\
& & & $\ic' \Longrightarrow (\ie_1' \cap \ie_2) = (\ie_1' \cap \ie_2')$ \\
& & & $\ic' \Longrightarrow (\ie_1 \cap \ie_2) = (\ie_1' \cap \ie_2)$ \\
& & & $\ic' \Longrightarrow (\ie_1 \cap \ie_2') = (\ie_1' \cap \ie_2')$ \\ \hline
\end{tabular}
\end{center} \caption{\textbf{XCombine} for Arbitrary Invariants}\label{xsum1}
\end{table}

\begin{table}[htb]
\begin{center}
\begin{tabular}{|c|c|c|c|} \hline
$\Re_1$ & $\Re_2$ & \textbf{simplify}($\ic_1, \ic_2$) & derived\_inv \\ \hline
\hline
* & * & \NIL & \NIL \\ \hline 
$\subseteq$ & $\subseteq$ & $\ic'$ & $\ic' \Longrightarrow (\ie_1 \cup \ie_2)
\subseteq (\ie_1' \cup \ie_2')$ \\ 
& & & $\ic' \Longrightarrow (\ie_1 \cup \ie_2) \subseteq (\ie_1 \cup \ie_2')$ \\    
& & & $\ic' \Longrightarrow (\ie_1' \cup \ie_2) \subseteq (\ie_1' \cup \ie_2')$
 \\   
& & & $\ic' \Longrightarrow (\ie_1 \cup \ie_2) \subseteq (\ie_1' \cup \ie_2)$ \\ 
& & & $\ic' \Longrightarrow (\ie_1 \cup \ie_2') \subseteq (\ie_1' \cup \ie_2')$
 \\ \hline 
$\subseteq$ & = & $\ic'$ & $\ic' \Longrightarrow (\ie_1 \cup \ie_2)
\subseteq (\ie_1' \cup \ie_2')$ \\ 
& & & $\ic' \Longrightarrow (\ie_1 \cup \ie_2) = (\ie_1 \cup \ie_2')$ \\
& & & $\ic' \Longrightarrow (\ie_1' \cup \ie_2) = (\ie_1' \cup \ie_2')$ \\
& & & $\ic' \Longrightarrow (\ie_1 \cup \ie_2) \subseteq (\ie_1' \cup \ie_2)$ \\ 
& & & $\ic' \Longrightarrow (\ie_1 \cup \ie_2') \subseteq (\ie_1' \cup \ie_2')$
 \\ \hline
= & $\subseteq$ & $\ic'$ & $\ic' \Longrightarrow (\ie_1 \cup \ie_2)
\subseteq (\ie_1' \cup \ie_2')$ \\ 
& & & $\ic' \Longrightarrow (\ie_1 \cup \ie_2) \subseteq (\ie_1 \cup \ie_2')$ \\
& & & $\ic' \Longrightarrow (\ie_1' \cup \ie_2) \subseteq (\ie_1' \cup \ie_2')$
 \\
& & & $\ic' \Longrightarrow (\ie_1 \cup \ie_2) = (\ie_1' \cup \ie_2)$
 \\
& & & $\ic' \Longrightarrow (\ie_1 \cup \ie_2') = (\ie_1' \cup \ie_2')$ \\ \hline 
= & = & $\ic'$ & $\ic' \Longrightarrow (\ie_1 \cup \ie_2) = (\ie_1' \cup \ie_2
')$ \\ 
& & & $\ic' \Longrightarrow (\ie_1 \cup \ie_2) = (\ie_1 \cup \ie_2')$ \\
& & & $\ic' \Longrightarrow (\ie_1' \cup \ie_2) = (\ie_1' \cup \ie_2')$ \\
& & & $\ic' \Longrightarrow (\ie_1 \cup \ie_2) = (\ie_1' \cup \ie_2)$ \\
& & & $\ic' \Longrightarrow (\ie_1 \cup \ie_2') = (\ie_1' \cup \ie_2')$ \\ \hline
\end{tabular}
\end{center}\caption{\textbf{XCombine} for Arbitrary Invariants}
\end{table}\label{xsum2}

\nop{
We strongly believe that our main theorem and corollary also hold for
this extended version.  We have chosen to employ the
\textbf{Combine\_1} algorithm rather than its extended version in the
experiments.  Since our invariant sets contained general invariants,
this led to a sound but incomplete version of the
\textbf{Compute-Derived-Invariants} algorithm.
}

 \section{Deployment Phase}
 \label{sec:deployment}
 Once an agent is \emph{up and running}, it is continuously confronted
 with requests for its services. One crucial observation is that there
 might be enormous overlap among these requests. These overlaps can be
 exploited if a given set $\mathcal{C}$ of code call conditions are
 merged in a way that executes common portions only once. 
 However, in order to exploit commonalities, we must first
 determine the type of those commonalities, that is we must first identify  
 code call conditions (1) that are \emph{equivalent} to other code call
 conditions, (2) that are \emph{implied} by other code call conditions and (3)
 that \emph{overlap} with other code call conditions. Moreover, given two code
 call conditions $ccc_1$ and $ccc_2$, it might be the case that they are 
 neither equivalent, nor implied, nor overlapped. On the other hand, parts 
 of $ccc_1$ and $ccc2$ maybe equivalent, implied or overlapped. 
 We also want to exploit such cases. This gives rise to the following 
 definition:

\begin{definition}[Sub-Code Call Condition]\label{subccc}
Let $ccc$ = $\chi_1 \& \chi_2 \& \ldots \& \chi_n$ be a code call condition.
$ccc^j:= \, \chi_{i_1} \& \chi_{i_2} \& \ldots \& \chi_{i_j}$, for 
$1 \leq i_1,\ldots i_j \leq n$ and $i_l \neq i_k \ \forall  1 \leq l,k \leq j$
is called a \emph{sub-code call condition} of $ccc$.
\end{definition}

\begin{example}
Let $ccc_1$ = $\chi_1 \& \chi_2 \& \chi_3 \& \chi_4 \& \chi_5$.
Then, $\chi_1 \& \chi_2 \& \chi_3$, $\chi_1 \& \chi_3 \& \chi_5$
and ~$\chi_2 \& \chi_5$ are some sub-code call conditions of $ccc_1$.
\end{example}

Note that a code call condition with $k$ atomic/(in)equality code call
conditions has $2^k$ different sub-code call conditions.  We are now
ready to define equivalent, implied and overlapping sub-code call
conditions $\chi, \chi^{'}$. To do so, we
need to fix the variable(s) in each $ccc$ to which we want to project
(see Definition~\ref{denot}: the sequence $\langle v_{i_1}, \ldots ,
v_{i_k}\rangle$ of variables occurring in $\chi_1$, and the sequence
$\langle v_{i_1}^{'}, \ldots , v_{i_k}^{'}\rangle$ of variables occurring in
$\chi_2$ are important). Often the sequences consist just of one
single variable: this is the case when there is only one non-base
variable occurring in the ccc's. In that case we do not explicitly
mention the sequences.

\begin{definition}[Equivalent (Sub-) CCC]
  Two (sub-) code call conditions $\chi_1$ and $\chi_2$ are said to be
  \emph{equivalent} w.r.t.~the sequences $\langle v_{i_1}, \ldots ,
  v_{i_k}\rangle$ and $\langle v_{i_1}^{'}, \ldots , v_{i_k}^{'}\rangle$,
  denoted by $\chi_1 \equiv \chi_2$, if and only if for all states $S$
  of the agent and all assignments $\mathbf{\theta}$, it is the case
  that
\[ [\chi_1]_{S,\mathbf{\theta},\langle v_{i_1}, \ldots ,
   v_{i_k}\rangle} = [\chi_2]_{S,\mathbf{\theta},\langle v_{i_1}^{'}, \ldots ,
   v_{i_k}^{'}\rangle}.\]
\end{definition}

In the case of equivalent ccc's, we only need to execute one of the sub-code
call conditions. We can use the cached solutions for the other
sub-code call condition. 

\begin{example}
  The ccc $\IN{\var{C}}{\cc{excel}{chart}{\textit{excelFile}, \var{FinanceRec},
      \textit{day}}}$ is equivalent wrt.~the sequences 
$\langle C\rangle $, $\langle C^{'}\rangle $
  to the ccc $\IN{\var{C'}}{\cc{excel}{chart}{\textit{excelFile}, \var{Rec},
      \textit{day}}}$, since the two code call conditions unify with the mgU
  $\gamma = [\var{FinanceRec}/\var{Rec}]$.
\end{example}

\begin{definition}[Implied (Sub-) CCC]
  A (sub-) code call condition $\chi_1$ is said to \emph{imply} another
  (sub-) code call condition $\chi_2$ wrt.~the sequences $\langle
  v_{i_1}, \ldots , v_{i_k}\rangle$ and $\langle v_{i_1}^{'}, \ldots ,
  v_{i_k}^{'}\rangle$, denoted by $\chi_1 \to \chi_2$, if and only if
  for all states $S$ of the agent and all assignments
  $\mathbf{\theta}$, it is the case that
\[ [\chi_1]_{S,\mathbf{\theta},\langle v_{i_1}, \ldots ,
  v_{i_k}\rangle} \subseteq  [\chi_2]_{S,\mathbf{\theta},\langle v_{i_1}^{'},
  \ldots , v_{i_k}^{'}\rangle},\]
and it is not the case that  $\chi_1 \equiv \chi_2$.
\end{definition}

In the case of implied ccc's, we execute and cache the solutions of
$\chi_2$. In order to evaluate $\chi_1$, all we need to do is 
to use the cached results to restrict the solution set of $\chi_2$.

\begin{example}
  The code call condition $\IN{\var{T_1}}{\cc{spatial}{range}{\textit{map}1, 5,
      5, 30}}$ implies the code call condition
  $\IN{\var{T_2}}{\cc{spatial}{range}{\textit{map}1, 5, 5, 50}}$, because all
  the points that are within $30$ units of the point $(5,5)$ are also
  within $50$ units of $(5,5)$. As mentioned above, in this case, we suppress
  the sequences $\langle T_1 \rangle$ and $\langle T_2
  \rangle$.
\end{example}

\begin{definition}[Overlapping (Sub-) CCC]
  Two (sub-) code call conditions $\chi_1$ and $\chi_2$ are said to be
  \emph{overlapping} wrt.~the sequences $\langle v_{i_1}, \ldots ,
  v_{i_k}\rangle$ and $\langle v_{i_1}^{'}, \ldots , v_{i_k}^{'}\rangle$,
  denoted by $\chi_1 \perp \chi_2$, if and only if for some states $S$
  of the agent and for some assignments $\mathbf{\theta}$, it is the
  case that 
\[ [\chi_1]_{S,\mathbf{\theta},\langle v_{i_1}, \ldots ,
  v_{i_k}\rangle} \cap [\chi_2]_{S,\mathbf{\theta},\langle v_{i_1}^{'},
  \ldots , v_{i_k}^{'}\rangle} \neq \emptyset,\]
 and
  neither $\chi_1 \to \chi_2$ nor $\chi_2 \to \chi_1$.
\end{definition}

In the case of overlapping ccc's, we execute and cache the solutions of
$\chi_3$, where $\chi_3$ is a code call condition the  solution of which
is set  
equal to the union of the solution sets of $\chi_1$ and $\chi_2$. In order to
evaluate both $\chi_1$ and $\chi_2$, we need to access the cache and 
restrict the solution set of $\chi_3$ to those of $\chi_1$'s and $\chi_2$'s
solution sets. Note that the definition of overlapping ccc's
requires that the intersection of the solution sets of
$\chi_1$ and $\chi_2$ be non-empty for \emph{some} state of the agent.
This implies there might be states of the agent, where the intersection
is empty. However, the solution set of $\chi_3$ in such a case still
contains the solutions to $\chi_1$ and $\chi_2$.

\begin{example}
  The code call condition
  $\IN{\var{T_1}}{\cc{rel}{rngselect}{\textit{emp}, \textit{age}, 25,
      35}}$ overlaps with the code call condition
  $\IN{\var{T_2}}{\cc{rel}{rngselect}{\textit{emp}, \textit{age}, 30,
      40}}$, because all employees between the ages 30 and 35 satisfy
  both code call conditions.
\end{example}

In order to identify various relationships between code call conditions,
we use the derived invariants that are computed at development phase.
We are now faced with the following problem:

\begin{definition}[Common sub-ccc identification problem]
  Given a set of code call conditions $\mathcal{C}$=$\{ccc_1,
  ccc_2,\ldots, C=ccc_n\}$, and a set of derived invariants,
  $\cali^*$, find all sub-code call conditions of $\bigwedge_{i=1}^n
  ccc_i$ that are
\begin{itemize}
\item equivalent with respect to $\cali^*$,
\item imply one another with respect to $\cali^*$,
\item overlap with each other with respect to $\cali^*$.
\end{itemize}
\end{definition}

The brute-force solution to the above problem is to choose two code call
conditions, $ccc_i$ and $ccc_j$ from $\mathcal{C}$, then traverse the list of
invariants, $\cali^*$, and apply each invariant to various sub-code 
call conditions of $ccc_i$ and $ccc_j$. The algorithm
\textbf{Brute-Force-CSI}, given in Figure~\ref{alg:brute}, implements
this approach.

\begin{figure}[htb]
\alg{
\begin{tabbing}
\par\noindent\textbf{Brute-Force-CSI($\mathcal{C}$,$\cali^*$)}\\
/*  $O$ = $\{ (\chi_i, \chi_j, \chi_k) \mid \chi_i \perp \chi_j ~and
~\chi_i \to \chi_k ~and ~\chi_j \to \chi_k \}$ \=*/ \kill
/* \=\textbf{Input}: ~~~\=$\mathcal{C}$ = $\{ccc_1, ccc_2,\ldots, ccc_n\}$ \ \ \ \=*/ 
\\
/*\>\> $\cali^* = \{\iv_1, \iv_2, \ldots, \iv_m\}$ \>*/ \\
/* \> \textbf{Output}: \> $Eq$ = $\{ (\chi_i, \chi_j) \mid \chi_i \equiv \chi_j\}$ \>*/ \\
/* \>\> $I$ = $\{ (\chi_i, \chi_j) \mid \chi_i \to \chi_j\}$\>*/ \\
/* \>\> $O$ = $\{ (\chi_i, \chi_j, \chi_k) \mid \chi_i \perp \chi_j ~and
~\chi_i \to \chi_k ~and ~\chi_j \to \chi_k \}$ />*/
\end{tabbing}

$SC$ := $\bigwedge_{i=1}^n C_i$ \\
$SC^p$ := all sub-code call conditions of $SC$ \\
\kw{for all} $\chi_i \in$ $SC^p$ \kw{do} \\
\pq \kw{for all} $\chi_j \neq \chi_i \in$ $SC^p$ \kw{do} \\
\pqq \kw{for all} $\iv \in \cali^*$ \kw{do} \\
\pqqq  ApplyInvariant($\iv, \chi_i, \chi_j, Eq, I, O$) \\
\pqqq  ApplyInvariant($\iv, \chi_j, \chi_i, Eq, I, O$) \\
\kw{Return} ($Eq, I, O$) \\
\kw{End-Algorithm}
}
\caption{\textbf{Brute-Force CSI} Algorithm}
\label{alg:brute}
\end{figure}

The \textbf{Brute-Force-CSI} algorithm makes use of an \textbf{ApplyInvariant}
routine which takes as input an invariant and two sub-code call conditions,
as well as the equivalent, implied and overlapped sub-code call conditions
sets. It applies the invariant to the sub-code call conditions, and inserts
the relationship entailed by the invariant into the respective set.  
This routine is given in Figure \ref{alg:appinv}. Note that we need to call
\textbf{ApplyInvariant} twice with different relative orders for $\chi_i$
and $\chi_j$.

\begin{figure}[thb]
\alg{
\par\noindent\textbf{ApplyInvariant($\iv, \chi_i, \chi_j, Eq, I, O$)}\\

\kw{if} ($iv$ is of the form $\ic \Longrightarrow ie_1 = ie_2$) and \\
\pq ($\exists \theta$, such that $\chi_i = (ie_1)\theta$ and
$\chi_j = (ie_2)\theta$ and $(\ic)\theta$ = true) \kw{then} \\
\pqq $Eq$ = $Eq$ $\cup \{(\chi_i, \chi_j)\}$  ~// $\chi_i \equiv \chi_j$\\
\kw{else if} ($\iv$ is of the form $\ic \Longrightarrow ie_1 \subseteq ie_2$)
and \\
\pqqq ~~($\exists \theta$, such that $\chi_i = (ie_1)\theta$ and
$\chi_j = (ie_2)\theta$ and $(\ic)\theta$ = true) \kw{then} \\
\pqqqq ~~$I$ = $I$ $\cup \{(\chi_i,\chi_j)\}$  ~// $\chi_i \to \chi_j$ \\
\kw{else if} ($\iv$ is of the form $\ic \Longrightarrow (ie_1 \cup ie_2) =
ie_3$) and \\
\pqqq ~~($\exists \theta$, such that $\chi_i = (ie_1)\theta$ and
$\chi_j = (ie_2)\theta$ and $(\ic)\theta$ = true) \kw{then} \\
\pqqqq ~~$\chi_k = (ie_3)\theta$ \\
\pqqqq ~~$O$ = $O$ $\cup \{(\chi_i, \chi_j, \chi_k) \}$ ~// $\chi_i \perp \chi_j$\\
\kw{Return} \\
\kw{End-Algorithm}
}
\caption{\textbf{ApplyInvariant} Routine}
\label{alg:appinv}
\end{figure}

Assuming \textbf{ApplyInvariant} takes constant time to execute,
the complexity of the \textbf{Brute-Force-CSI} algorithm is $O(m*2^{2k})$, where 
$m$ is the number of invariants in $\cali^*$ and $k$ is the number of
atomic/(in)equality code call conditions in $\mathcal{SC}$. 
However, one important observation is that we do not 
have to apply \emph{each} invariant to \emph{all}
possible sub-code call conditions.
An invariant expression can only unify with a (sub-) code call condition if 
both contain ``similar'' (sub-) code call conditions. 
The performance of the \textbf{Brute-Force-CSI} algorithm can be significantly
improved by making use of this observation.
But, before describing this improved CSI algorithm, let us first define 
similar (sub-) code call conditions.

\begin{definition}[Similar (sub-) code call conditions]
Two (sub-) code all conditions $\chi_1$ and $\chi_2$ are said to 
be \emph{similar} if one of the following holds:
\begin{itemize}
\item Both $\chi_1$ and $\chi_2$ are atomic code call conditions
  of the form \IN{\cdot}{\cc{d}{f}{\cdot}}.
\item Both $\chi_1$ and $\chi_2$ are equality/inequality code call conditions.
\item $\chi_1$ is of the form $\chi_{11} \& \chi_{12}$ and $\chi_2$ is
  of the form $\chi_{21} \& \chi_{22}$, and $\chi_{11}$ is similar to
  $\chi_{21}$ and $\chi_{12}$ is similar to $\chi_{22}$.
\end{itemize}
\end{definition}

\begin{figure}[htb]
\alg{
\begin{tabbing}
\pq \par\noindent\textbf{Improved-CSI($C$,$\cali^*$)}\\
/* \=Input: ~~~\=$\mathcal{C}$ = $\{ccc_1, ccc_2,\ldots, ccc_n\}$ \ \ \ \= \\
/* \>\> $\cali^* = \{\iv_1, \iv_2, \ldots, \iv_m\}$ \>*/ \\
/* \> Output: \> $Eq$ = $\{ (\chi_i, \chi_j) \mid \chi_i \equiv \chi_j\}$\>*/ \\
/* \>\> $I$ = $\{ (\chi_i, \chi_j) \mid \chi_i \to \chi_j\}$\>*/ \\
/* \>\> $O$ = $\{ (\chi_i, \chi_j, \chi_k) \mid \chi_i \perp \chi_j and
~\chi_i \to \chi_k ~and ~\chi_j \to \chi_k \}$ */ \\
\end{tabbing}

(1) \pq $\{ G_1, G_2, \ldots , G_l\}$ := \textbf{Classify($C$)}; \\
(2) \pq \kw{for all} $G_i$ for $i=1, \ldots l$ \kw{do} \\
(3) \pqq $\cali = \{ \iv \mid \iv ~contains ~similar ~sub-code ~call 
~conditions ~with ~G_i\}$ \\
(4) \pqq \kw{for all} $\chi_j \in ~G_i$ \kw{do} \\
(5) \pqqq \kw{for all} $\chi_k \neq \chi_j \in G_i$ \kw{do} \\
(6) \pqqqq \kw{for all} $\iv \in \cali$ \kw{do} \\
(7) \pqqqqq ApplyInvariant($\iv, \chi_j, \chi_k, Eq, I, O$) \\
(8) \pqqqqq ApplyInvariant($\iv, \chi_k, \chi_j, Eq, I, O$) \\
(9) \pq \kw{Return} ($Eq, I, O$) \\
(10) \pq \kw{End-Algorithm}
}
\caption{\textbf{Improved CSI} Algorithm}
\label{alg:improved}
\end{figure}

The \textbf{Improved-CSI} algorithm is given in
Figure~\ref{alg:improved}.  Lines (1) and (3) of the algorithm need
further explanation.  In order to facilitate fast unification of
sub-code call conditions with invariant expressions, the
\textbf{Classify($C$)} routine in the \textbf{Improved-CSI} algorithm
organizes sub-code call conditions into groups such that each group
contains similar sub-code call conditions.  Example \ref{ex:grp}
demonstrates how \textbf{Classify($C$)} works.

\begin{figure}[htb]
\begin{center}
\input{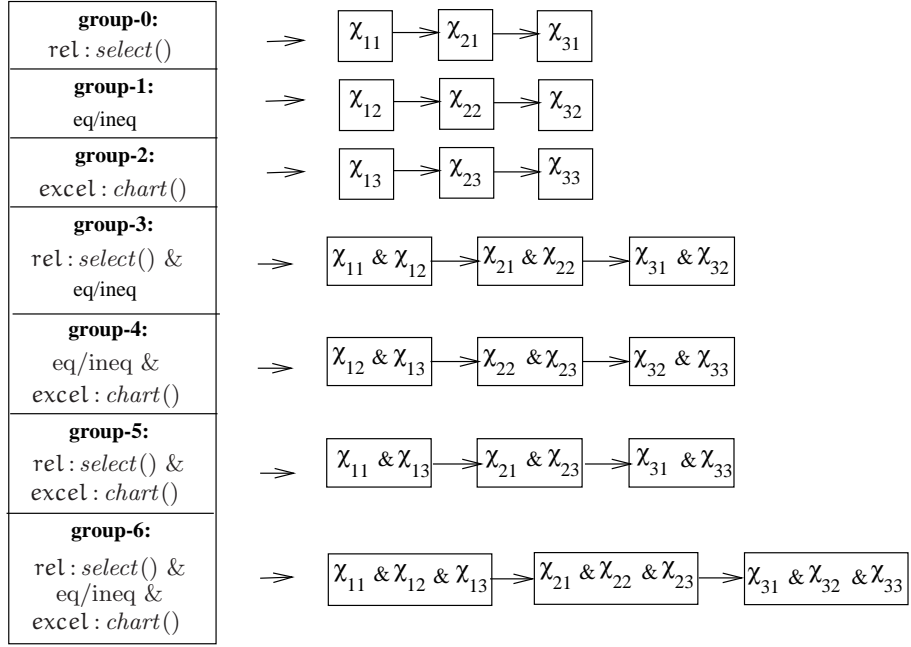}
\caption{Organization of Sub-code Call Conditions of Example \ref{ex:grp}}
\label{classes}
\end{center}
\end{figure}

\begin{example}
\label{ex:grp}
Consider the following code call conditions:
\[\begin{array}{lll}
\chi_{11} &= &\IN{FinanceRec}{\cc{rel}{select}{\textit{financeRel}, \textit{sales}, \str{$\geq$}, 10K}}\\
\chi_{12} &= &\var{FinanceRec.date} \geq \str{6/6/2000} \\
\chi_{13} &= &\IN{C}{\cc{excel}{chart}{\textit{excelFile}, FinanceRec, \textit{day}}}\\
\chi_{21} &= &\IN{FinanceRec}{\cc{rel}{select}{\textit{financeRel}, \textit{sales}, \str{$\geq$}, 20K}}\\
\chi_{22} &= &\var{FinanceRec.date} = \str{7/7/2000} \\
\chi_{23} &= &\IN{C}{\cc{excel}{chart}{\textit{excelFile}, FinanceRec, \textit{day}}}\\
\chi_{31} &= &\IN{FinanceRec}{\cc{rel}{select}{\textit{financeRel}, \textit{sales}, \str{$\geq$}, 30K}} \\
\chi_{32} &= &\var{FinanceRec.date} = \str{7/7/2000} \\
\chi_{33} &= &\IN{C}{\cc{excel}{chart}{\textit{excelFile}, FinanceRec,\textit{month}}}.
\end{array}
\]
Let $ccc_1 = \chi_{11} \& \chi_{12} \& \chi_{13}$, 
$ccc_2 = \chi_{21} \& \chi_{22} \& \chi_{23}$ and
$ccc_3 = \chi_{31} \& \chi_{32} \& \chi_{33}$. Figure \ref{classes} shows
how sub-code call conditions of $ccc_1$, $ccc_2$ and $ccc_3$ are grouped.
\end{example}

Line (3) of the algorithm identifies a subset $\cali \subseteq
\cali^*$ of invariants that are applicable to a given group of
sub-code call conditions. In order to speed up this task, the
invariants are stored in a hash table based on the
\IN{\cdot}{\cc{d}{f}{\cdot}}'s they contain. Given a group of sub-code call
conditions, we apply only those invariants which contain similar
sub-code call conditions (lines (7) and (8)).  The
\textbf{Improved-CSI} algorithm also uses \textbf{ApplyInvariant} to
compute various relationships.  However, the number of times it is
invoked is much smaller than the number of times it is invoked in
\textbf{Brute-Force-CSI} algorithm.  Example \ref{ex:csi} demonstrates
how \textbf{Improved-CSI} algorithm works.

\begin{example}
\label{ex:csi}
The algorithm first processes group-0 (Figure \ref{classes}).  It
identifies all invariants containing
\IN{\cdot}{\cc{rel}{select}{\cdot}} code calls. Then, it tries to apply
  each of those invariants to various combinations of this group. For
  example, the following invariant will unify with pairs of code call
  conditions in this group: \[\begin{array}{c}\var{Rel}=\var{Rel'}
  \,\land \var{Attr}=\var{Attr'} \,\land \var{Op}=
  \var{Op'}=\str{\ensuremath{\geq}} \,\land \var{Val}> \var{Val'}\\
  \Longrightarrow \\\IN{X}{\cc{rel}{select}{Rel,Attr,Op,V}} \subseteq
  \IN{Y}{\cc{rel}{select}{Rel',Attr',Op',V'}}.
\end{array}\]
 As a result of the
  application of this invariant the following relationships are found:
\[\begin{array}{|l|}\hline
\\
     \IN{FinanceRec}{\cc{rel}{select}{\textit{financeRel}, \textit{sales}, \str{\ensuremath{\geq}} 20K}}
     \ \to\\
     \IN{FinanceRec}{\cc{rel}{select}{\textit{financeRel}, \textit{sales}, \str{\ensuremath{\geq}}, 10K}}
     \\ \hline
     \IN{FinanceRec}{\cc{rel}{select}{\textit{financeRel}, \textit{sales}, \str{\ensuremath{\geq}}, 30K}} \to\\
     \IN{FinanceRec}{\cc{rel}{select}{\textit{financeRel}, \textit{sales}, \str{\ensuremath{\geq}}, 10K}}
     \\ \hline \IN{FinanceRec}{\cc{rel}{select}{\textit{financeRel}, \textit{sales},
         \str{\ensuremath{\geq}}, 30K}}
     \to\\

\IN{FinanceRec}{\cc{rel}{select}{\textit{financeRel}, \textit{sales}, \str{\ensuremath{\geq}}
    20K}}\\ 
\\\hline
\end{array}
\] 

The same procedure is applied to group-1 resulting in the discovery of the
following relationships:
\[\begin{array}{c}
 \var{FinanceRec.date} = \str{7/7/2000} \to  \var{FinanceRec.date} \geq \str{6/6/2000} \\
 \var{FinanceRec.date} = \str{7/7/2000} \equiv  \var{FinanceRec.date} = \str{7/7/2000}\\
\end{array}\]

As a result of processing group-2, the following relationship is found:
\[\begin{array}{c}
\IN{C}{\cc{excel}{chart}{\textit{excelFile}, FinanceRec, \textit{day}}}\\ \equiv \\
\IN{C}{\cc{excel}{chart}{\textit{excelFile}, FinanceRec, \textit{day}}}.
 \end{array}\]
 
 We process the other groups similarly. When processing
 group-5, we only apply invariants containing both
 \IN{\cdot}{\cc{rel}{select}{\cdot}} and
   \IN{\cdot}{\cc{excel}{chart}{\cdot}} code calls. Finally, the
     following relationships are found:
\[\begin{array}{|l|l|l|}\hline
\chi_{21} \to \chi_{11} & \chi_{31} \to \chi_{11} & \chi_{31} \to \chi_{21} \\
\chi_{22} \equiv \chi_{32} & \chi_{22} \to \chi_{12} & \chi_{32} \to \chi_{12} \\
\chi_{13} \equiv \chi_{23} & & \\
\chi_{21} \& \chi_{22} \to \chi_{11} \& \chi_{12} & 
\chi_{31} \& \chi_{32} \to \chi_{11} \& \chi_{12} &
\chi_{31} \& \chi_{32} \to \chi_{21} \& \chi_{22} \\
\chi_{22} \& \chi_{23} \to \chi_{12} \& \chi_{13} & & \\
\chi_{21} \& \chi_{23} \to \chi_{11} \& \chi_{13} & & \\
\chi_{21} \& \chi_{22} \& \chi_{23} \to \chi_{11} \& \chi_{12} \& \chi_{13} & &.
\\ \hline
\end{array}
\]
\end{example}

\noindent
It is important to note that in the above algorithms the derived
invariants computed during the development phase are used to match the
sub-code call conditions.  This assumes that the derived invariants
are \emph{complete}, that is they contain all possible relationships
derivable from $\cali$. However, this may be too costly to compute.
Moreover, we may end up storing a lot of invariants which never match
with any of the sub-code call conditions. One solution to this problem
is to restrict the length of invariant expressions in the derived
invariants. However, in that case we need to perform some inferencing
at deployment to make sure that we compute all sub-code call condition
relationships.

Hence, in the case of \emph{incomplete} derived invariants, we also
need to perform a second phase where we use the inference rules in
Table~\ref{inf:rul1} to deduce further relationships.

\begin{table}[htb]
\begin{center}
\begin{tabular}{|c|} \hline
if $\chi_i \equiv \chi_j$ and $\chi_k \equiv \chi_l$ then $(\chi_i \&\ \chi_k)
\equiv (\chi_j \&\ \chi_l)$ \\ \hline
if $\chi_i \equiv \chi_j$ and $\chi_k \to \chi_l$ then $(\chi_i \&\ \chi_k)
\to (\chi_j \&\ \chi_l)$ \\ \hline
if $\chi_i \to \chi_j$ and $\chi_k \to \chi_l$ then $(\chi_i \&\ \chi_k) \to
(\chi_j \&\ \chi_l)$ \\ \hline
\end{tabular}
\caption{Inference Rules Used in \textbf{Improved-CSI} Algorithm}
\label{inf:rul1}
\end{center}
\end{table}

Once the agent identifies equivalent, implied and overlapping
sub-code call conditions in $\mathcal{C}$, it merges those sub-code
call conditions to decrease execution costs. In the next section
we will describe how to merge a set of sub-code call conditions.

 \subsection{Merging Code Call Conditions}
\label{obtain}
In this section, we first describe a technique for evaluating costs of
code call conditions.  We then describe two algorithms---the
\textbf{DFMerge} and the \textbf{BFMerge} algorithms---which are used
to process the set $\mathcal{C}$ = $\{ccc_1,ccc_2,...,ccc_n\}$ of code
call conditions.  Both of these algorithms are parameterized by a
\emph{selection} strategy.  Later, in our experiments, we will try
multiple alternative selection strategies in order to determine which
ones work the best.  We will also compare the performance of the
\textbf{DFMerge} and \textbf{BFMerge} algorithms so as to assess the
efficiency of computation of these algorithms.
 \subsubsection{Cost Estimation for Code Call Conditions}
 In this section, we describe how to estimate the cost of merged code
 call conditions for a set $\mathcal{C}$ = $\{ccc_1,ccc_2,...,ccc_n\}$ of
 code call conditions. We assume that there is a cost
 model that can  assess costs of individual code call conditions.
 Such costing mechanisms have been already developed for heterogeneous
 sources by \cite{du92, adali96, disco:cost, tork99}.  Using this,
 we may state the cost of a single code call condition.


\begin{definition}[Single Code Call Condition Cost]
  The cost of a code call condition ccc is defined as: $\cost(ccc) =
  \sum_{\chi_i \in ccc} \cost(\chi_i)$ where $\cost(\chi_i)$ is the cost of
  executing the atomic or equality/inequality code call condition
  $\chi_i$. Note that the cost of $\chi_i$ may include a variety of
  parameters such as disk/buffer read time, network retrieval time,
  network delays, etc.
\end{definition}

We may now extend this definition to describe the coalesced
cost of executing two code call conditions $ccc_k$ and $ccc_{k+1}$.

\begin{definition}[Coalesced cost]
  The coalesced cost of executing code call conditions $ccc_k$ and
  $ccc_{k+1}$ by exploiting equivalent, implied and overlapped sub-code call
  conditions of $ccc_k$ and $ccc_{k+1}$ is defined as:
\begin{eqnarray*}
\coacost (ccc_k,ccc_{k+1}) &  = &  \cost(ccc_k) + \cost(ccc_{k+1}) - 
\gain(ccc_k,ccc_{k+1})  
\end{eqnarray*}
where $\gain(ccc_k,ccc_{k+1})$ is the cost of the savings obtained by sharing
sub-code call conditions between $ccc_k$ and $ccc_{k+1}$. 
 \end{definition}
 
 We are now left with the problem of defining the concept of \gain
used above.

 \begin{definition}[Gain of two sub-ccc's]
   Suppose $\chi_i$ and $\chi_j$ are sub-code call
   conditions in $ccc_k$ and $ccc_{k+1}$, respectively, and $\cali$ is a
   set of invariants. Then, the gain of executing $\chi_i,\chi_j$ is
   defined as:
   
   {\small \[\gain(\chi_i,\chi_j) = \left \{ \begin{array}{ll}
         \cost(\chi_i) &  \mbox{if $\cali\models \chi_i \equiv \chi_j$} \\
        \cost(\chi_i)-\cost(eval(\chi_i,\chi_j)) & \mbox{if $\cali\models \chi_i
           \to \chi_j$} \\
         exp_k &\begin{tabular}[t]{@{}l@{\ }l}
                                if &$\cali\models \chi_i \perp  \chi_j$ and
           $\cali\models \chi_i \to \chi_k$
           and\\
            &$\cali\models \chi_j \to \chi_k$
            \end{tabular}\\
         0 & \mbox{otherwise}
\end{array} \right. \] }
where $exp_k =
\cost(\chi_i)+\cost(\chi_j)-\cost(\chi_k)-
\cost(\eval(\chi_i,\chi_k))- \cost(eval(\chi_j,\chi_k))$ and
$\eval(\chi_i,\chi_j)$ is the task of executing
code call condition $\chi_i$ by using the results of code call condition
$\chi_j$.
\end{definition}


An explanation of the above definition is important.  If code call
conditions $\chi_i$ and $\chi_j$ are equivalent, then we only need to
execute one of them, leading to a saving of $\cost(\chi_i)$.  If $\chi_i
\to \chi_j$ (i.e.  $\chi_j$'s answers include those of $\chi_i$) then we can
first evaluate $\chi_j$, and then select the answers of $\chi_i$ from the
answers to $\chi_j$.  A third possibility is that $\chi_i$ and $\chi_j$
overlap, and there exists a code call condition $\chi_k$ such that
$\chi_k$ is implied by both $\chi_i,\chi_j$.  In this case, we can compute
$\chi_k$ first, and then use the result to select the answers of
$\chi_i,\chi_j$.  The cost of this is
$\cost(\chi_k)+\cost(\eval(\chi_i,\chi_k))+ \cost(\eval(\chi_j,\chi_k))$.  As
the cost of executing $\chi_i,\chi_j$ sequentially is
$\cost(\chi_i)+\cost(\chi_j)$, the gain is computed by taking the
difference, leading to the third expression.
We now define the gain for two code call conditions
in terms of the gains of their sub-code call conditions involved.

 \begin{definition}[Gain of two code call conditions]
   The gain for $ccc_k$ and $ccc_{k+1}$ is defined as:
\begin{eqnarray*}
\gain(ccc_k,ccc_{k+1}) & = & \sum_{\chi_i \in ccc_k, \chi_j \in ccc_{k+1}} 
\gain(\chi_i,\chi_j).
\end{eqnarray*}
\end{definition}

 \begin{example}
   Consider the following code call conditions: 
\[\begin{array}{ll}
\chi_1:&
   \IN{FinanceRec}{\cc{rel}{select}{\textit{financeRel}, \textit{sales}, \str{\ensuremath{\geq}},
       20K}},\\
\chi_2:& \IN{C}{\cc{excel}{chart}{C, FinanceRec, \textit{day}}},\\
\chi_3: & \IN{FinanceRec}{\cc{rel}{select}{\textit{financeRel},
       \textit{sales}, \str{\ensuremath{\geq}}, 10K}}
\end{array}\]
 Let $ccc_1 = \chi_1 \,\&\, \chi_2$ and $ccc_2 = \chi_3$.
   It is evident that $\text{ANS}(\chi_1) \subseteq \text{ANS}(\chi_3)$.  Suppose further that
   the costs of these code call conditions are given as: \cost($\chi_1$)
   = 25, \cost($\chi_2$) = 10, \cost($\chi_3$) = 15 and \cost(\eval($\chi_1,
   \chi_3$))=10. Then, \gain($ccc_1,ccc_2$) = $\sum_{\chi_i \in ccc_1, \chi_j \in ccc_2}
   \gain(\chi_i,\chi_j)$ = \gain($\chi_1, \chi_3$) because \gain($\chi_2,
   \chi_3$)=0, as there is no relation between code call conditions
   $\chi_2$ and $\chi_3$.  As $\chi_1 \to \chi_3$, \gain($\chi_1, \chi_3$) =
   \cost($\chi_1$)- \cost(\eval($\chi_1, \chi_3$))= 25 - 10 =15. Then, the
   coalesced cost of $ccc_1$ and $ccc_2$ is given by, $\coacost(ccc_1,ccc_2) =
   \cost(ccc_1) + \cost(ccc_2) - \gain(ccc_1,ccc_2$) = (25 + 10) + 15 - 15 =
   35.
\end{example}

 \subsubsection{Merging Code Call Conditions}
 We now develop two algorithms that produce a
 \emph{global} merged code call evaluation graph for a set,
 $\mathcal{C}$ = $\{ccc_1,ccc_2,...,ccc_n\}$ of code call conditions.
 These algorithms use the cceg representation of each code call
 condition $ccc_i$, and merge two graphs at a time until all graphs
 are merged into a final global code call evaluation graph.  They make
 use of a \textbf{merge} routine which merges code call evaluation
 graphs by using the $Eq, I$ and $O$ sets generated by the
 \textbf{Improved-CSI} algorithm. 


While merging two code call evaluation graphs, the
\textbf{merge} routine may need to delete some nodes from the ccegs.
Recall that in a cceg, a node represents
an atomic/(in)equality code call condition.
The following procedure is applied recursively to delete a node $\chi_i$
from a code call evaluation graph:
\begin{enumerate}
\item First the node $\chi_i$ is
removed.
\item Then all incoming edges $(\chi_j,\chi_i)$ and all outgoing
edges $(\chi_i,\chi_l)$ are deleted.
\item If any of the nodes $\chi_j$, encountered in the
previous step, has no outgoing edges, then node $\chi_j$
is also deleted recursively.
\end{enumerate}

The \textbf{merge} routine uses a
set of three transformations which we define now.  The first
transformation takes a set of graphs of equivalent code call conditions
and creates a single graph.

\begin{definition}[Equivalence Transformation, T1]
Let $\mathcal{C}$ = $\{C_1, C_2, \ldots, C_m\}$   be of code call
conditions. Let  $\mathcal{CCEG}$ = $\{cceg(C_1), cceg(C_2), \ldots, cceg(C_m)\}$
be their code call evaluation graphs.
Given a set $Eq$ of equivalent code call conditions, which are sub-cccs
of the $C_i$'s, the equivalence transformation \textbf{T1}
is defined as follows:
\begin{description}
\item[T1:] 
\begin{tabbing} 
$cceg$($\mathcal{C}$) = $(\bigcup_{1 \leq i \leq m}V_i,
~~\bigcup_{1 \leq i \leq m}E_i)$ \\
$Eq'$ := \=$\{(\chi_i, \chi_j) \mid$\=$(\chi_i, \chi_j) \in Eq \text{ and }
\nexists (\chi_i', \chi_j') \in Eq \text{ such that $\chi_i$ is a sub-ccc}$\\
\>\> of $\chi_i'$ and $\chi_j$ is a sub- ccc of $\chi_j'$\,$\}$ \\
\mbox{}\\
\kw{for all}\ $(\chi_i, \chi_j) \in Eq'$ \kw{do} \\
\> \kw{if} \=$\gain$($\chi_i,\chi_j) > 0$ ~\kw{then} \\
\>\>delete all the nodes corresponding to atomic cccs in
$\chi_i$\\
\>\> from $cceg$($\mathcal{C}$) recursively \\
\>\>delete all outgoing edges $\langle \chi_i,\chi_k \rangle \in$
$cceg$($\mathcal{C}$) \\
\>\>create the edges $\langle \chi_j,\chi_k \rangle \in$
$cceg$ ($\mathcal{C}$)
\end{tabbing}
\end{description}
\end{definition}

The second transformation \textbf{(T2)} below takes a
set of graphs of code call conditions, together with a
set of known implications between sub-code call conditions
of these code call conditions.  Using these known implications,
it merges these graphs into one.

\begin{definition}[Implication Transformation, T2]
Let  $\mathcal{C}$ = $\{C_1, C_2, \ldots, C_m\}$   be of code call
conditions. Let $\mathcal{CCEG}$ = $\{cceg(C_1), cceg(C_2), \ldots, cceg(C_m)\}$
be their code call evaluation graphs.
Given a set $I$ of implied code call conditions, which are sub-cccs
of the $C_i$'s, the implication transformation \textbf{T2}
is defined as follows:
\begin{description}
\item[T2:] 
\begin{tabbing} 
$cceg$($\mathcal{C}$) = $(\bigcup_{1 \leq i \leq m}V_i,
~~\bigcup_{1 \leq i \leq m}E_i)$ \\
$I'$ := \=$\{(\chi_i, \chi_j) \mid$\=$(\chi_i, \chi_j) \in I \text{
  and }
\nexists (\chi_i', \chi_j') \in I \text{ such that $\chi_i$ is a sub-ccc}$\\
\>\> of $\chi_i'$ and $\chi_j$ is a sub- ccc of $\chi_j'$\,$\}$ \\
\mbox{}\\
\kw{for all}\  $(\chi_i, \chi_j) \in I'$\  \kw{do} \\
\> \kw{if}\  $\gain$\=($\chi_i,\chi_j) > 0$\  \kw{then} \\
\>\> delete all incoming edges $\langle \chi_l,\chi_i \rangle \in$
$cceg$($\mathcal{C}$) to $\chi_i$ recursively \\
\>\> create the edge $\langle \chi_j,\chi_i \rangle \in$ $cceg$($\mathcal{C}$)\\
\>\> set \cost($\chi_i$) to \cost(\eval($\chi_i,\chi_j$))
\end{tabbing}
\end{description}
\end{definition}

The third  transformation \textbf{(T3)} below takes a
set of graphs of code call conditions, together with a
set of known overlaps between sub-code call conditions
of these code call conditions.  Using these known overlaps,     
it merges these graphs into one.

\begin{definition}[Overlap Transformation, T3]
We consider the set $\mathcal{C}$ = $\{C_1, C_2, \ldots, C_m\}$   of code call
conditions. Let $\mathcal{CCEG}$ = $\{cceg(C_1), cceg(C_2), \ldots, cceg(C_m)\}$
be their code call evaluation graphs.
Given a set $O$ of overlapping code call conditions, which are sub-cccs
of $C_i$'s, the overlap transformation \textbf{T3} is defined as follows:
\begin{description}
\item[T3:] 
\begin{tabbing} 
$cceg$($\mathcal{C}$) = $(\bigcup_{1 \leq i \leq m}V_i,
~~\bigcup_{1 \leq i \leq m}E_i)$ \\
$O'$ := \=$\{(\chi_i, \chi_j, \chi_k) \mid$\=$(\chi_i, \chi_j) \in O \text{
  and }
\nexists (\chi_i', \chi_j', \chi_k') \in O \text{ such that $\chi_i$ is a sub-ccc}$\\
\>\> of $\chi_i'$ and $\chi_j$ is a sub- ccc of $\chi_j'$\,$\}$ \\
\mbox{}\\
\kw{for all}\  $(\chi_i, \chi_j) \in O'$\  \kw{do} \\
\> \kw{if}\  $\gain$\=($\chi_i,\chi_j) > 0$ \  \kw{then} \\
\>\> create a node $\chi_k \in$ $cceg$($\mathcal{C}$)\\
\>\> create edges $\langle \chi_k, \chi_i \rangle \in$ $cceg$($\mathcal{C}$)
and $\langle \chi_k, \chi_j \rangle \in$ $cceg$($\mathcal{C}$)\\
\>\> delete all incoming edges $\langle \chi_l,\chi_i \rangle \in$
$cceg$($\mathcal{C}$) to $\chi_i$ recursively \\
\>\> delete all incoming edges $\langle \chi_m,\chi_j \rangle \in$
$cceg$($\mathcal{C}$) to $\chi_j$ recursively \\
\>\> create edges $\langle \chi_l,\chi_k \rangle \in$ $cceg$($\mathcal{C}$)
and $\langle\chi_m,\chi_k\rangle \in$ $cceg$($\mathcal{C}$)\\
\>\> set \cost($\chi_i$) to \cost(\eval($\chi_i,\chi_k$)) \\
\>\> set \cost($\chi_j$) to \cost(\eval($\chi_j,\chi_k$))
\end{tabbing}
\end{description}
\end{definition}


The merge routine merely applies the above three transformations 
sequentially in the order \textbf{T1),(T2),(T3)}.
\begin{definition}[The \textbf{Merge} Routine]
The \textbf{merge} routine takes as input a set of code call evaluation
graphs, and the sets of equivalent, implied and overlapped sub-code
call conditions, and uses \textbf{T1, T2} and \textbf{T3} to produce
a single code call evaluation graph. It is given by the following:
\begin{center}
\textbf{merge}($\mathcal{CCEG}$, $Eq, I, O$) = 
$\mathbf{T3}( \mathbf{T2}( \mathbf{T1}$($\mathcal{CCEG}$, $Eq), I), O)$. 
\end{center}
\end{definition}
 
The merge routine works as follows: First, it gets the sets of
equivalent, implied and overlapped sub-code call conditions from the
\textbf{Improved-CSI} algorithm. Then, it applies the
\textbf{merge}-transformations in the order: \textbf{T1, T2, T3}.  The
intuition behind this order is the fact that the maximum gain is
obtained by merging equivalent code call conditions.

The merge routine can be utilized with any search paradigm (e.g.
depth-first search, dynamic programming, etc.) to obtain an algorithm
which creates a ``global'' code call evaluation graph. In
Figures~\ref{dfmerge} 
and~\ref{bfmerge}, we provide two algorithms that use
the \textbf{merge} routine to create a global code call evaluation
graph. Both algorithms merge two graphs at a time until a single graph
is obtained. The \textbf{DFMerge} algorithm starts with the empty
graph, and chooses the next ``best'' code call evaluation graph to
merge with the current global code call evaluation graph.  This
process is iteratively executed.  On the other hand, the
\textbf{BFMerge} algorithm picks the ``best'' pair of code call
evaluation graphs to merge from the ToDo list, which initially
contains all code call evaluation graphs.  Upon merging, the merged
code call evaluation graph replaces the two code call evaluation
graphs being merged. This process is executed iteratively till
only one code call evaluation graph remains in the ToDo list.

\begin{figure}[htb]
\alg{
\begin{tabbing}
\textbf{DFMerge}($\mathcal{CCEG}, Eq, I, O$) \\ 
/* \=\textbf{Input}: $\mathcal{CCEG} = \{cceg_1,...,cceg_n\}$ \=*/\\
/*\>\textbf{Output}: a global cceg \>*/ \\
$\ $\\
$ToDo :=\, \mathcal{CCEG}$; \\
$currentGraph :=\, \text{selectNext}(ToDo, \NIL);$ \\
delete $currentGraph$ from $ToDo$; \\
\kw{while} ($ToDo$ is not empty) \kw{do} \\
\> $nextGraph :=\, \text{selectNext}(ToDo, currentGraph);$ \\
\> delete $nextGraph$ from $ToDo$; \\
\> $currentGraph :=\, \textbf{merge}(\{currentGraph, nextGraph\}, Eq, I, O$); \\
\kw{Return} $currentGraph$; \\
\kw{End-Algorithm}
\end{tabbing}
}
\caption{\protect\textbf{DFMerge} Algorithm}
\label{dfmerge}
\end{figure}

\begin{figure}[htb]
\alg{
\begin{tabbing}
\textbf{BFMerge}($\mathcal{CCEG}, Eq, I, O$) \\
/* \=\textbf{Input:} $\mathcal{CCEG} = \{cceg_1,...,cceg_n\}$ \=*/\\
/*\>\textbf{Output:} a global cceg \>*/ \\
$\ $\\
$ToDo :=\, \mathcal{CCEG}$; \\
\kw{while} (card($ToDo$) $> 1$ ) \kw{do} \\
\pq ($cceg_i,cceg_j$) :=\, selectNextPair($ToDo$); \\
\pq delete $cceg_i,cceg_j$ from $ToDo$; \\
\pq $newGraph$ :=\, merge($\{cceg_i,cceg_j\}, Eq, I, O$); \\
\pq insert $newGraph$ into $ToDo$; \\
\kw{Return}  $cceg_i \in ToDo$; \\
\kw{End-Algorithm}
\end{tabbing}
}
\caption{\textbf{BFMerge} Algorithms}
\label{bfmerge}
\end{figure}

The success of both the \textbf{DFMerge} and \textbf{BFMerge} algorithms
depends very much on how the next ``best'' merge candidates) are
selected.  Below, we present three alternative strategies for doing
this which we have used in our experiments.

 \begin{description}
 \item[Strategy 1:]\mbox{}\\[-.1cm]
\begin{description}
\item[DFMerge:] Choose the graph which has the largest number of
  equivalent code call conditions with the \emph{currentGraph}.
\item[BFMerge:] Choose a pair of graphs which have the largest number
of equivalent code call conditions.
\end{description}
\item[Strategy 2:]\mbox{}\\[-.1cm]
\begin{description}
\item[DFMerge:] Choose the graph which has the largest number of
  equivalent, implied or overlapped code call conditions in common with
  the \emph{currentGraph}.
\item[BFMerge:] Choose a pair of graphs which have the largest number
  of equivalent, implied or overlapped code call conditions between the
  two of them.
\end{description}
\item[Strategy 3:]\mbox{}\\[-.1cm]
\begin{description}
\item[DFMerge:] Choose the graph which leads to the greatest gain with
  the the \emph{currentGraph}.
\item[BFMerge:] Choose the pair of graphs the associated gain of which 
  is
  maximal.
\end{description}
\end{description}

\subsubsection{Executing The Global CCEG}

The final problem that needs to be addressed is to find an execution 
order for the global code call evaluation graph. Any topological
sort of the global cceg is a valid execution order. However, there might
be several topological sorts that can be obtained from the global cceg,
and some of them might be preferable to others. For example, a topological
sort that gives preferences to certain nodes, i.e. outputs them
earlier in the sequence, might be desirable. In order to find
such an execution order, we compute weights for topological sorts.

\begin{definition}[Weight of a topological sort]
Let $\pi$ be a topological sort, and $\weight_{\pi(i)}$ be the weight
of the $i$th node in the topological sort. If we have $n$ total
nodes, the weight of $\pi$, denoted by $\weight(\pi)$, is given by
\[ \weight(\pi) = \sum_{i=1}^n i*\weight_{\pi(i)} \]
\end{definition}

Any topological sort that minimizes $\weight(\pi)$ gives  a desirable
execution order. Besides, we can implement various strategies
with this function simply by assigning weights accordingly. For example,
if we want to favor nodes that output results, we can assign larger weights
to such nodes. In order to find the topological sort with the minimum
$\weight(\pi)$, we use a modified topological sort algorithm which is
given in Figure \ref{alg:tsort}.

\begin{figure}[htb]
\alg{
\begin{tabbing}
\par\noindent \textbf{FindExecutionOrder($cceg$)} \\
/* \=Output: a topological sort $\pi$ that minimizes $\weight(\pi)$ \ \=*/ \kill
/*\>\textbf{Input}:  global $cceg$ \>*/\\
/*\>\textbf{Output}: a topological sort $\pi$ that minimizes $\weight(\pi)$ \>*/ \\

$D$ := $\{ v \mid v \text{ has indegree } 0\}$ \\
\kw{while} $D$ is not empty \kw{do} \\
\> $v'$ := node with the heighest weight in $D$,\\
\> output $v'$, \\
\> remove $v'$ from $D$, \\
\> delete all outgoing edges of $v'$,\\
\> $D$ := $D \cup \{ v \mid v \text{ has in-degree } 0, ~v ~\notin ~D\}$\\
\kw{End-Algorithm}
\end{tabbing}
}
\caption{Modified Topological Sort Algorithm That Finds the Minimal $\weight(\pi)$}
\label{alg:tsort}
\end{figure}

\section{Experiments}
\label{sec:exps}
We ran various sets of experiments on a Sun Ultra1 machine with 320 MB
memory running Solaris 2.6. In the first set of these experiments, we
study the execution time of the \textbf{Create-cceg} algorithm. Specifically,
we evaluate the performance of this algorithm with varying
number of dependencies and conjunctions in the code call conditions.
In the second set of the experiments, we study the execution time of the
development phase component. In
particular, we study the trade-offs involved in the generic \chkimp
~Algorithm.  In the last set of experiments, we demonstrate the
efficiency of the \textbf{Improved-CSI} Algorithm, as well as
the merging algorithms. We compare the performance of the merging
algorithms (with different strategies) with the $A^*$ algorithm of
\cite{sellis88}.  Our implementation of the development phase and
deployment phase components involved over $9,500$ lines of \emph{C++} code.

\subsection{Performance Evaluation of the \textbf{Create-cceg} Algorithm}
\label{exp:cceg}

To evaluate the performance of the \textbf{Create-cceg} algorithm, 
we generated several code call conditions, with
varying number of conjuncts and number of dependencies.
In the first set of experiments, we kept the number of
dependencies constant and varied the number of conjuncts from 5 to 40.
We repeated the same experiments when
10, 15, 20 and 25 dependencies are present. 
For each combination of number of dependencies and conjuncts,
we created 500 code call conditions and recorded the average running time.  
Figure \ref{conj} shows the results. As seen from the figure,
the execution time increases linearly with the number of conjuncts.
The \textbf{Create-cceg} algorithm is extremely fast,
taking only 14 milliseconds for code call conditions involving 40 
conjunctions and 25 dependencies. 

\begin{figure*} [htb]
\begin{center}
\psfig{figure=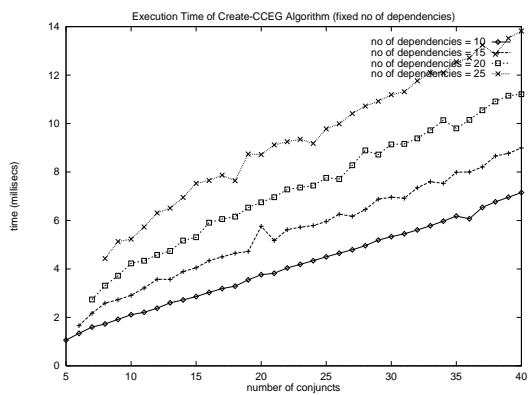,width=2.8in, height=2in}
\caption{Execution Time of \textbf{Create-cceg} (constant number of dependencies)}
\label{conj}
\end{center}
\end{figure*}

In the second, set of experiments, we kept the number of conjuncts constant,
and varied the number of dependencies from 10 to 50. 
We ran four experiments with 10, 20, 30 and 40 number of conjuncts.  Again,
we generated 500 code call conditions for each combination and used the 
average running time. The results are given in Figure \ref{dep}.
Again, the execution time increases linearly with the number of 
dependencies.

\begin{figure*} [htb]
\begin{center}
\psfig{figure=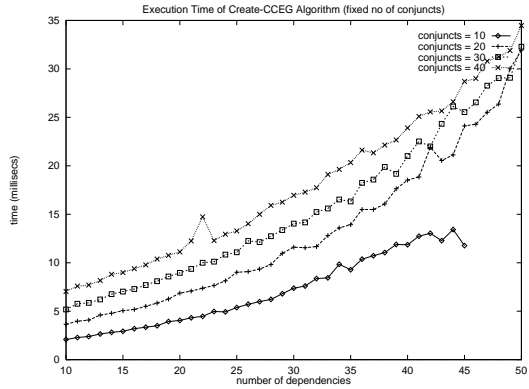,width=2.8in, height=2in}
\caption{Execution Time of \textbf{Create-cceg} (constant number of conjuncts)}
\label{dep}
\end{center}
\end{figure*}

\subsection{Performance Evaluation of the 
Development Phase Component}
\label{exps:compile}
In order to evaluate the performance of the development phase
component, we conducted a set of experiments which use the
\chkent~algorithm described in Section~\ref{sec:development} and different
instances of the \chkimp ~Algorithm.  We varied the \emph{threshold}
and the \textbf{Axiomatic Inference System} used by \chkimp.  The
instances we used are described in Table \ref{table:inst}.  As the
instance number increases, the complexity of the $\chkimp$ ~Algorithm
also increases.

\begin{table}[htb]
{\small
\begin{center}
$\begin{array}{|l|c|c|}\hline
\textbf{Instance} & \textbf{Threshold} & \textbf{Axiomatic Inference System}\\ \hline
\text{Instance $0$} & \infty & \begin{array}[t]{lll}
\chi &\subseteq& \chi\\
\chi \cap \chi' &\subseteq& \chi\\
\chi &\subseteq& \chi \cup \chi'
\end{array} \\ \hline
\text{Instance }i &  i & \text{All rules in Appendix~\ref{rules}.}\\ \hline
\text{Instance }\omega & \infty & \text{All rules in Appendix~\ref{rules}.}\\ \hline
\end{array}$
\end{center}
}
 \caption{Instances of the \chkimp Algorithm}
\label{table:inst}
\end{table}

We ran a set of experiments with two different data sets, namely
spatial domain invariants and the relational domain invariants, which
are given in Appendix \ref{invs}.  For each instance of the algorithm
we ran the development phase component several times until we get an
accuracy of $3 \%$, with $3\%$ confidence interval.  Figure \ref{compile}
shows the execution time of the \textbf{Compute-Derived-Invariants}
algorithm for these two data sets.  As the only difference is the
\chkimp ~Algorithm instance employed, the x-axis is labeled with those
instances.

\emph{Note that the $x$-axis used a logarithmic-scale} and hence, we may
conclude that execution time increases linearly with the instance
number, until instance $4096$, and increases exponentially after that.
However, we have observed that all instances starting from instance $4$,
produced the same final set of \textbf{Derived Invariants}, $18$
invariants for the spatial domain, and $15$ invariants for the
relational domain.  For the relational domain invariants, the
execution-time increases more rapidly than the spatial case. The
observed time increase is due to the time spent in detecting failure.
Memory overflows prevented us from running experiments with larger
\emph{threshold} values.

\begin{figure*} [htb]
\begin{center}
\psfig{figure=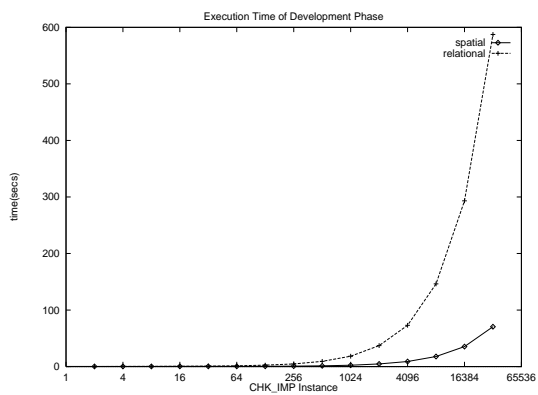,width=2.8in, height=2in}
\caption{Execution Time of \textbf{Compute-Derived-Invariants}}
\label{compile}
\end{center}
\end{figure*}

\subsection{Performance Evaluation of the Deployment Phase Component}
For the performance evaluation of the deployment phase component, we
ran experiments to evaluate both the execution times of the
merging algorithms and the net savings obtained by the algorithms.  We
will describe the experimental setting in detail in the following:

In the experiments we assume a \ag{hdb} agent that accesses relational
and spatial (PR-quadtree) data sources.  We have built cost estimation
modules for these two sources where the cost calculations are similar
to those of \cite{salz}.  We also built an agent cost module which
coordinates with the above two modules to estimate the cost of a code
call condition.  The individual cost estimation modules report the
cost and the cardinality of their code call conditions to the agent
cost module.  The agent cost model also includes network costs.  For
the experiments, it is assumed that the data sources and the agent are
on a fast Ethernet LAN. We created a synthetic database schema given
below, and used the cost estimates in the experiments.

\begin{center}
  {\tt supplier(sname, pno, quantity) }\\
  {\tt product(pno, price, color) } \\
  {\tt map(name, x-location, y-location)}\\
  {\tt purchases(customer\_name, pno)} \\
\end{center}

%

We used the ccc templates given in Table \ref{templs} in the
experiments. In Table \ref{templs}, $Op = \{ \leq, =, \geq \}$.
Note that the last entry in Table~\ref{templs} involves only
relational data sources.

\begin{table}[htb]
\begin{center}
\begin{tabular}{|p{4.2in}|}\hline
Code Call Condition Template \\ \hline
$\IN{T1}{\cc{rdb1}{select2}{supplier, pno, \str{=}, Val1, qty, Op1, Val2}}$ \&\
$\IN{P}{\cc{quadtree}{range}{\textit{map}, X, Y, Rad}}$ \&\ $\mbox{=(T1.sname, P.name)}$
\&\ $\IN{T2}{\cc{rdb2}{select}{product, price, Op2, Val3}}$ \&\
$\mbox{=(T1.pno, T2.pno)}$ \\ \hline
$\IN{T1}{\cc{rdb1}{select2}{supplier, pno, \str{=}, Val1, qty, Op1, Val2}}$ \&\
$\IN{P}{\cc{quadtree}{range}{\textit{map}, X, Y, Rad}}$ \&\ $\mbox{=(T1.sname, P.name)}$
\&\ $\IN{T2}{\cc{rdb2}{rngselect}{product, price, Val3, Val4}}$ \&\
$\mbox{=(T1.pno, T2.pno)}$ \\ \hline
$\IN{T1}{\cc{rdb1}{rngselect}{supplier, qty, Val1, Val2}}$ \&\
$\mbox{=(T1.pno, Val3)}$ \&\ $\IN{P}{\cc{quadtree}{range}{\textit{map}, X, Y, Rad}}$
\&\ $\mbox{=(T1.sname, P.name)}$
\&\ $\IN{T2}{\cc{rdb2}{select}{product, price, Op2, Val3}}$ \&\
$\mbox{=(T1.pno, T2.pno)}$ \\ \hline
$\IN{T1}{\cc{rdb1}{rngselect}{supplier, qty, Val1, Val2}}$ \&\
$\mbox{=(T1.pno, Val3)}$ \&\
$\IN{T2}{\cc{rdb2}{select}{product, price, Op2, Val4}}$ \&\
$\mbox{=(T1.pno, T2.pno)}$ \&\
$\IN{T3}{\cc{rdb3}{rngselect}{purchases, pno, Val5, Val6}}$ \&\
$\mbox{=(T1.pno, T3.pno)}$ \\ \hline
\end{tabular}
\caption{Query Templates Used in the Experiments}
\label{templs}
\end{center}
\end{table}

By changing constants in these template code call conditions, we have
created various commonality relationships. We have constructed the
following three types of code call condition sets by using the above
templates. 

\begin{itemize}
\item[] \emph{Type 1:} Such sets of code call conditions only contain
    equivalent code call conditions.
\item[] \emph{Type 2:} 
Such sets of code call conditions only contain
both equivalent and implied code call conditions.
\item[] \emph{Type 3:}
Such sets of code call conditions contain
equivalent,implied and overlapping code call conditions.
\end{itemize}

Before describing the experiments, let us first define the metrics we
use in these sets of experiments.

\begin{definition}[Savings Percentage]
  Let $\ccost$ be the initial total cost of the set
  of code call conditions, i.~e., the sum of the individual code call
  condition costs, $\fincost$ be the cost of the global merged code
  call condition produced by the merging algorithm, $\ccccost$ be the
  execution time of the \textbf{Improved-CSI} algorithm and
  $\mergecost$ be the execution time of the merge algorithm employed.
  Then, the savings percentage achieved by the merge algorithm is given
  by:
\[ \text{savings percentage} = \frac{\ccost - \fincost - \ccccost 
  - \mergecost} {\ccost} \]
\end{definition}

We try to capture the net benefit of merging the code call conditions
with the savings percentage metric. Moreover, in order to remedy
the difference between high-cost code call conditions and low-cost
code call conditions, we normalize the savings percentage metric.

 \begin{definition}[Sharing factor]
   Let $\mathcal\mathcal{C}$ = $ \{C_1,..,C_N\}$ be the set of given
   code call conditions. Let $[\chi_1],...,[\chi_m]$ be equivalence
   relations, where each $[\chi_i]$ contains a set of equivalent code
   call conditions and $card([\chi_i]) \geq 2, i=1,..,m$. Let $I = \{
   \chi_i \mid such ~that ~\chi_i \notin [\chi_j]$, j=1,..,m, and
   there exists at least one $\chi_k$, such that $\chi_i \to \chi_k
   \}$. And finally let $0 = \{ (\chi_i,\chi_j,\chi_k) \mid such ~that
   ~\chi_i, \chi_j \notin [\chi_k], k=1,..,m, and ~\chi_i, \chi_j
   \notin I, and ~\chi_i \longleftrightarrow \chi_j, \chi_k \to \chi_i
   and ~\chi_k \to \chi_j \}$.

Then, the sharing factor of this set of code call conditions is given by:
\[ \frac{\sum_{i=1}^{m}card([\chi_i])*card(\chi_i) + \sum_{\chi_i \in I} card(\chi_i) + \sum_{(\chi_i,\chi_j,\chi_k) \in O} card(\chi_k)}{\sum_{i=1}^{N} \sum_{\chi_j \in C_i} card(\chi_j)} \]
\end{definition}

The sharing factor basically gives the percentage of data objects
shared among code call conditions. The intuition behind this metric is
that we expect to see an increasing benefit from merging as the
sharing among the code call conditions increases. In this metric, we
try to avoid counting the cardinality of any code call condition more
than once, so that the sharing factor is between $0\%$ and $100\%$.

In order to compare our algorithms with a well known algorithm 
\cite{ssn94} for
merging multiple \emph{relational database} only  queries using the
$A^*$ algorithm, 
we implemented an adapted version of the $A^*$ of \cite{ssn94}.
We used an improved version of their heuristic function.
We adapted our \textbf{Improved-CSI} algorithm
to work with the $A^*$ algorithm. We enumerated the most
promising $8$ execution plans for each individual ccc and input those
plans to the $A^*$ algorithm.

\cite{sellis90} also uses similar measures. In their case, they only
have equivalent relationships, hence the sharing factor metric is
trivially calculated. In their version of the savings percentage
metric, they only measure the difference between initial cost and the
final cost obtained by merging, and fail to take into account the cost
of achieving that savings. Our experiments show that although the
$A^*$ algorithm finds better global results, the cost of obtaining
those results is so prohibitively high that the $A^*$ algorithm
is often infeasible to use in practice.

In all of the experiments, the algorithms are run several times
to obtain results that are accurate within plus or minus $3\%$, with
a $3\%$ confidence interval.

\subsubsection{The Execution Time of the \textbf{Improved-CSI} Algorithm}

\begin{figure*} [htb]
\begin{center}
\psfig{figure=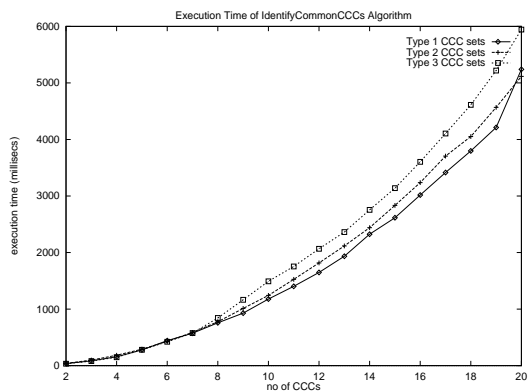,width=2.8in, height=2in}
\caption{Execution Time of of \textbf{Improved-CSI}}
\label{cti}
\end{center}
\end{figure*}

The \textbf{Improved-CSI} algorithm has been ran with the three
types of ccc sets. 
Figure \ref{cti} shows the execution times of the
algorithm as the number of ccc's in the set increases. As seen from the
figure, although the execution time is exponential with a small slope,
it is in the order of seconds. It takes only 6 seconds
for the \textbf{Improved-CSI} algorithm to find all relationships
in a set containing 20 queries. Moreover, the execution time
increases as more types of relationships exist in the ccc sets. It has
the highest execution time for Type 3 ccc sets, and the lowest
execution time for Type 1 ccc sets.

\subsubsection{Savings Achieved by the Merge Algorithms}

In these experiments, we investigate the net savings the merge
algorithms achieve for our three different types of ccc sets, as well
as for ccc sets involving only relational sources.  We have 10
ccc's in each set. The reason for this is that the $A^*$ algorithm
exhausts memory for ccc sets having more than 11-12 ccc's.

Figures \ref{exp1}, \ref{exp2} and \ref{exp4} show the savings
percentage achieved for Type 1, 2 and 3 ccc sets, respectively.  As
seen from Figure \ref{exp1}, the $A^*$ algorithm performs as well as
our merge algorithms once the sharing factor exceeds approximately
30\%. We have not been able to run the $A^*$ algorithm for low sharing
factors because of the memory problem.  The $A^*$ algorithm has an
effective heuristic function for equivalent ccc's, hence it is able to
obtain high quality plans in a very short time.  However, as seen from
the figure, our merge algorithms are also able to achieve the same
level of savings.

\begin{figure*} [htb]
\begin{center}
\psfig{figure=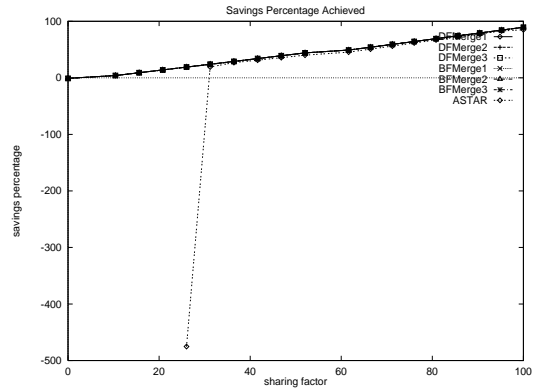,width=2.8in, height=2in}
\caption{Net savings achieved with Type 1 ccc Sets}
\label{exp1}
\end{center}
\end{figure*}

\begin{figure*} [htb]
\begin{center}
\psfig{figure=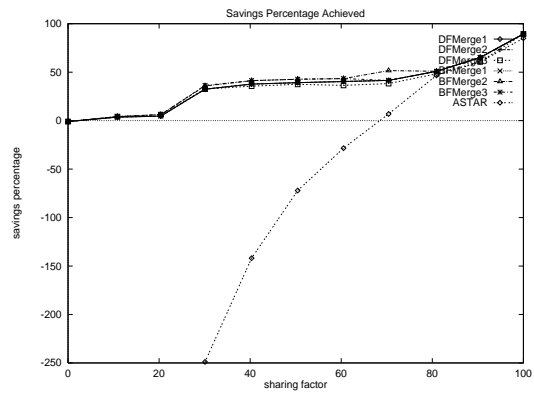,width=2.8in, height=2in}
\caption{Net savings achieved with Type 2 ccc Sets}
\label{exp2}
\end{center}
\end{figure*}

Figure \ref{exp2} shows the results when there are both equivalent and
implied ccc relationships. This time the heuristic function of the
$A^*$ algorithm is not as effective as with Type 1 ccc sets, and the
net savings it achieves are negative until very high sharing factors.
Although the $A^*$ algorithm finds low cost global execution plans,
the execution time of the algorithm is so high that the net savings are
negative.  Our merge algorithms achieve very good net savings
percentages. All the selection strategies perform almost equally well,
with \textbf{BFMerge3} performing slightly better.

Figure \ref{exp4} shows the net savings obtained when all three types
of relationships exist in the ccc sets. Note that the $A^*$ algorithm
only considers equivalent and implied relationships. The results are
very similar to the previous experiment. Again, our merge algorithms
perform much better than the $A^*$ algorithm. Our different select
strategies have similar performances, with \textbf{BFMerge3} performing the 
best.

As the $A^*$ algorithm was devised only for relational data sources,
we designed another experiment involving only relational data
sources. In this type of ccc sets, we only allowed equivalent
relationships, as the $A^*$ algorithm performed best with equivalent
ccc's. Figure \ref{exp3} shows the net savings achieved in this case.
As seen from the figure, our algorithms perform as well as the $A^*$
algorithm for sharing factor greater than 30\%, and better for the
rest.

\begin{figure*} [htb]
\begin{center}
\psfig{figure=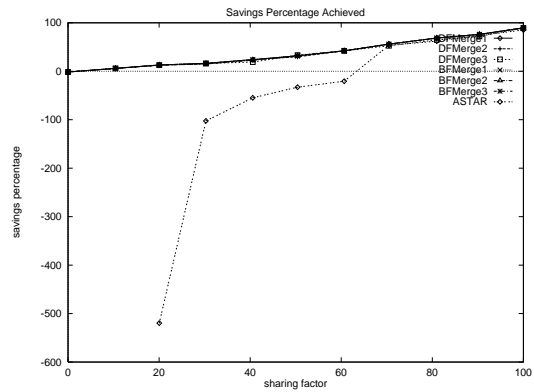,width=2.8in, height=2in}
\caption{Net savings achieved with Type 3 ccc Sets}
\label{exp4}
\end{center}
\end{figure*}

\begin{figure*} [htb]
\begin{center}
\psfig{figure=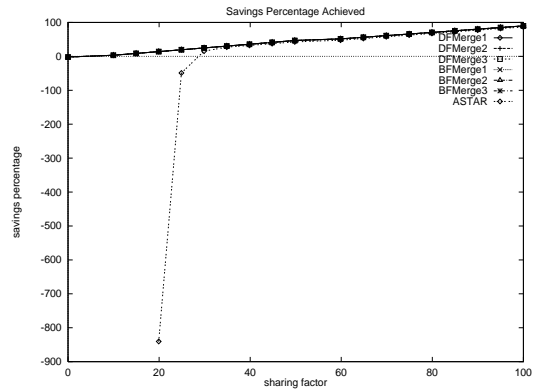,width=2.8in, height=2in}
\caption{Net savings achieved with relational sources}
\label{exp3}
\end{center}
\end{figure*}
 
These results suggest that although our algorithms explore a smaller
search space with respect to the $A^*$ algorithm, the savings we
obtain in practice are as good as that of the $A^*$ algorithm, and the high
execution cost of the $A^*$ algorithm is prohibitive.

\subsubsection{Execution Times of Merge Algorithms}

\begin{figure*} [htb]
\begin{center}
\psfig{figure=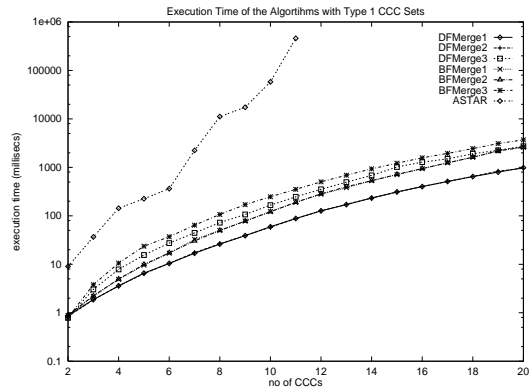,width=2.8in, height=2in}
\caption{Execution Time of Merge Algorithms with Type 1 ccc Sets}
\label{id}
\end{center}
\end{figure*}
 
\begin{figure*} [htb]
\begin{center}
\psfig{figure=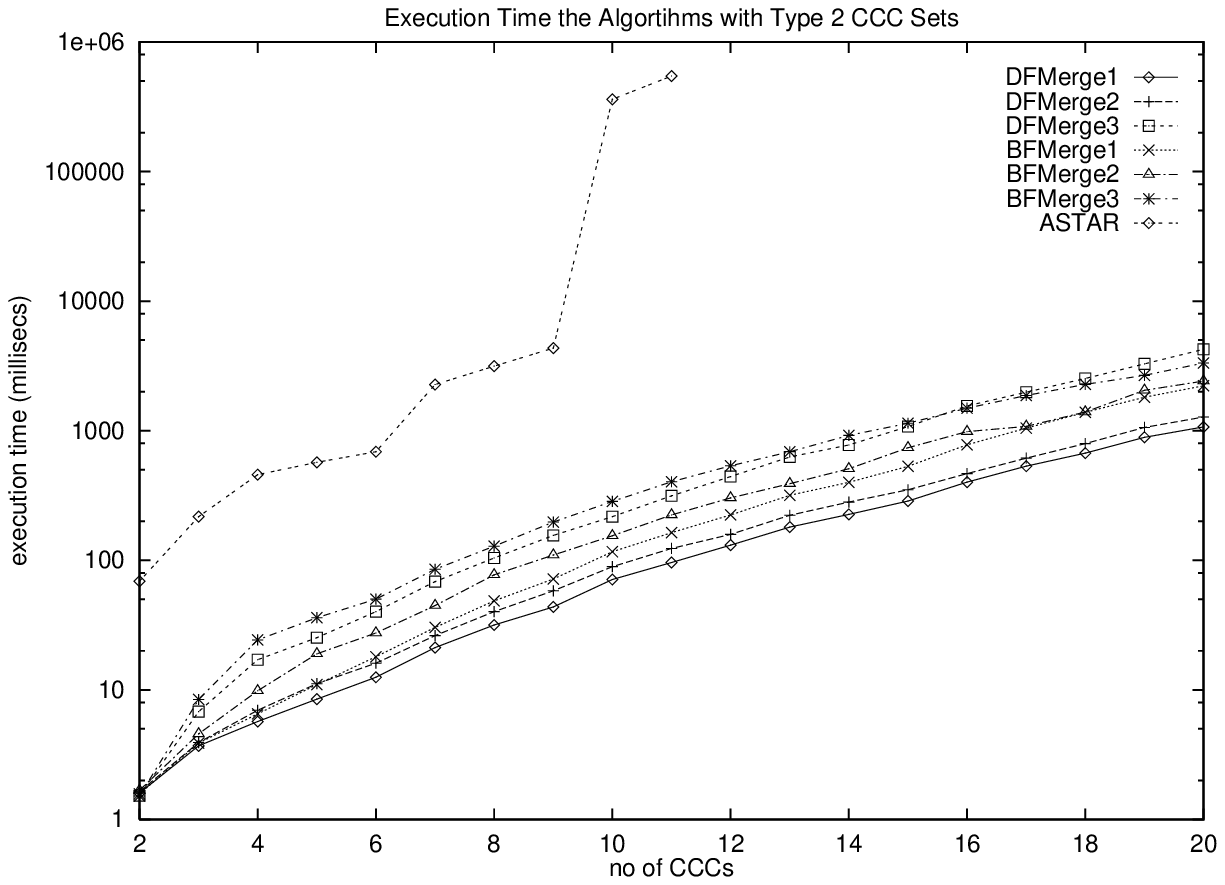,width=2.8in, height=2in}
\caption{Execution Time of Merge Algorithms with Type 2 ccc Sets}
\label{imp}
\end{center}
\end{figure*}

In these experiments, we studied the execution times of
our Merge algorithms and the $A^*$ algorithm.  Figures \ref{id},
\ref{imp} and \ref{ovp} show the execution times for Type 1, Type 2
and Type 3 ccc sets as the number of ccc's in the sets increases. \emph{
\textbf{Note
that the y-axes in the figures have logarithmic scale.}}  As seen from
the figures, the $A^*$ algorithm has double-exponential execution
time, and it cannot handle ccc sets having more than 10-11 ccc's, as it
exhausts memory. The results show that our algorithms run (1)
\emph{1300} to \emph{5290} times faster than the $A^*$ algorithm for
Type 1 ccc sets, (2) \emph{1360} to \emph{6280} times faster than the
$A^*$ algorithm for Type 2 ccc sets, and (3) \emph{100} to \emph{350}
times faster than the $A^*$ algorithm for Type 3 ccc sets.

The execution times of our Merge algorithms are exponential, but 
in the order of milliseconds, taking less than a second for even
20 ccc's. Among our algorithms, \textbf{BFMerge3} has the highest execution
time, as it uses an expensive heuristic and explores a relatively
larger search space than the \textbf{DFMerge} algorithms. \textbf{DFMerge3} 
has the next highest execution time, and \textbf{DFMerge1} has the lowest. 
One important observation is that although \textbf{BFMerge3} and 
\textbf{DFMerge3} use a relatively expensive and more
informed heuristic, and therefore have higher execution times,
and find better global execution
plans, they achieve the same level of net savings with the other
strategies. Hence, the increased cost induced by these two strategies
are not offset by the net savings they achieve.

\begin{figure*} [htb]
\begin{center}
\psfig{figure=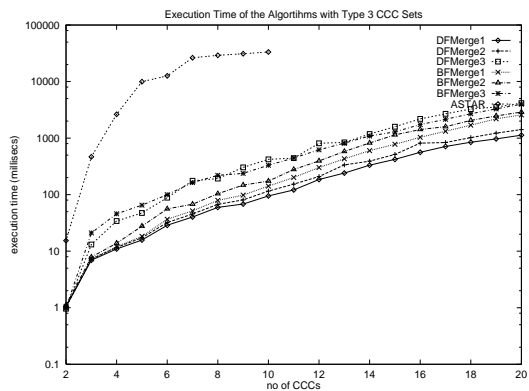,width=2.8in, height=2in}
\caption{Execution Time of \textbf{Merge} Algorithms with Type 3 ccc Sets}
\label{ovp}
\end{center}
\end{figure*}

 \begin{figure*} [htb]
\begin{center}
\psfig{figure=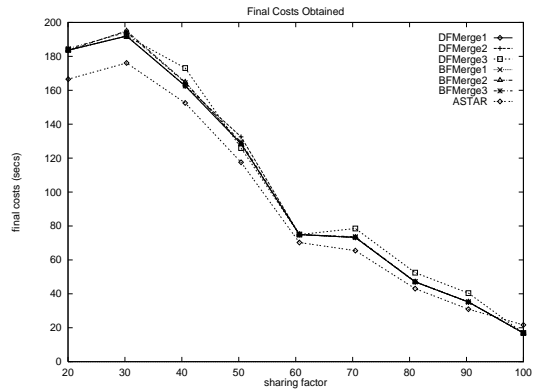,width=2.8in, height=2in}
\caption{Final Plan Costs Generated for Type 3 ccc Sets}
\label{final}
\end{center}
\end{figure*}

\subsubsection{Final Cost of Plans Generated by the \textbf{Merge} Algorithms}
 
As the $A^*$ algorithm examines an exhaustive search space, we 
studied the quality of plans generated by our \textbf{Merge}
algorithms and the $A^*$ algorithm to determine how suboptimal our
final plans are. For this purpose, we examined the final costs of
the plans for our Type 3 ccc sets.  Figure \ref{final} shows the
estimated execution costs of the final plans generated by the
\textbf{Merge} algorithms.  As seen from the figure, the $A^*$
algorithm almost always finds better plans than our algorithms.
However, the time it spends in finding those quality plans is not
offset by the net savings it achieves. Although our algorithms explore
only a restricted search space, the results show that they are able to
compute plans whose costs are \emph{at most 10}\% more than the plans
produced by the $A^*$ algorithm.  From these results, we can conclude
that our algorithms are both feasible and practical.

\section{Related Work}
\label{sec:related}
Our work has been influenced by and is related to various areas of research.
Over the last decade, there has been increasing interest in building
information agents that can access a set of diverse data sources.
These systems include \hermes~\cite{adali96},
\textsl{SchemaSQL}~\cite{laks-96, laks-99},
\tsimmis \cite{chaw-etal-94,moli-etal-97}, \sims~\cite{aren-etal-93},
\textsl{Information Manifold}~\cite{levy96, manifold},
The \textsl{Internet Softbot}
\cite{etzi-weld-94}, \textsl{InfoSleuth}~\cite{bayardo},
\textsl{Infomaster}~\cite{infomaster}, and \textsl{ARIADNE}~\cite{ariadne}.
Although all these systems provide mechanisms to optimize
individual requests, the only one which addresses
the problem of optimizing overall agent performance is
\textsl{ARIADNE}.

In \cite{ashish-98, ashish-etal-99}, the authors
propose techniques to selectively materialize data to improve the
performance of subsequent requests. They use the LOOM \cite{loom}
knowledge representation language for modeling data
and maintain an ontology of classes of information sources in LOOM.
They determine
what to materialize by examining previous user requests as follows.
They first look at the constraints imposed by user queries and create
subclasses in the ontology corresponding to these restrictions.
They then try to merge subclasses whenever possible. After all user queries
have been examined, they sort these subclasses according to the frequency
of requests and materialize subclasses from this list
until the space reserved for materialization is exhausted. They repeat
this process in fixed intervals. Their idea is similar to
previous semantic caching ideas \cite{as95,adali96,dar-96}.
In semantic caching, the cache is organized into semantic
regions instead of pages. When a new query arrives, the contents of the cache
is examined to determine what portion of the data requested in the
query is present in the cache. A query is then created to retrieve the
rest of the data from disk. The problem with semantic caching is that
containment checking is hard  and having a large number
of semantic regions creates performance problems.

Both \cite{ashish-etal-99} and \cite{dar-96} process
one query at a time, and try to reduce the execution time by using caches,
whereas we examine a set of requests (in our framework, agents
can be built on top of legacy software code bases such as PowerPoint,
Excel, route planners, etc. which may not support a database style
query language) and try to optimize
the overall execution time of this set of requests
by exploiting the commonalities between them.
Since we process a set of requests simultaneously, we
cache the results of a code call
condition evaluation only if another 
code call condition  in this set can
make use of the cached results. On the other hand, in \cite{ashish-etal-99}
caching decisions are based on user request histories.
The advantage of their approach is that they can make use
of the cache for a longer period of time, while in our case
the cache contents are valid during the execution of a
particular set of code call conditions.
When we process the next batch of code call conditions,
we discard the contents of the cache. On the other hand, the disadvantage
of history based caching is that it cannot rapidly adapt to changes
in interests. Nevertheless, we believe that incorporating more global
level caching techniques, like the ones in \cite{ashish-etal-99},
into our framework is a promising research area that is worth
pursuing. Another important
difference is that our results also include soundness
and completeness theorems.

The problem of simultaneously optimizing and merging a set of
queries has been studied within the context of relational and
deductive databases \cite{minker80, sellis88, ssn94, sellis90, fink82,
minker85}. \cite{minker80, sellis88, sellis90, ssn94} address
the problem of creating a globally optimal access plan for a set
of queries, provided that the common expressions among the queries
are given as input.
\cite{minker80} describe a branch-and-bound algorithm which searches
a state space in a depth-first manner to optimize a set of relational
expressions. Their algorithms are not cost-based, and hence
they may increase the total execution cost of the queries.  Moreover,
they only consider equivalence relationships, but not containment
relationships and they only deal with relational sources.
Furthermore, they do not deal with non database sources.

\cite{sellis88,ssn94,sellis90} propose exhaustive
algorithms to create a globally optimal execution plan for a set of
relational database queries. \cite{sellis90} show that the multiple-query
optimization (MQO) problem in relational databases is \NP-hard even when
only equivalence relationships are considered. Hence, exact algorithms for
MQO are not practical and therefore, approximations or heuristic
algorithms are worth pursuing.

\cite{sellis88} formulates the MQO problem as
a state search problem and uses the $A^*$ algorithm.
In their approach, a {\it state} is defined as an
n-tuple $\langle$ P$_{1{j_1}}$,P$_{2{j_2}}$,..P$_{n{j_n}} \rangle$, where
P$_{1{j_1}} \in \{NULL\} \cup P_i$ and $P_i$ is the set of possible access plans
for query $Q_i$. The initial state is the vector $\langle$ NULL, \ldots , NULL
$\rangle$, that is no access plan is chosen for any query.
A state transition chooses an access plan for the next query whose corresponding
access plan is NULL in the state vector.
The heuristic function proposed by \cite{sellis88} takes
only equivalence relationships into account.
\cite{ssn94} improves and extends this heuristic function by
incorporating implication relationships and by modifying the estimated costs.
This improved heuristic function provides a tighter bound
than the one proposed in \cite{sellis88}.

However, their approach requires enumeration
of all possible plans for each query, leading to a (theoretically) very large
search space. As a result, these algorithms have an exponential worst case
running time. Moreover, in a heterogeneous environment, it may not be
possible to assume that all query plans can be enumerated since
queries might have infinitely many access plans.
Furthermore, application program interfaces of
individual data sources and/or software packages may not
enumerate all such plans for requests shipped to them. This may be
because (i) their internal code does not
support it, or (ii) they are not willing to do so.

While \cite{sellis88,ssn94,sellis90} focus on only relational data sources,
we address the problem of optimizing a set of code call conditions
in agents which are built on top
of arbitrary data sources. For this purpose, we provide a
framework to define and identify
common subexpressions for arbitrary data sources. Moreover,
we do not need to enumerate all possible plans of a single query.
We have implemented an adapted version of the $A^*$ algorithm of
\cite{ssn94} and compared it with our merging algorithms. As the
results in Section \ref{sec:exps} show, our merging algorithms are
much faster than the $A^*$-based algorithm. As the $A^*$-based algorithm
examines a larger search space, it may find low-cost plans that our merging
algorithms may miss. However, the time it takes to find such good
plans is usually not offset by the savings it achieves.

\cite{fink82, minker85}, on the other hand, focus on
detecting common expressions among a set of queries in relational and
deductive databases. Since the notion of ``common subexpression''
varies for different data sources, the common expression
identification problem for agents is very different from those of
relational and deductive databases. Furthermore, they only consider
equivalence and containment relationships among queries when detecting common
subexpressions, whereas we also consider overlapping cases.

The only work that addresses heterogeneity and hence
is most closely related to ours is that of \cite{sv98}. 
The authors propose an architecture to process complex decision support
queries that access to a set of heterogeneous data sources. They
introduce {\it transient views}, which are materialized views that
exist during the execution of a query. \cite{sv98} describe algorithms
which analyze the query plan generated by an optimizer to identify
similar sub-plans, combine them into transient views and insert
filters for compensation. Moreover, \cite{sv98} presenst the implementation
of their algorithms within the context of \textsl{DataJoiner}'s
\cite{gupt-lin-94, datajoiner} query optimizer.
They try to optimize a complex decision support query by exploiting
common subexpressions within this single query, whereas we try to
simultaneously optimize a given set of requests. While
they examine relational-style operators in detecting common subexpressions,
we process any code call condition defined over arbitrary data sources not
just relational sources.
Moreover, they do not have a language
to describe equivalence and containment relationships for heterogeneous
data sources and hence these relationships are fixed apriori in
the optimizer code. On the other hand, we provide invariants to
describe relationships for heterogeneous data sources.
Our algorithms for merging multiple code call conditions take such
invariants and cost information into account when performing the merge.

Another area of research that is related to ours is partial evaluation
in logic programs \cite{leus-etal-98, lloy-shep-91, dech-etal-99}.
Partial evaluation takes a program and a goal and rewrites the program
by using a set of transformations to optimize its performance. The
rewritten program usually runs faster for the particular goal when SLD
or SLD-NF resolution is used for query processing. On the other hand,
our framework takes an agent program and a set of derived invariants,
and tries to optimize the agent program apriori, that is at
development time, prior to occurence of state changes.  An interesting
research problem in our framework may be the following: If a state
change can be encoded as a goal, then we can use partial evaluation
techniques to further optimize the rewritten agent program, as shown
in Figure~\ref{fig:pe}. We believe this problem needs further
attention and research.

\begin{figure*} [htb]
\begin{center}
\input{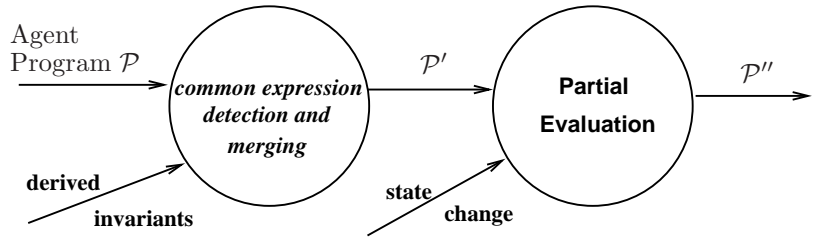}
\caption{Application of Partial Evaluation Techniques to Agent Programs}
\label{fig:pe}
\end{center}
\end{figure*}

Another area of research that is very closely related to ours is query
optimization in relational and deductive databases
\cite{Graefe,starburst,ioan-90,graefe:survey,ibaraki-84,kim-82,magic},
and in mediators \cite{adali96,manifold,garlic-97,ambite00,dgl-00}.
It is worth noting that this list is not exhaustive since over the
last decades, enormous effort has been devoted to the query
optimization problem.  Our work is orthogonal to techniques for
optimizing individual queries, as they can be incorporated into our
framework in numerous ways.  For example, individual requests might be
first optimized by using the techniques in \cite{manifold} or
\cite{ambite00}, then our techniques might be applied to the results.
However, our focus in this paper is on the simultaneous optimization
of a set of requests.

Finally, the problem of choosing appropriate materialized views to
answer queries is also related to our work and there exist
several papers in this area \cite{qian, lmss95,ckps95}.
\cite{lmss95} describes algorithms to determine
the portions of a query that can be expressed using the definitions of
materialized views. \cite{ckps95} identifies portions of a query that can be
answered using materialized views, and determine if it is efficient
to answer the query using the view. The focus of such techniques is
to efficiently compute the answers to a single query, whereas our focus
is to optimize the overall cost of a set of requests submitted to
a heavily loaded agent.

\section{Conclusion}
\label{sec:conc}

There is now an incredible increase in the amount of research being
conducted on software agents.  Software agents now provide a host of
web based services, ranging from creating personalized newspapers for
people, to building multimedia presentations.  In addition, agents for
corporate web sites often try to personalize the web site for a given
user by tracking histories of that user's interest.  Agents are also
being increasingly used in the aerospace and defense industries.

When an agent gets lots of requests within a short time frame,
the standard mechanism that most agent frameworks use is to queue
the requests in accordance with some queueing policy (e.g. LIFO, FIFO,
priority queue) and then service the requests one after the other.
This often leads to long wait times for requests that occur later on
in the queue.  
In this paper, we have shown how to improve the performance of an
agent by merging a set of requests and servicing the requests
simultaneously. We proposed a generic and customizable framework for
this purpose.  Our solution applies to agents that are built on top
of legacy code, which is certainly a practical assumption, as the success
of the agent endeavor rests on the ability to build on top of existing
data and software sources.
Our solution consists of two parts: 
\begin{enumerate}
\item[(1)] identifying ``commonalities'' among a set of code call
  conditions and
\item[(2)] computing a \emph{single} global execution plan that
  simultaneously optimizes the total expected cost of this set of code
  call conditions.
\end{enumerate}
We first provided a formal framework within which an agent developer
can specify what constitutes a ``common subexpression'' for a data
source via a set of structures, called \emph{invariants}.
\emph{Invariants} describe \emph{(1)} code call conditions that are
``equivalent'' to other code call conditions, \emph{(2)} code call
conditions that are ``contained'' in other code call conditions, and
\emph{(3)} code call conditions that overlap with other code call
conditions.  Moreover, such invariants may \emph{imply} other
invariants. We developed provably sound and complete algorithms to
take the initial set of invariants input by the developer and compute
\emph{all implied} invariants.

Second, we provided an architecture to merge multiple requests in
agents.  We provided algorithms to identify equivalent, implied and
overlapped code call conditions in any set ${\cal C}$. We then
proposed two heuristic based algorithms, \textbf{BFMerge} and
\textbf{DFMerge}, that take as input, the set of code call conditions,
and produce as output, a \emph{single} execution plan.  The merging
decisions are based on costs, hence the resulting global plan is
guaranteed to have a reduced cost.

We have experimentally shown that our algorithms achieve significant
savings. We have compared our merging algorithms with Sellis'
$A^*$-based algorithm (which applied to merging multiple requests
in the relational database case only)
and demonstrated that our algorithms almost
always outperform theirs.  We have shown that our merging algorithms
\emph{(1)} can handle \emph{more than twice} as many simultaneous code
call conditions as the $A^*$ algorithm and \emph{(2)} run \emph{100}
to \emph{6300} times faster than the $A^*$ algorithm and \emph{(3)}
produce execution plans the cost of which is \emph{at most 10\%} more
than the plans generated by the $A^*$ algorithm.

We conclude with a brief remark on an important piece of future work.
Eiter et. al. \cite{esr99} have developed a class of agents called
\emph{regular agents}.  In their framework, the semantics of an agent
is given by computing certain kinds of semantic constructs called
``status sets.'' When an agent experiences a state (which may occur, for
example, when it receives a message), the agent 
computes a new ``status set'' having some properties decribed in 
\cite{eite-etal-99a}.  This ``status set'' specifies what the agent
is supposed to do in order to respond to the state change. 
\cite{subetal99} shows that this framework is rich enough not only to
deal with reactive agent behavior \cite{kowa-sadri-99}, but also the so-called
autonomous agent behavior of the type described by Shoham \cite{shoh-93,shoh-99}.
Eiter et.~al.~\cite{esr99}'s regular agent framework reduce the problem
of computing ``status sets'' of regular agents to that of evaluating a
set of code call conditions.  The beauty of their result is that the
syntactic restrictions on regular agents makes it possible, to associate
with each agent, prior to deployment of the agent, a set of code call
conditions.  Whenever the agent needs to find a new status set in response
to a state change, it recomputes a new status set by evaluating this set of
code call conditions.  Hence, all the techniques described in this paper
may be used to optimize, once and for all, this set of code call conditions,
so that once the agent is deployed, this optimized set of code call
conditions is used by the agent for ``status set'' computations.  We
are pursuing this research avenue.

\bibliographystyle{esub2acm}
\bibliography{heavy}

\appendix

\section{Proofs of Theorems}
 \begin{proof}[of Theorem~\ref{minwitness}]\mbox{}\\
\begin{enumerate}
\item[($\Rightarrow$):] Suppose $\pi$ is a witness to the safety of $\chi$.  There
  are two cases:
\begin{enumerate}
\item[\emph{Case 1:}] Let $\chi_{\pi(i)}$ be an atomic code call
  condition of the form $\IN{\var{X_{\pi(i)}}}{cc_{\pi(i)}}$, then by
  the definition of safety, $\root{cc_{\pi(i)}} \subseteq RV_{\pi}(i)$, where
  $RV_{\pi}(i) =\{ \root{Y} \mid \exists j<i ~s.t. ~Y ~occurs ~in ~\chi_{\pi(j)}
  \}$, and either $\var{X_{\pi(i)}}$ is a root variable or
  $\root{X_{\pi(i)}} \in RV_{\pi}(i)$. Then, there exist $\chi_{\pi(j_1)}$,
  $\chi_{\pi(j_2)}$, $\ldots$, $\chi_{\pi(j_k)}$, $j_k <$ i , such that
  $\root{X_{\pi(j_k)}} \subseteq RV_{\pi}(i)$, and $\root{X_{\pi(j_k)}} \subseteq
  \root{cc_{\pi(i)}}$. But, then $\chi_{\pi(i)}$ is dependent on each of
  the $\chi_{\pi(j_1)}$, $\chi_{\pi(j_2)}$, $\ldots$, $\chi_{\pi(j_k)}$, $j_k <$ i
  by definition.  Hence, there exist edges \[(\chi_{\pi(j_1)},
  \chi_{\pi(i)}), (\chi_{\pi(j_2)}, \chi_{\pi(i)}), \ldots (\chi_{\pi(j_k)},
  \chi_{\pi(i)}).\] Therefore, $\chi_{\pi(j_1)}$, $\chi_{\pi(j_2)}$, $\ldots$,
  $\chi_{\pi(j_k)}$, $j_k <$ i precede $\chi_{\pi(i)}$, hence $\pi$ is also a
  topological sort of the cceg of $\chi$.
  
\item[\emph{Case 2:}] If $\chi_{\pi(i)}$ is an equality/inequality of the
  form \var{s_1} op \var{s_2}, then at least one of \var{s_1},
  \var{s_2} is a constant or a variable \var{S} such that $\root{S} \in
  RV_{\pi(i)}$.  Suppose at least one of \var{s_1}, \var{s_2 } is a
  variable. Then, there exists a $\chi_{\pi(j)}$, j $<$ i, such that
  $\root{S} \in \root{X_{\pi(j)}}$, as $root(X_{\pi(j)}) \subseteq RV_{\pi}(i)$.
  But, then $\chi_{\pi(i)}$ is dependent on $\chi_{\pi(j)}$ by definition,
  and there exists an edge $(\chi_{\pi(j)}, \chi_{\pi(i)})$ in the cceg of
  $\chi$. Hence, $\chi_{\pi(j)}$ precedes $\chi_{\pi(i)}$ in the topological
  sort of the cceg. If both \var{s_1} and \var{s_2} are constants,
  then their nodes have in-degree 0 in the cceg, and no code call
  condition needs to precede $\chi_{\pi(i)}$ in the topological sort
  order, i.e., they are unrestricted.  Therefore, $\pi$ is also a
  topological sort of the cceg of $\chi$.
\end{enumerate}
\item[($\Leftarrow$):] Suppose $\pi$ is a topological sort of the cceg of $\chi$.
  Let \[\chi_{\pi(i)}, \chi_{\pi(j_1)}, \chi_{\pi(j_2)}, \ldots, \chi_{\pi(j_k)}, \ 
  j_k < i\] be code call conditions such that there exist edges
  \[(\chi_{\pi(j_1)}, \chi_{\pi(i)}), (\chi_{\pi(j_2)}, \chi_{\pi(i)}), \ldots,
  (\chi_{\pi(j_k)}, \chi_{\pi(i)})\] in the cceg of $\chi$. Then, by
  definition each $\chi_{\pi(j_m)}$, $m=1,\ldots,k$, depends on $\chi_{\pi(i)}$.
  If $\chi_{\pi(i)}$ is an atomic code call condition of the form
  $\IN{\var{X_{\pi(i)}}}{cc_{\pi(i)}}$, then $\root{X_{\pi(j_m)}} \subseteq
  \root{cc_{\pi(i)}}, m = 1,\ldots, k$. As $\forall j_m, m=1,\ldots, k, j_m < i$,
  $\root{X_{\pi(j_m)}} \subseteq RV_{\pi(i)}$, by definition of $RV_{\pi(i)}$,
  hence $\root{cc_{\pi(i)}} \subseteq RV_{\pi(i)}$. On the other hand, if
  $\chi_{\pi(i)}$ is an equality/inequality of the form \var{s_1} op
  \var{s_2}, then either \var{s_1} is a variable and $\root{s_1} \in
  \root{X_{\pi(j_m)}}$, where $j_m \in \{j_1,\ldots, j_k\}$, or \var{s_2} is
  a variable and $\root{s_1} \in \root{X_{\pi(j_m')}}$, where $j_m' \in
  \{j_1,\ldots, j_k\}$, or both. But, $\root{X_{\pi(j_m)}} \subseteq RV_{\pi(i)}$
  $\forall j_m, m=1,\ldots, k, j_m < i$. Hence, $\root{s_1}, \root{s_2} \in
  \root{X_{\pi(j_m)}}$. If both \var{s_1} and \var{s_2} are constants,
  then they are unrestricted in the topological sort. Therefore, $\pi$
  is also a witness to the safety of $\chi$. \qed
\end{enumerate}
\end{proof}

 \begin{proof}[of Theorem~\ref{elimcondlists}]
The proof is by induction on the  structure of condition lists.
\begin{description}
\item[Base Cases:] Base cases are  when the condition list
consists of $t_1\: \op\: ~t_2$ where
  $\op \in \{ <, >, \leq, \geq, =  \}$ and each of $t_1$,
  $t_2$ is either a variable or a constant. We suppress the cases when
  both $t_1, t_2$ are constants: the relation either holds (in that
  case we an eliminate $t_1\: \op\: ~t_2$) or it does not (in that
  case we can eliminate the whole invariant).
\begin{description}
\item[$\mathbf{\op=\str{$\leq,\geq$}}$:] We have to consider  terms of the form 
$t_1 \leq  t_2$ (resp. $t_1 \geq  t_2$) and distinguish the following  cases.
For each case we define expressions $\ie_1', \ie_2'$ 
such that
$ \text{true}  \Longrightarrow ~~\ie_1' ~~\Re  ~~\ie_2'$ is equivalent to 
$t_1 \leq  t_2 \Longrightarrow ~~\ie_1 ~~\Re  ~~\ie_2$.
\begin{enumerate}
\item \emph{$t_2$ is a constant $a$:} Then $t_1$ is a variable. We
  modify $\ie_1,\ie_2$ by introducing a new variable $\var{X_{new}}$
  and adding the following ccc to all subexpressions of $\ie_1,\ie_2$
  containing $t_1$
\[\IN{t_1}{\cc{ag}{subtraction}{a,\var{X_{new}}}}\:\&\:
\IN{1}{\cc{ag}{geq\_0}{\var{X_{new}}}}. \] We note that $t_1$ now
becomes an auxiliary variable and $\var{X_{new}}$ is a base variable.

$\mathfrak{Trans}(\ic,\ie_i)$ is defined to be the modified 
$\ie_i$ just described.
   
\item \emph{$t_1$ is a constant $a$:} Then $t_2$ is a variable.  We
  modify $\ie_1,\ie_2$ by introducing a new variable $\var{X_{new}}$
  and adding the following ccc to all subexpressions of $\ie_1,\ie_2$
  containing $t_2$
\[\IN{t_2}{\cc{ag}{addition}{a,\var{X_{new}}}} \:\&\:
\IN{1}{\cc{ag}{geq\_0}{\var{X_{new}}}}. \] Again, $t_2$ becomes an
auxiliary variable and $\var{X_{new}}$ is a base variable.

$\mathfrak{Trans}(\ic,\ie_i)$ is defined to be the modified 
$\ie_i$ just described.

\item \emph{Both $t_1, t_2$ are variables:} We modify $\ie_1,\ie_2$ by
  introducing a new variable $\var{X_{new}}$ and adding the following
  ccc to all subexpressions of $\ie_1,\ie_2$ containing $t_2$
\[\IN{t_2}{\cc{ag}{addition}{\var{t_1},\var{X_{new}}}}\:\&\:
\IN{1}{\cc{ag}{geq\_0}{\var{X_{new}}}}. \]
Again, $t_2$  becomes an auxiliary variable and
$\var{X_{new}}$ is a base variable. 

$\mathfrak{Trans}(\ic,\ie_i)$ is defined to be the modified 
$\ie_i$ just described.
\end{enumerate}
The case $\geq$ is completely analogous: just switch $t_1$ with $t_2$.
Note that the above covers all possible cases, as any variable in the
condition list must be a base variable (see Definition~\ref{basevar}).
\item[$\mathbf{\op=\str{<,>}}$:] Analogous to the previous case,
  just replace ``$\cc{ag}{geq\_0}{\var{X_{new}}}$'' by 
``$\cc{ag}{ge\\_0}{\var{X_{new}}}$''
\item[$\mathbf{\op=\str{=}}$:]  If in $t_1 = t_2$ the term $t_1$
  is a variable, then we replace each occurrence of $t_1$ in
  $\ie_1,\ie_2$ by $t_2$. If $t_1$ is a constant and $t_2$ is a
  variable, replace each occurrence of $t_2$ in
  $\ie_1,\ie_2$ by $t_1$.
\end{description}
\item[Inductive Step:] As the condition list is just a conjunction of
  the cases mentioned above, we can apply our modifications of $\ie_1, 
  \ie_2$ one after another. Once all modifications have been
  performed, we arrive at an equivalent formula of the form
\[\text{true}  \Longrightarrow ~~\mathfrak{Trans}(\ic,\ie_1)
 ~~\Re  ~~\mathfrak{Trans}(\ic,\ie_2) \ \qed\]
\end{description}
\end{proof}

\begin{proof}[of Corollary~\ref{cor:trans}]\mbox{}\\
($\Rightarrow$) : Let $\iv : \ic \Longrightarrow ~~\ie_1 ~~\Re ~~\ie_2$ be an invariant. 
We can assume that \ic is in DNF:  $C_1 \lor C_2 \lor \ldots \lor C_m$. Thus we
can write   $\iv$ as follows: 
\[\{ C_i \Longrightarrow ~~\ie_1 ~~\Re ~~\ie_2 \mid 1 \leq i \leq m\} \]
Let $\iv\theta$ be any ground instance of $\iv$.
If $(S,\theta)\models \iv$, then either 
$(C_1 \lor C_2 \lor \ldots \lor C_m)\theta$ evaluates to false in state $S$, or
$(\ie_1)\theta ~~\Re ~~(\ie_2)\theta$ is true in $S$. Assume that
$(C_1 \lor C_2 \lor \ldots \lor C_m)\theta$ evaluates to false, then each 
$(C_i)\theta$ has to be false in $S$. Hence,
\[ (S, \theta) \models (C_i \Longrightarrow ~~\ie_1 ~~\Re ~~\ie_2) for 1 \leq i \leq m\]
Assume $(C_1 \lor C_2 \lor \ldots \lor C_m)\theta$ evaluates to true in $S$.
Then there exists at least one $(C_i)\theta$ that evaluates to true in state
$S$. Let $T = \{(C_j)\theta \mid 1 \leq j \leq m\}$ be the set of conjunctions 
that are true in $S$. As all other $(C_i)\theta ~\notin T$ evaluates to 
false, $(S, \theta) \models (C_i \Longrightarrow \ie_1 ~\Re ~\ie_2) ~for 
~1 \leq i \leq m, ~and ~(C_i)\theta ~\notin ~T$. But 
$(S, \theta) \models \iv$, hence
$(\ie_1)\theta ~\Re ~(\ie_2)\theta$ is true in $S$. As a result,
$(S, \theta) \models (C_j \Longrightarrow ~~\ie_1 ~~\Re ~~\ie_2) ~for
~1 \leq j \leq m ~and (C_j)\theta \in T$.

Since,
each $C_i \Longrightarrow ~\ie_1 ~\Re ~\ie_2$ is an ordinary invariant
the result follows from Theorem \ref{elimcondlists}.\\

($\Leftarrow$) : Assume that $(\forall C_i, 1 \leq i \leq m) ~~(S,\mathbf{\theta})\models
\text{true} \Longrightarrow \mathfrak{Trans}(C_i,\ie_1) ~~\Re ~\mathfrak{Trans}(C_i,
\ie_2)$ and suppose $(S, \theta) \models (C_1 \lor C_2 \lor \ldots \lor C_m)$.  Then by
Theorem \ref{elimcondlists}, $(\forall C_i, 1 \leq i \leq m)$ $(S,\theta)\models (C_i \Longrightarrow
\ie_1 ~\Re ~\ie_2)$. There exists at least one $(C_j)\theta$ which
evaluates to true in $S$. But then, $(\ie_1)\theta ~~\Re ~~(\ie_2)\theta$ is
true in state $S$.  Hence, $(S, \theta) \models \iv$. \qed
\end{proof}

 \begin{proof}[of Lemma~\ref{reduction}]
\[\begin{array}{c}
\chkimp(\ie_1,\ie_2)\\  \iff\\ \text{ for all states $S$ and all assignments $\mathbf{\theta}$:  $[\ie_1]_{S,\mathbf{\theta}} \subseteq [\ie_2]_{S,\mathbf{\theta}}$}\\
\iff \\
\text{for all states $S$ and all assignments $\mathbf{\theta}$: $\text{true}  \Longrightarrow \ie_1\mathbf{\theta} \subseteq \ie_2\mathbf{\theta}$}\\
\iff \\
 \chktaut(\text{true}
  \Longrightarrow \ie_1 \subseteq \ie_2).
\end{array}\]

\emph{(2)} follows from Theorem~\ref{elimcondlists} and Corollary
\ref{cor:trans}. Note that it also holds for invariants of the form
$\ic \Rightarrow \ie_1= \ie_2$ because they can be written as two separate
invariants: ``$\ic \Rightarrow \ie_1\subseteq \ie_2$'' and ``$\ic \Rightarrow \ie_1\supseteq \ie_2$''.
\emph{(3)} is immediate by the very definition. \qed
\end{proof}

 \begin{proof}[of Proposition~\ref{undec}]
   We show that the containment problem~\cite{ullman89} in the
   relational model of data is an instance of the problem of checking
   implication between invariant expressions. The results follow then
   from Lemma~\ref{reduction} and the fact, that the containment
   problem in relational databases is well known to be undecidable.

To be more precise, we use the results in \cite{calv-pods98}, where it
has been shown that in the relational model of data, the
\emph{containment of conjunctive queries containing inequalities} is
undecidable.  It remains to show that our implication check problem
between invariant expressions can be reduced to this problem.

Let \cc{relational}{query}{Q} be a code call that takes as input an
arbitrary set of subgoals corresponding to the conjunctive query $Q$
 and returns as output the result of
executing $Q$.

Let $Q_1$ and $Q_2$ be arbitrary conjunctive queries which may contain
inequalities.  we define
\[ \ie_1 = \cc{relational}{query}{Q_1}, \ \ \ie_2 =
\cc{relational}{query}{Q_2}.\] 
Then, clearly
\[\chkimp(\ie_1,\ie_2)= \true  \ \ \iff \ \  Q_1 \subseteq Q_2.\]
Hence the implication check problem
is also undecidable. \qed
\end{proof}

 \begin{proof}[of Proposition~\ref{conp}]
Clearly, by Lemma~\ref{reduction},  it suffices to prove the
proposition for \chkimp.

 For an invariant expression \ie, the set of all
  substitutions $\theta$ such that $\ie\theta$ is ground, is finite
(because of our assumption about finiteness of the domains of all datatypes).
  Thus, our atomic code call conditions $\IN{obj}{\cc{ag}{f}{args}}$
  can all be seen as propositional variables.  Therefore, using this
  restriction, we can view our formulae as \emph{propositional}
  formulae and a \emph{state} corresponds to a \emph{propositional
    valuation}.

  With this restriction, our problem is certainly in \coNP, because
  computing $[\ie]_{S,\mathbf{\theta}}$ is nothing but evaluating a
  propositional formula (the valuation corresponds to the state $S$).
  Thus ``$[\ie_1]_{S,\mathbf{\theta}} \subseteq [\ie_2]_{S,\mathbf{\theta}}$ for
    all $S$ and all assignments $\mathbf{\theta}$'' translates to checking
  whether a propositional formula is a tautology: a problem known to
  be in \coNP.
  
  To show completeness, we use the fact that checking whether $C$ is a
  logical consequence of $\{ C_2,\ldots ,C_n\}$ (where $C$ is an
  arbitrary clause and $\{ C_2,\ldots ,C_n\}$ an arbitrary consistent
  set of clauses) is well-known to be \coNP-complete.
  
  We prove our proposition by a polynomial reduction of
  \emph{implication between atomic invariant expressions} to this problem.

  Let $\ie$ be an atomic invariant expression, i.e.~an atomic code
  call condition: it takes as input, a set of clauses, and returns as
  output, all valuations that satisfy that set of clauses. Let
  $\text{ANS}(\ie(\{ C\}))$ denote the set of results of evaluating
  $\ie$ on $C$ with respect to a state $S$.  Then \[\text{ANS}(\ie(\{
  C\}))\subseteq \text{ANS}(\ie(\{ C_2,\ldots ,C_n\})) \ \iff \ \{C_2,\ldots ,C_n\} \models
  C.\] Hence, checking whether an arbitrary atomic invariant
  expression $\ie_1$ implies another atomic invariant expression
  $\ie_2$ is \coNP hard. \qed
\end{proof}

 
\begin{lemma}[Translation into predicate logic]\label{trans}
There is a translation \transz from simple invariants
$\textsf{INV}_{\text{simple}}$
into predicate logic
with equality such that the following holds
\[\cali \models \iv \text{\  \iff \ } \trans{\cali} \cup T_{ord} \models
\trans{\iv},\]
where $T_{ord}$ is the theory of strict total orders $<$ and $a\leq b$,
(resp.~$a\geq b$), is an abbreviation for ``$a<b \, \lor a=b$'', (resp.~``$a>b
\, \lor a=b$'').

Moreover, a simple invariant ``$\ic_1 \Longrightarrow ~\ie_1 \subseteq \ie_1'$''
is translated into a formula of the form
\[ \underline{\forall} \,(\ic_1  \to  \forall x (pred_{\langle d_1,f_1\rangle }(\ldots,x) \to 
(pred_{\langle d_2,f_2\rangle }(\ldots,x))) \]
where $\underline{\forall}$ denotes the universal closure with respect to
all remaining variables. This is a universally quantified formula.
 \end{lemma}
\begin{proof}
We translate  each simple invariant to a predicate logic
formula by induction on the structure of the invariant.

\begin{description}
\item[Code Calls:] For each $n$-ary code call $\cc{d}{f}{\ldots}$ we introduce a
  $(n+1)$-ary predicate $pred_{\langle d,f\rangle }(\ldots,\cdot )$. Note that we
  interpret $\cc{d}{f}{\ldots}$ as a set of elements. The additional
  argument is used for containment in this set.
\item[Atomic ccc's:] We then replace each simple invariant expression
\[\IN{X}{\cc{d_1}{f_1}{\ldots}}\subseteq \IN{Y}{\cc{d_2}{f_2}{\ldots}}\] by the
universal closure (with respect to all base variables) of the formula
\[\forall x (pred_{\langle d_1,f_1\rangle }(\ldots,x) \to pred_{\langle d_2,f_2\rangle }(\ldots,x))\]
\item[Simple Ordinary Invariants:] A simple ordinary invariant of 
the form 
\[ \ic_1 \Longrightarrow ~\ie_1 \subseteq \ie_1'\]
is translated into
\[ \underline{\forall} \,(\ic_1  \to  \forall x (pred_{\langle d_1,f_1\rangle }(\ldots,x) \to 
(pred_{\langle d_2,f_2\rangle }(\ldots,x))) \]
where $\underline{\forall}$ denotes the universal closure with respect to
all remaining variables.

\item[Simple Invariants:] A simple invariant of the form
\[ (C_1 \lor C_2 \lor \ldots C_m) \Longrightarrow ~\ie_1 \subseteq \ie_1'\]
is translated into the following $m$ statements ($1\leq i\leq m$)
\[  ~~\underline{\forall} \,(C_i  \to  \forall x (pred_{\langle d_1,f_1\rangle } (\ldots,x) \to (pred_{\langle d_2,f_2\rangle }(\ldots,x))) \]
where $\underline{\forall}$ denotes the universal closure with respect to
all remaining variables.
\end{description}

Note that according to the definition of a simple ordinary invariant and
according to the definition of a code call condition (in front of
Example~\ref{ex1}), $\ic_1$ and the $C_i$  are conjunctions
 of equalities $s=t$ and  
inequalities $s\leq t$,  $s\geq  t$, $s<t$, $s>t$  where $s,t$ are
real numbers or variables. 

The statement \[\cali \models \iv \text{\  \iff \ } \trans{\cali} \cup T_{ord} \models \trans{\iv}\]
is easily proved by structural induction on simple invariants and
condition lists.  
\end{proof}

 \begin{proof}[of Lemma~\ref{subsumption}]
We use the translation of Lemma~\ref{trans}.

The assumption $\not \models \iv_2$ expresses that there is a state $S_0$ and
a substitution $\mathbf{\theta}$ of the base variables in $\iv_2$ such
that $S_0 \models \ic_2\mathbf{\theta}$ and there is an object $a$ such that
$S_0\models pred_{\langle d_2,f_2\rangle }(\ldots, a)\mathbf{\theta}$ and $S_0\not \models pred_{\langle
  d_2',f_2'\rangle }(\ldots, a)\mathbf{\theta}$.

As $\iv_1$ entails $\iv_2$, $\iv_1$ is not satisfied by $S_0$. Thus
there is $\mathbf{\theta}'$ such that $S_0 \models \ic_1\mathbf{\theta}'$ and
there is an object $a'$ with $S_0\models pred_{\langle d_1,f_1\rangle }(\ldots,
a')\mathbf{\theta}$ and $S_0\not \models pred_{\langle d_1',f_1'\rangle }(\ldots,
a')\mathbf{\theta}$.

Now suppose $\langle d_1',f_1'\rangle\neq \langle d_2',f_2'\rangle$. Then we simply modify
the state $S_0$ (note a state is just a collection of ground code call
conditions) so that $S_0\models pred_{\langle d_1',f_1'\rangle }(\ldots, a')\mathbf{\theta}$.
We do this for all $\mathbf{\theta}'$ that are counterexamples to the
truth of $\iv_1$.  Because $\langle d_1',f_1'\rangle\neq \langle d_2',f_2'\rangle$, this
modification does not affect the truth of $S_0\models pred_{\langle d_2,f_2\rangle
  }(\ldots, a)\mathbf{\theta}$ and $S_0\not \models pred_{\langle d_2',f_2'\rangle }(\ldots,
a)\mathbf{\theta}$. But this is a contradiction to our assumption that
$\iv_1$ entails $\iv_2$.  Thus we have proved: $\langle d_1',f_1'\rangle= \langle
d_2',f_2'\rangle$.

Similarly, we can also modify $S_0$ by changing the extension of 
$pred_{\langle d_1,f_1\rangle }(\ldots, a')\mathbf{\theta}$ and guarantuee that $\iv_1$ holds 
in  $S_0$. So we also get a contradiction as long as  $\langle d_1,f_1\rangle\neq
\langle d_2,f_2\rangle$. Therefore we have proved that  $\langle d_1,f_1\rangle=
\langle d_2,f_2\rangle$.

Our second claim follows trivially from $\langle d_1',f_1'\rangle= \langle
d_2',f_2'\rangle$, and $\langle d_1,f_1\rangle= \langle d_2,f_2\rangle$. \qed
\end{proof}

 \begin{proof}[of Lemma~\ref{lemma1}]
  Let $\iv_1 : \ic_1 \Longrightarrow ~\ie_1 ~\Re_1 ~\ie_1'$ and
  $\iv_2 : \ic_2 \Longrightarrow ~\ie_2 ~\Re_2 ~\ie_2'$. Then by the
  computation performed by the \textbf{Combine\_1} algorithm, the
  derived invariant has the following form\[ \iv :
  \textbf{simplify}(\ic_1 \,\land \, \ic_2) \Longrightarrow ~\ie_1 ~\Re
  ~\ie_2',\] where $\Re$ is determined by Table~\ref{summary}.  If
  $\textbf{simplify}(\ic_1 \,\land \ic_2) = \emph{false}$ we are done.
  In this case, there is no state $S$ satisfying a ground instance of
  $\ic_1 \,\land \ic_2$. 

 We assume that we are given a state  $S$
   of the agent that satisfies $\iv_1$, $\iv_2$ and \cali. Let
  $\iv_1\Theta$ and $\iv_2\Theta$ be any ground instances of $\iv_1$ and
  $\iv_2$.  Then, either $\ic_1(\Theta)$ evaluates to false, or
  $\ie_1(\Theta) ~\Re_1 ~\ie_1'(\Theta)$ is true in $S$.  Similarly, either
  $\ic_2(\Theta)$ is false or $\ie_2(\Theta) ~\Re_2 ~\ie_2'(\Theta)$
  is true in $S$.

  If either $\ic_1(\Theta)$ or $\ic_2(\Theta)$ evaluates to false, then
  $(\ic_1 \,\land \ic_2)(\Theta)$ also evaluates to false, and $\iv$ is
  also satisfied. Let's assume both $\ic_1(\Theta)$ and $\ic_2(\Theta)$
  evaluate to true. Then so does $(\ic_1 \,\land \ic_2)(\Theta)$, and
  both $\ie_1(\Theta) ~\Re_1 ~\ie_1'(\Theta)$ and $\ie_2(\Theta) ~\Re_2 ~\ie_2'(\Theta)$
  are true in $S$, as $S$ satisfies both $\iv_1$ and $\iv_2$. If $\Re =
  \str{=}$, then both $\Re_1 = \str{=}$ and $\Re_2 = \str{=}$, 
$\ie_1' \to\ie_2$
  and $\ie_2 \to\ie_1'$ (in all states satisfying \cali and $\iv_1, \iv_2$). Then, we have $\ie_1
  = \ie_1' = \ie_2 = \ie_2'$, hence $\ie_1(\Theta) ~= ~\ie_2(\Theta)$ is true
  in $S$, and $\Theta$ satisfies $\iv$. If $\Re = \str{\ensuremath{\subseteq}}$, then $\ie_1' ~\to
  ~\ie_2$ (in all states satisfying \cali and $\iv_1, \iv_2$), and we have $\ie_1 ~\Re_1 ~\ie_1',~\ie_1'\subseteq ~\ie_2, ~\ie_2 ~\Re_2
  ~\ie_2'$, and $\ie_1(\Theta) ~\subseteq
  ~\ie_2(\Theta)$.  As $\iv$ is satisfied by any $S$ that also satisfies
  both $\iv_1$, $\iv_2$ and \cali, we have $\{\iv_1,\iv_2 \}\, \cup
  \cali  \models \iv$. \qed
\end{proof}

\begin{proof}[of Lemma~\ref{lemma1-3}]
Let $\iv_1 : \ic_1 \Longrightarrow ~\ie_1 ~\Re_1 ~\ie_1'$ and
  $\iv_2 : \ic_2 \Longrightarrow ~\ie_2 ~\Re_2 ~\ie_2'$. Then, either
  \textbf{Combine\_3} returns \NIL or the derived invariant
  has the following form\[ \iv :
  \textbf{simplify}(\ic_1 \, \lor \ic_2) \Longrightarrow ~\ie_1 ~\Re_1
  ~\ie_1'.\] In the latter case, $\ie_1 = \ie_2$, $\Re_1 = \Re_2$ and
   $\ie_1' = \ie_2'$ as implied by the \textbf{Combine\_3} algorithm.

  We assume that we are given a state  $S$ of the agent that satisfies
  both $\iv_1$ and $\iv_2$. Let $\iv_1\Theta$ and $\iv_2\Theta$ be any 
  ground instances of $\iv_1$ and $\iv_2$.  Then, either $\ic_1(\Theta)$ 
  evaluates to false, or $\ie_1(\Theta) ~\Re_1 ~\ie_1'(\Theta)$ is true in 
  $S$. Similarly, either $\ic_2(\Theta)$ is false or 
  $\ie_1(\Theta) ~\Re_1 ~\ie_1'(\Theta)$ is true in $S$. We have four 
  possible cases. 

\begin{enumerate}
\item[] \emph{Case 1:} Both $\ic_1(\Theta)$ and $\ic_2(\Theta)$ evaluate
to false. Then $(\ic_1 ~\lor ~\ic_2)$ also evaluates to false, and
$\iv$ is also satisfied.
\item[] \emph{Case 2:} $\ic_1(\Theta)$ evaluates to false and 
$\ic_2(\Theta)$ evaluates to true. Since $S \models \iv_2$,
$\ie_1(\Theta) ~\Re_1 ~\ie_1'(\Theta)$ is true in $S$. Then $S$ also
satisfies $\iv$.
\item[] \emph{Case 3:} $\ic_1(\Theta)$ evaluates to true and 
$\ic_2(\Theta)$ evaluates to false. In this case, 
$\ie_1(\Theta) ~\Re_1 ~\ie_1'(\Theta)$ is true in $S$, since
$S$ satisfies $\iv_1$. Hence $S$ also satisfies $\iv$.
\item[] \emph{Case 4:} Both $\ic_1(\Theta)$ and $\ic_2(\Theta)$ evaluate
to true. Again, since $S$ satisfies both $\iv_1$ and $\iv_2$,
$\ie_1(\Theta) ~\Re_1 ~\ie_1'(\Theta)$ is true in $S$ and $\iv$ is 
also satisfied. \qed
\end{enumerate}
\end{proof}

 \begin{proof}[of Proposition~\ref{theo3}]
  Suppose $X_1 \subseteq X_2$ and $\iv \in C_{\cali}(X_1)$.  We need to show
  that $\iv \in C_{\cali}(X_2)$. By definition of $C_{\cali}$, there
  are five possible cases:
\begin{enumerate}
\item[] \emph{Case 1:} $\iv \in \cali$, hence $\iv \in C_{\cali}(X_2)$
  by definition of $C_{\cali}$.
  
\item [] \emph{Case 2:} $\iv \in X_1$. As $X_1 \subseteq X_2$, $\iv \in X_2$.
  Hence, $\iv \in C_{\cali}(X_2)$ by definition of $C_{\cali}$.
  
\item[] \emph{Case 3:} $\iv$ = \textbf{Combine\_1}($\iv_1, \iv_2, \cali$) where $\iv_1,
  \iv_2 \in \cali \cup X_1$. But then $\iv_1, \iv_2 \in \cali \cup X_2$ as
  $X_1 \subseteq X_2$. Hence, $\iv \in C_{\cali}(X_2)$. 
\item[] \emph{Case 4:} $\iv$ = \textbf{Combine\_2}($\iv_1, \iv_2$) 
where $\iv_1, \iv_2 \in \cali \cup X_1$. But then 
$\iv_1, \iv_2 \in \cali \cup X_2$ as
  $X_1 \subseteq X_2$. Hence, $\iv \in C_{\cali}(X_2)$. 
\item[] \emph{Case 5:} $\iv$ = \textbf{Combine\_3}($\iv_1, \iv_2$) 
where $\iv_1, \iv_2 \in \cali \cup X_1$. But then $\iv_1, \iv_2 
\in \cali \cup X_2$ as $X_1 \subseteq X_2$. 
Hence, $\iv \in C_{\cali}(X_2)$.  \qed
\end{enumerate}
\end{proof}

\begin{proof}[of Lemma~\ref{theo5}]
  Let $\iv \in C_{\cali}(C_{\cali} \uparrow^\omega)$. We need to show $\iv \in
  C_{\cali} \uparrow^\omega$. By the definition of $C_{\cali}$, there are three
  possible cases:
\begin{enumerate}
\item[] \emph{Case 1:} $\iv \in \cali$, then $\iv \in C_{\cali} \uparrow^\omega$
  by the definition of $C_{\cali}$.
\item[] \emph{Case 2:} $\iv \in C_{\cali} \uparrow^\omega$, which is trivial.
\item[] \emph{Case 3:} $\iv$ = \textbf{Combine\_1}($\iv_1, \iv_2,\cali$) 
(or $\iv$ = \textbf{Combine\_2}($\iv_1, \iv_2$) or 
$\iv$ = \textbf{Combine\_3}($\iv_1, \iv_2$)) such that
  $\iv_1, \iv_2 \in \cali \cup (C_{\cali} \uparrow^\omega)$.  There exists a
  smallest integer $k_i$ (i=1,2) such that $\iv_i \in \cali \cup
  (C_{\cali} \uparrow^{k_i})$. Let $k := max(k_1, k_2)$. Then, $\iv_1, \iv_2 \in
  \cali \cup (C_{\cali} \uparrow^k)$. By definition of $C_{\cali}$ and as
  $\cali \subseteq C_{\cali} \uparrow^k$, $\iv \in C_{\cali} \uparrow^{(k+1)}$. Hence, $\iv
  \in C_{\cali} \uparrow^\omega$. \qed
\end{enumerate}
\end{proof}

 \begin{proof}[of Lemma~\ref{theo6}]
 Suppose $\iv \in C_{\cali} \uparrow^\omega$. Then, there exists a smallest
  integer $k$, such that $\iv \in C_{\cali} \uparrow^k$. The proof is by
  induction on $k$. Let the inductive hypothesis be defined as $\forall k^{\prime}:\,
  1\leq k^{\prime} \leq k$, if $\iv \in C_{\cali} \uparrow^k$, then $\cali \models \iv$.
\begin{description}
\item[Base Step:] $k=1$, $\iv \in C_{\cali} \uparrow^1$, then there
  are four possible cases: \emph{(1)} $\iv \in \cali$, hence $\cali \models
  \iv$, \emph{(2)} $\iv = \textbf{Combine\_1}(\iv_1, \iv_2, \cali)$ where
  $\iv_1, \iv_2 \in \cali$.  As $\iv_1, \iv_2 \in \cali$, $\cali \models
  \iv_1, \iv_2$. Then, by Lemma \ref{lemma1}, $\{\iv_1, \iv_2\} \models
  \iv$. Therefore, $\cali \models \iv$. \emph{(3)} $\iv$ = 
  \textbf{Combine\_2} ($iv_1, \iv_2$) where $\iv_1, \iv_2 \in \cali$.
  Since $\iv_1, \iv_2 \in \cali$, $\cali \models \iv_1, \iv_2$. Then, by 
  Lemma \ref{lemma1-2}, $\{\iv_1, \iv_2\} \models \iv$. Therefore, 
  $\cali \models \iv$. \emph{(4)} $\iv$ = \textbf{Combine\_3} 
  ($iv_1, \iv_2$) where $\iv_1, \iv_2 \in \cali$. As 
  $\iv_1, \iv_2 \in \cali$, $\cali \models \iv_1, \iv_2$. 
  Then, by Lemma \ref{lemma1-3}, $\{\iv_1, \iv_2\} \models
  \iv$. Therefore, $\cali \models \iv$. 
\item[Inductive Step:] $k > 1$. Let $\iv \in C_{\cali} \uparrow^k$.  Then,
  there exist $\iv_1, \iv_2 \in C_{\cali} \uparrow^ (k-1)$, such that
  $\iv$ is derived by one of \textbf{Combine\_1}, \textbf{Combine\_2} or
  \textbf{Combine\_3} operators. That is, either 
  $\iv=\textbf{Combine\_1}(\iv_1, \iv_2, \cali)$, or
  $\iv=\textbf{Combine\_2}(\iv_1, \iv_2)$, or
  $\iv=\textbf{Combine\_3}(\iv_1, \iv_2)$. Because this is the
  only possibility, as $\iv \notin C_{\cali} \uparrow^j, j < k$, by definition
  of $k$. By the inductive hypothesis $\cali \models \iv_1$ and $\cali \models
  \iv_2$.  By Lemma~\ref{lemma1}, $\{\iv_1, \iv_2\} \models \iv$. Hence,
  $\cali \models \iv$. \qed
\end{description}
\end{proof}
\begin{lemma}\label{predlemma} We consider predicate logic with equality and a binary
  predicate symbol $<$. The language also contains arbitrary constants 
  and parameters from the reals.

We consider a special class of formulae, namely 
universally quantified formulae of the form
$\ic \to P_i(\underline{t}) \to P_j(\underline{t'})$, where
$\underline{t}, \underline{t'}$ are tuples consisting of variables,
constants and parameters, $P_i$ are predicate symbols and \ic is an
invariant condition involving equality, $<$, variables, constants and
parameters. We call this class \iv-formulae. Let $T$ be a set of
\iv-formulae.

The proof system consisting of \textbf{(R0)}:
$\frac{\phi(\underline{x})}{\phi(\underline{x})\theta}$, where $\theta$ is any
substitution for the variables in the tupel $\underline{x} $ and $\phi$
is an \iv-formula, and the two inference rules \textbf{(R1)} and
\textbf{(R2)} below is complete for the class of \iv-formulae: For
each formula $\ic \to P_i(\underline{t}) \to P_j(\underline{t'})$ which
follows from $T$, there is an instance of a derived formula which is
identical to it. And each derived formula also follows from $T$.
\[\begin{array}{lll}
\textbf{(R1)} & 
\begin{array}{clc}
\ic_1 & \to &  P_1(\underline{t_1}) \to P_1'(\underline{t_1'})\\
\ic_2 & \to &  P_2(\underline{t_2}) \to P_2'(\underline{t_2'})\\ \hline
 \textbf{simplify}((\ic_1 \land \ic_2)\theta)  &  \to &  P_1(\underline{t_1})\theta \to P_2'(\underline{t_2'})\theta
\end{array} & \ \ \begin{array}{l}
              \text{where $\theta$ is such that}\\
             P_1'(\underline{t_1'})\theta=P_2(\underline{t_2})\theta\\
\end{array}
\end{array}\]
\[\begin{array}{lll}
\textbf{(R2)} & 
\begin{array}{clc}
\ic_1 & \to &  P_1(\underline{t_1}) \to P_1'(\underline{t_1'})\\
\ic_2 & \to &  P_2(\underline{t_2}) \to P_2'(\underline{t_2'})\\ \hline
 \textbf{simplify}((\ic_1 \lor \ic_2)\theta\gamma)  &  \to &
 P_1(\underline{t_1})\theta\gamma  \to P_1'(\underline{t_1'})\theta\gamma 
\end{array} & \ \ \begin{array}{l}
               \text{where }
  P_1(\underline{t_1})\theta=P_2(\underline{t_2})\theta,\\
  P_1' (\underline{t_1'})\gamma =P_2' (\underline{t_2'})\gamma\\
\end{array}
\end{array}\]
The \textbf{simplify} routine simplifies invariant conditions
(containing the binary symbol $<$) wrt.~the theory of real numbers in
the signature $<$, $=$, and arbitrary constants and parameters in the reals.
\end{lemma}
\begin{proof}
The correctness of the system is obvious, as all rules have this property. 

The completeness follows by adapting the classical completeness proof
of first-order logic and taking into account the special form of the
\iv-formulae.  Let
\[\varphi : \ \ic^*  \to   P^*(\underline{t^*}) \to P^{\prime*}(\underline{t^{\prime*}})\]
be a formula that follows from $T$ .
Then the set
\[T \cup \, \{\underline{\exists} \, (\ic^* \land P^*(\underline{t^*}) \land \lnot
P^{*\prime}(\underline{t^{*\prime}}))\}\]
is unsatisfiable (because  $T\models \varphi$). Therefore it suffices to show the 
following claim:
\begin{quote}
\emph{ Given a set $T\cup \{ \varphi  \}$ of
\iv-formulae, whenever $T\not \vdash \varphi $, then $T\cup \{ \lnot \varphi  \}$ is
satisfiable.}
\end{quote}
Because then the assumption $T\not \vdash  \varphi'$ leads to a contradiction.
Therefore, taking into account \textbf{(R0)}, we can conclude that
at least an \iv-formula of the form 
\[\varphi': \ \ic  \to   P(\underline{t}) \to P'(\underline{t'}),\]
with  $\varphi=\varphi'\theta$ must be derivable.

The claim can be shown by establishing that each consistent
set $T\cup \{ \lnot \varphi  \}$  containing \iv-formulae and their negations,
can be extended to a \emph{maximally consistent} set
$\Phi_T$ which \emph{contains witnesses}.\footnote{This is analogous to the classical Henkin proof of
the completeness of first-order logic. 
In our case the theory $T$ in
question contains only finitely many free variables which simplifies
the original proof.}  For such sets, the following holds: \emph{(1)}
$\phi_{T\cup \{\lnot \varphi \}}\vdash \gamma 
 $ implies $\gamma \varphi \in\phi_{T\cup \{\lnot \varphi \}}$, \emph{(2)} for all $\gamma $: $\gamma  \in\phi_{T\cup \{\lnot
   \varphi \}}$ or  $\lnot \gamma 
\not \in\phi_{T\cup \{\lnot \varphi \}}$, and \emph{(3)} $\exists x\gamma  \in\phi_{T\cup \{\lnot \varphi \}}$ implies that there is a
term $t$ with $\gamma [\frac{t}{x}]\in \phi_{T\cup \{\lnot \varphi \}}$. These
properties induce in a natural way an interpretation which is a model
of $\phi_{T\cup \{\lnot \varphi \}}$.
\end{proof}
 \begin{proof}[of Theorem~\ref{theo7}]
   The proof is by reducing the statement into predicate logic using
   Lemma~\ref{trans}. We are then in a situation to apply
   Lemma~\ref{predlemma}. Note that the inference rules of
   Lemma~\ref{predlemma} act on \iv-formulae exactly as
   \textbf{Combine1} and \textbf{Combine3} on invariants. Therefore
   there is a bijection between proofs in the proof system described
   in Lemma~\ref{predlemma} and derivations of invariants using
   \textbf{Combine1} and \textbf{Combine3}.  \qed

\end{proof}
 \begin{proof}[of Corollary~\ref{corol1}]
  We are reducing the statement to Theorem~\ref{theo7}. We transform
  each invariant with $\Re = \str{=}$, into two separate invariants
  with $\Re = \str{\ensuremath{\subseteq}}$.  
  
  If $\iv$ is of the form $\ic \Rightarrow \ie_1 \subseteq \ie_2$, we
  are done, because
\begin{enumerate}
\item the set of transformed invariants is equivalent to the original
  ones, and
\item although deriving invariants with $\Re = \str{=} $ is possible
  (such invariants are contained in the set \text{Taut} and new ones
  will be generated by \textbf{Combine\_1}\footnote{see the first line 
    in Table~\ref{summary}} and by \textbf{Combine\_3}), for all such invariants
  we have also both their $\subseteq$ counterparts (this can be easily 
  shown by induction).
\end{enumerate}

Let's suppose therefore that $\iv$ has the form $\ic \Rightarrow
\ie_1 = \ie_2$. We know that both $\ic \Rightarrow \ie_1 \subseteq
\ie_2$ and $\ic \Rightarrow \ie_1 \supseteq \ie_2$ are entailed by
\cali and we apply Theorem~\ref{theo7} to these cases. We can assume
wlog that none of these two invariants is a tautology (otherwise we
are done).

Thus there are $\iv^{\prime}$ (for $\subseteq$) and
$\iv^{\prime\prime}$ (for $\supseteq$). We apply
Lemma~\ref{subsumption} and get that $\iv^{\prime}$
(resp.~$\iv^{\prime\prime}$) has the form $\ic^{\prime} \Rightarrow
\ie_1 \subseteq \ie_2$ (resp.~$\ic^{\prime\prime} \Rightarrow \ie_1
\supseteq \ie_2$). By symmetry $\ic^{\prime}$ is equivalent (in fact, 
by using a deterministic strategy it can be made identical) to
$\ic^{\prime\prime}$. Thus by our \textbf{Combine\_2}, there is also
a derived invariant of the form
\[\ic^{\prime} \Rightarrow \ie_1 = \ie_2,\] 
and this derived invariant clearly entails $\ic \Rightarrow
\ie_1 = \ie_2$ (because $\iv^{\prime}$ entails $\ic \Rightarrow
\ie_1 \subseteq \ie_2$ and $\iv^{\prime\prime}$ entails $\ic \Rightarrow
\ie_1 \supseteq \ie_2$).
\end{proof}
 \begin{proof}[of Corollary~\ref{cor2}]\mbox{}\\
\begin{description}
\item[(1):]
The proof is by induction on the iteration of the
while loop in the \textbf{Compute-Derived-Invariants} algorithm. Let the
inductive hypothesis be $\forall i \geq 0$ ~if $\iv$ is inserted into $X$ in
iteration i, then $\cali \models \iv$.
\begin{enumerate}
\item[] \emph{Base Step:} For $i=0$, $\iv \in \cali$, $\iv \to \iv$, ~hence
  $\cali \models \iv$.
\item[] \emph{Inductive Step:} Let $\iv$ be inserted into $X$ in
  iteration $i > 0$, and $\iv$ = \textbf{Combine\_1}($\iv_1, \iv_2, \cali$), where $\iv_1$
  and $\iv_2$ are inserted into $X$ at step $(i-1)$ or earlier. Then, by
  the inductive hypothesis, $\cali \models \iv_1$ and $\cali \models \iv_2$.  By
  Lemma~\ref{lemma1}, $\{\iv_1,\iv_2\} \models \iv$, hence $\cali \models \iv$.
\end{enumerate}
\item[(2):] First note that the \textbf{Compute-Derived-Invariants}
  algorithm computes and returns $C_{\cali} \uparrow^\omega$.  The result
  follows from Theorem~\ref{theo7} and Corollary~\ref{corol1}.  \qed
\end{description}
\end{proof}

\section{Axiomatic Inference System}
\label{rules}

\alg{{\small
\begin{tabular}{@{}p{3.4in}|p{2in}} 
\begin{tabular}{p{3.3in}}
\textbf{Equivalence Rules} \\  \\
$A \cup A = A \cap A = A$ \\ 
$A \cup \emptyset = A$ ~~and ~$A \cap \emptyset = \emptyset$ \\ 
$(A \cup B) \cup C = A \cup (B \cup C)$ ~~and
~$(A \cap B) \cap C = A \cap (B \cap C)$ \\ 
$A \cup B = B \cup A$ ~~and ~$A \cap B = B \cap A$ \ \\ 
$A \cup (B \cap C) = (A \cup B) \cap (A \cup C)$   \\ 
$A \cap (B \cup C) = (A \cap B) \cup (A \cap C)$ \\ 
$\lnot(\lnot(A)) = A$ \\ 
$\lnot(A \cup B) = \lnot A \cap \lnot B$ ~~and
~$\lnot{(A \cap B)} = \lnot{A} \cup \lnot{B}$ \\ 
$A \cup (A \cap B) = A$ ~~and
~$A \cap (A \cup B) = A$  \\ 
\end{tabular}
&
\begin{tabular}{p{3.2in}}
\textbf{Inference Rules} \\  \\
$A \subseteq A$ \\ 
$(A \cap B) \subseteq A$ \\ 
$A \subseteq (A \cup B)$ \\ 
$((A \cup B) \cap \lnot B) \subseteq A$  \\ 
$ A \subseteq ((A \cap B) \cup \lnot B)$ \\ 
\begin{tabular}{@{}l@{}}
if $A \subseteq B$ ~and $B \subseteq C$\\
then $A \subseteq C$
\end{tabular} \\ 
\begin{tabular}{@{}l@{}}
if $A \subseteq B$ ~and $C \subseteq D$\\
then $(A \cap C) \subseteq 
(B \cap D)$ \end{tabular}\\ 
\begin{tabular}{@{}l@{}}if $A \subseteq B$ ~and $C \subseteq D$\\
then
$(A \cap C) \subseteq (B \cup D)$  \end{tabular}\\ 
\end{tabular}
\end{tabular}
}}

\section{Invariants For the Spatial and Relational Domains}
\label{invs}

\begin{table}[hbt]
{\small
\begin{center}
\begin{tabular}{||l||} \hline \hline
\begin{tabular}{@{}c@{}}
\var{T}=\var{T'} $\,\land$ \var{L}=\var{L'} $\,\land$ \var{R}=\var{R'} $\Longrightarrow$ \\
\IN{Y}{\cc{spatial}{vertical}{R,L,R}} =
\IN{Y}{\cc{spatial}{vertical}{T',L',R'}}
 \end{tabular} \\ \hline 
\begin{tabular}{@{}c@{}}
 \var{T}=\var{T'} $\,\land$ \var{L}=\var{L'} $\,\land$ \var{R}$<$\var{R'}  $\Longrightarrow$\\
\IN{Y}{\cc{spatial}{vertical}{R,L,R}} $\subseteq$ \IN{Y}{\cc{spatial}{vertical}{T',L',R'}} \end{tabular} \\ \hline
\begin{tabular}{@{}c@{}}
\var{T}=\var{T'} $\,\land$ \var{R}=\var{R'} $\,\land$ L$>$L' $\Longrightarrow$\\
\IN{Y}{\cc{spatial}{vertical}{R,L,R}} $\subseteq$ \IN{Y}{\cc{spatial}{vertical}{T',L',R'}} \end{tabular} \\ \hline
\begin{tabular}{@{}c@{}}
\var{T}=\var{T'} $\,\land$ R$<$R' $\,\land$ L$>$L'  $\Longrightarrow$\\
\IN{Y}{\cc{spatial}{vertical}{T,L,R}} $\subseteq$ \IN{Y}{\cc{spatial}{vertical}{T',L',R'}} \end{tabular} \\ \hline
\begin{tabular}{@{}c@{}}
\var{T}=\var{T'} $\,\land$ \var{B}=\var{B'} $\,\land$ \var{U}=\var{U'} $\Longrightarrow$\\
\IN{Y}{\cc{spatial}{vertical}{T,B,U}}  =
\IN{Y}{\cc{spatial}{vertical}{T',B',U'}} \end{tabular} \\ \hline  
\begin{tabular}{@{}c@{}}
\var{T}=\var{T'} $\,\land$ \var{B}=\var{B'} $\,\land$ \var{U}$<$\var{U'}  $\Longrightarrow$\\
\IN{Y}{\cc{spatial}{vertical}{T',B',U'}} $\subseteq$ \IN{Y}{\cc{spatial}{vertical}{T',B',U'}}  \end{tabular}\\ \hline
\begin{tabular}{@{}c@{}}
\var{T}=\var{T'} $\,\land$ \var{U}=\var{U'} $\,\land$ \var{B}$>$\var{B'}  $\Longrightarrow$\\
\IN{Y}{\cc{spatial}{vertical}{T,B,U}} $\subseteq$ 
\IN{Y}{\cc{spatial}{vertical}{T',B',U'}}  \end{tabular} \\ \hline
\begin{tabular}{@{}c@{}}
\var{T}=\var{T'} $\,\land$ \var{U}$<$\var{U'} $\,\land$ \var{B}$>$\var{B'} $\Longrightarrow$\\
\IN{Y}{\cc{spatial}{vertical}{T,B,U}} $\subseteq$ 
\IN{Y}{\cc{spatial}{vertical}{T',B',U'}}  \end{tabular} \\ \hline 
\begin{tabular}{@{}c@{}}
\var{T}=\var{T'} $\,\land$ \var{X}=\var{X'} $\,\land$ \var{Y}=\var{Y'} $\,\land$ \var{Rad}=\var{Rad'} $\Longrightarrow$\\
\IN{Z}{\cc{spatial}{range}{T,X,Y,Rad}} =
\IN{W}{\cc{spatial}{range}{T',X',Y',Rad'}}  \end{tabular} \\ \hline
\begin{tabular}{@{}c@{}}
\var{T}=\var{T'} $\,\land$ \var{X}=\var{X'} $\,\land$ \var{Y}=\var{Y'} $\,\land$ \hbox{\var{Rad} $<$ \var{Rad'}} 
$\Longrightarrow$\\ \IN{Z}{\cc{spatial}{range}{T,X,Y,Rad}} $\subseteq$ 
\IN{W}{\cc{spatial}{range}{T',X',Y',Rad'}}  \end{tabular} \\ \hline 
\end{tabular}
\caption{Invariants for the spatial domain (1)}
\end{center}
}
\end{table}

\begin{table}[hb]
{\small
\begin{center}
\begin{tabular}{||l||} \hline \hline
\begin{tabular}{@{}c@{}}
\var{T}=\var{T'} $\,\land$ \var{R}$\leq$\var{R'} $\,\land$ \var{L}$\leq$\var{L'} $\,\land$ \var{L'}$\leq$\var{R} $\Longrightarrow$ \\[.2cm]
\IN{Y}{\cc{spatial}{vertical}{T,L,R}} $\cup$  \IN{Y}{\cc{spatial}{vertical}{T',L',R'}}\\ =\\ \IN{Y}{\cc{spatial}{vertical}{T,L,R'}}  \end{tabular}\\ \hline
\begin{tabular}{@{}c@{}}
\var{T}=\var{T'} $\,\land$ \var{R}$\geq$\var{R'} $\,\land$ \var{L}$\geq$\var{L'} $\,\land$ \var{L}$\leq$\var{R'} $\Longrightarrow$ \\[.2cm]
\IN{Y}{\cc{spatial}{vertical}{R,L,R}} $\cup$
\IN{Y}{\cc{spatial}{vertical}{T',L',R'}}\\ =\\
\IN{Y}{\cc{spatial}{vertical}{T,L',R}}  \end{tabular} \\ \hline 
\begin{tabular}{@{}c@{}}
\var{T}=\var{T'} $\,\land$ U$\leq$U' $\,\land$ B$\leq$B' $\,\land$ B'$\leq$U $\Longrightarrow$ \\[.2cm]
\IN{Y}{\cc{spatial}{horizontal}{T,B,U}} $\cup$  \IN{Y}{\cc{spatial}{horizontal}{T',B',U'}}\\ =\\ \IN{Y}{\cc{spatial}{horizontal}{T',B,U'}} \end{tabular} \\ \hline
\begin{tabular}{@{}c@{}}
\var{T}=\var{T'} $\,\land$ \var{U}$\geq$\var{U'} $\,\land$ \var{B}$\geq$\var{B'} $\,\land$ \var{B}$\leq$\var{U'} $\Longrightarrow$ \\[.2cm]
\IN{Y}{\cc{spatial}{horizontal}{T,B,U}} $\cup$ \IN{Y}{\cc{spatial}{horizontal}{T',B',U'}}\\ =\\ \IN{Y}{\cc{spatial}{horizontal}{T',B',U}} \end{tabular} \\ \hline 
\end{tabular}
\caption{Invariants for the spatial domain (2)}
\end{center}
}
\end{table}

\begin{table}[hb]
{\small
\begin{center}
\begin{tabular}{||@{\,}l@{\,}||} \hline \hline
\begin{tabular}{@{}c@{}}
\var{Rel}=\var{Rel'} $\land$ \var{Attr}=\var{Attr'} $\land$ \var{Op}=\var{Op'} $\land$ \var{V}=\var{V'}
$\Longrightarrow$\\ \IN{X}{\cc{rel}{select}{Rel,Attr,Op,V}} = 
\IN{Y}{\cc{rel}{select}{Rel',Attr',Op',V'}} \end{tabular}  \\ \hline
\begin{tabular}{@{}c@{}}
\var{Rel}=\var{Rel'} $\land$ \var{Attr}=\var{Attr'} $\land$ \var{Op}=\var{Op'}=\str{\ensuremath{\leq}} $\land$ V$<$V'
$\Longrightarrow$\\ \IN{X}{\cc{rel}{select}{Rel,Attr,Op,V}} $\subseteq$
\IN{Y}{\cc{rel}{select}{Rel',Attr',Op',V'}} \end{tabular} \\ \hline
\begin{tabular}{@{}c@{}}
\var{Rel}=\var{Rel'} $\land$ \var{Attr}=\var{Attr'} $\land$ \var{Op}=\var{Op'}=\str{\ensuremath{\geq}} $\land$ V$>$V'
$\Longrightarrow$\\ \IN{X}{\cc{rel}{select}{Rel,Attr,Op,V}} $\subseteq$
\IN{Y}{\cc{rel}{select}{Rel',Attr',Op',V'}} \end{tabular} \\ \hline
\begin{tabular}{@{}c@{}}
\var{Rel}=\var{Rel'} $\land$ \var{Attr}=\var{Attr'} $\land$ \var{V1}=\var{V1'} $\land$ \var{V2}=\var{V2'}
$\Longrightarrow$\\ \IN{X}{\cc{rel}{rngselect}{Rel,Attr,V1,V2}} =
\IN{Y}{\cc{rel}{rngselect}{Rel',Attr',V1',V2'}} \end{tabular} \\ \hline
\begin{tabular}{@{}c@{}}
\var{Rel}=\var{Rel'} $\land$ \var{Attr}=\var{Attr'} $\land$
V1$\geq$\var{V1'} $\land$ \var{V2}$\leq$\var{V2'}
$\Longrightarrow$ \\ \IN{X}{\cc{rel}{rngselect}{Rel,Attr,V1,V2}}  $\subseteq$
\IN{Y}{\cc{rel}{rngselect}{Rel',Attr',V1',V2'}} \end{tabular} \\ \hline
\begin{tabular}{@{}c@{}}
\var{Rel}=\var{Rel'} $\land$ \var{Attr}=\var{Attr'} $\land$
V1$\leq$\var{V1'} $\land$ \var{V2}$\leq$\var{V2'}
$\land$ \var{V1'}$\leq$ V2 $\Longrightarrow$ \\[.2cm]
\IN{X}{\cc{rel}{rngselect}{Rel,Attr,V1,V2}}
$\cup$ \IN{Y}{\cc{rel}{rngselect}{Rel',Attr',V1',V2'}}\\ =\\
\IN{Z}{\cc{rel}{rngselect}{Rel,Attr,V1,V2'}} \end{tabular} \\ \hline
\begin{tabular}{@{}c@{}}
\var{Rel}=\var{Rel'} $\land$ \var{Attr}=\var{Attr'} $\land$ V1$\geq$\var{V1'} $\land$ \var{V2}$\geq$\var{V2'}
$\land$ V1$\leq$\var{V2'} $\Longrightarrow$ \\[.2cm]
\IN{X}{\cc{rel}{rngselect}{Rel,Attr,V1,V2}} $\cup$
\IN{Y}{\cc{rel}{rngselect}{Rel',Attr',V1',V2'}}\\
 =\\
\IN{Z}{\cc{rel}{rngselect}{Rel,Attr,V1',V2}} \end{tabular} \\ \hline 
\hline 
\end{tabular}
\caption{Invariants for the relational domain (1)}
\end{center}
}
\end{table}

\begin{table}[htb]
{\small
\begin{center}
\begin{tabular}{||l||} \hline \hline
\begin{tabular}{@{}c@{}}
\var{Rel}=\var{Rel'} $\land$ \var{Attr1}=\var{Attr1'} $\land$
\var{Attr2}=\var{Attr2'} $\land$ \var{V1'}=\var{V1}  $\land$ \\
 \var{V2}=\var{V2'} $\land$ \var{V3'}=\var{V3} $\land$ \var{V4}=\var{V4'}
$\Longrightarrow$\\[.2cm] \IN{X}{\cc{rel}{rngselect}{Rel,Attr1,V1,V2}} $\cap$
\IN{Y}{\cc{rel}{rngselect}{Rel,Attr2,V3,V4}}\\ =\\
\IN{Z}{\cc{rel}{rngselect}{Rel',Attr1',V1',V2'}} $\cap$
\IN{W}{\cc{rel}{rngselect}{Rel,Attr2,V3',V4'}} \end{tabular} \\ \hline 
\begin{tabular}{@{}c@{}}\var{Rel}=\var{Rel'} $\land$ \var{Attr1}=\var{Attr1'} 
$\land$ \var{Attr2}=\var{Attr2'} $\land$ \var{V1'}$\leq$\var{V1} \\
$\land$ \var{V2}$\leq$\var{V2'} $\land$ \var{V3'}$\leq$\var{V3} $\land$ 
\var{V4}$\leq$\var{V4'}
$\Longrightarrow$\\[.2cm] \IN{X}{\cc{rel}{rngselect}{Rel,Attr1,V1,V2}} $\cap$
\IN{Y}{\cc{rel}{rngselect}{Rel,Attr2,V3,V4}}\\ $\subseteq$\\
\IN{Z}{\cc{rel}{rngselect}{Rel',Attr1',V1',V2'}} $\cap$
\IN{W}{\cc{rel}{rngselect}{Rel,Attr2,V3',V4'}} \end{tabular}\\ \hline 
\begin{tabular}{@{}c@{}}
\var{Rel}=\var{Rel'} $\land$ \var{Attr1}=\var{Attr1'} $\land$ 
\var{Attr2}=\var{Attr2'} $\land$ \var{V1'}$\leq$\var{V1} $\land$\\
\var{V2'}$\leq$\var{V2} $\land$ \var{V3}$\leq$\var{V3'} $\land$
 \var{V4}$\leq$\var{V4'} $\land$ \var{V1}$\leq$\var{V2'} $\land$
\var{V3'}$\leq$\var{V4} $\Longrightarrow$ \\[.2cm]
(\IN{X}{\cc{rel}{rngselect}{Rel,Attr1,V1,V2}} $\cap$
\IN{X}{\cc{rel}{rngselect}{Rel,Attr2,V3,V4}}) $\cup$\\
(\IN{X}{\cc{rel}{rngselect}{Rel',Attr1',V1',V2'}} $\cap$
\IN{X}{\cc{rel}{rngselect}{Rel,Attr2,V3',V4'}})\\
 $\subseteq$\\
\IN{X}{\cc{rel}{rngselect}{Rel,Attr1,V1',V2}} $\cap$
\IN{X}{\cc{rel}{rngselect}{Rel,Attr2,V3,V4'}}  \end{tabular} \\ \hline
\begin{tabular}{@{}c@{}}
\var{Rel}=\var{Rel'} $\land$ \var{Attr1}=\var{Attr1'} $\land$ \var{Attr2}=\var{Attr2'} $\land$ \var{V1}$\leq$\var{V1'} \\
$\land$  \var{V2}$\leq$\var{V2'} $\land$ \var{V3}$\leq$\var{V3'} $\land$ 
\var{V4}$\leq$\var{V4'} $\land$ \var{V1'}$\leq$\var{V2} $\land$ \var{V3'}$\leq$\var{V4} $\Longrightarrow$ \\[.2cm]
(\IN{X}{\cc{rel}{rngselect}{Attr1,V1,V2}} $\cap$
\IN{Y}{\cc{rel}{rngselect}{Rel,Attr2,V3,V4}}) $\cup$\\
(\IN{Z}{\cc{rel}{rngselect}{Rel',Attr1',V1',V2'}} $\cap$
\IN{W}{\cc{rel}{rngselect}{Rel,Attr2,V3',V4'}})\\  $\subseteq$\\
\IN{X'}{\cc{rel}{rngselect}{Rel,Attr1,V1,V2'}} $\cap$
\IN{Y'}{\cc{rel}{rngselect}{Rel,Attr2,V3,V4'}}  \end{tabular} \\ \hline
\begin{tabular}{@{}c@{}}
\var{Rel}=\var{Rel'} $\land$ \var{Attr1}=\var{Attr1'} $\land$ \var{Attr2}=\var{Attr2'} $\land$ V1$\leq$\var{V1'} $\land$ \\
 \var{V2}$\leq$\var{V2'} $\land$
\var{V3'}$\leq$\var{V3} $\land$ \var{V4'}$\leq$\var{V4}
$\land$ \var{V1'}$\leq$V2 $\land$ \var{V3}$\leq$\var{V4'}
$\Longrightarrow$\\[.2cm]
(\IN{X}{\cc{rel}{rngselect}{Rel,Attr1,V1,V2}} $\cap$
\IN{Y}{\cc{rel}{rngselect}{Rel,Attr2,V3,V4}}) $\cup$\\
(\IN{Z}{\cc{rel}{rngselect}{Rel',Attr1',V1',V2'}} $\cap$
\IN{W}{\cc{rel}{rngselect}{Rel,Attr2,V3',V4'}}) \\ $\subseteq$\\
\IN{X'}{\cc{rel}{rngselect}{Rel,Attr1,V1,V2'}} $\cap$
\IN{Y'}{\cc{rel}{rngselect}{Rel,Attr2,V3',V4}}  \end{tabular} \\ \hline
\begin{tabular}{@{}c@{}}
 \var{Rel}=\var{Rel'} $\land$ \var{Attr1}=\var{Attr1'} $\land$
 \var{Attr2}=\var{Attr2'} $\land$ \var{V1'}$\leq$\var{V1} $\land$ \\
 \var{V2'}$\leq$\var{V2} $\land$ \var{V3'}$\leq$\var{V3} $\land$
 \var{V4'}$\leq$\var{V4} $\land$ \var{V1}$\leq$\var{V2'} $\land$
 \var{V3}$\leq$\var{V4'}
$\Longrightarrow$\\[.2cm]
(\IN{X}{\cc{rel}{rngselect}{Rel,Attr1,V1,V2}}  $\cap$
\IN{Y}{\cc{rel}{rngselect}{Rel,Attr2,V3,V4}}) $\cup$\\
(\IN{Z}{\cc{rel}{rngselect}{Rel',Attr1',V1',V2'}} $\cap$
\IN{W}{\cc{rel}{rngselect}{Rel,Attr2,V3',V4'}})\\ $\subseteq$ \\
\IN{X'}{\cc{rel}{rngselect}{Rel,Attr1,V1',V2}} $\cap$
\IN{Y'}{\cc{rel}{rngselect}{Rel,Attr2,V3',V4}}  \end{tabular}
 \\ \hline \hline
\end{tabular}
\caption{Invariants for the relational domain (2)}
\end{center}
}
\end{table}

\end{document}
